\documentclass[amsmath,amssymb]{revtex4}

\makeatletter
\def\p@section{}
\def\p@subsection{}
\def\p@subsubsection{}
\makeatother

\usepackage{graphicx}
\usepackage{dcolumn}
\usepackage{bm}
\usepackage[loose]{units}
\usepackage{epsfig}
\usepackage{xcolor}
\usepackage{tikz}

\newcommand{\ie}{\emph{i.e.}}

\newcommand{\revision}[1]{\textcolor{black}{#1}}
\newcounter{revisioncommentno}
\newcommand{\SecondRevision}[1]{\textcolor{black}{#1}}
\newcounter{SecondRevisioncommentno}
\newcommand{\revisionB}[1]{\textcolor{black}{#1}}
\newcounter{revisionBcommentno}
\newcommand{\revisionAB}[1]{\textcolor{black}{#1}}
\newcounter{revisionABcommentno}
\begin{document}

\preprint{}

\title{A Moving Boundary Flux Stabilization Method for Cartesian Cut-Cell Grids using Directional
Operator Splitting}

\author{W.P.~Bennett$^1$} \email{wpb22@cam.ac.uk} \thanks{Corresponding Author}
\author{N.~Nikiforakis$^1$} \email{nn10005@cam.ac.uk}
\author{R.~Klein$^2$} \email{rupert.klein@math.fu-berlin.de}
\affiliation{$^1$University of Cambridge,\\ Laboratory for Scientific
  Computing, Cavendish Laboratory,\\ Cambridge CB3 0HE, UK\\ $^2$Freie
Universit\"{a}t Berlin, Institut f\"{u}r Mathematik, Arnimallee 6, 14195 Berlin }

\begin{abstract}
An explicit moving boundary method for the numerical solution of time-dependent hyperbolic 
conservation laws on grids produced by the intersection of complex geometries with a regular Cartesian 
grid is presented. As it employs directional operator splitting, implementation of the scheme is 
rather straightforward. Extending the method for static walls from Klein et al., Phil.\ Trans.\ 
Roy.\ Soc., A {\bfseries 367}, no.\ 1907, 4559--4575 (2009), the scheme calculates fluxes needed 
for a conservative update of the near-wall cut-cells as linear combinations of ``standard fluxes''
from a one-dimensional extended stencil. Here the standard fluxes are those obtained without regard to 
the small sub-cell problem, and the linear combination weights involve detailed information regarding 
the cut-cell geometry. This linear combination of standard fluxes stabilizes the updates such that the 
time-step yielding marginal stability for arbitrarily small cut-cells is of the same order as that for 
regular cells. Moreover, it renders the approach compatible with a wide range of existing numerical 
flux-approximation methods. The scheme is extended here to time dependent rigid boundaries by 
reformulating the linear combination weights of the stabilizing flux stencil to account for the
time dependence of cut-cell volume and interface area fractions. The two-dimensional tests discussed
include advection in a channel oriented at an oblique angle to the Cartesian computational mesh, cylinders
with circular and triangular cross-section passing through a stationary shock wave, a piston moving through an open-ended shock tube, and the flow
around an oscillating NACA 0012 aerofoil profile.\\ 


\end{abstract}

\maketitle

\section{INTRODUCTION}

Embedded boundary (or immersed boundary) methods comprise a wide range of techniques for discretising 
complex geometries `embedded' into a simple Cartesian mesh. Cartesian mesh-based solvers can yield 
significant advantages over body-fitted and unstructured mesh methods in terms of mesh generation 
simplicity, reduction in cell skewness around complex geometries, automated mesh generation and 
improved parallelisation. This is particularly true when considering the movement of solid boundaries 
over time, where non-Cartesian mesh-based alternatives can require expensive regeneration of the mesh 
at each time step. 

\SecondRevision{The field of immersed boundary methods in 
general is large. An exhaustive review of this field is beyond the scope of the current work. We therefore confine ourselves here to a brief overview of the field, 
identifying some of the main approaches and their recent variants in order to highlight the broad diversity of methods available and to provide a wider 
context within which to place our cut-cell method.
We justify the brevity of this overview by noting that comprehensive reviews of immersed boundary methods are available by 
Mittal \& Iaccarino~\cite{mittal2005immersed}, Sotiropoulos \& Yang~\cite{Sotiropoulos20141} and recently by Maxey~\cite{maxey2017simulation}.} 

\SecondRevision{Immersed boundary methods can be categorised using a variety of taxonomies. Here we adopt the approach used by 
Sotiropoulos \& Yang~\cite{Sotiropoulos20141}, broadly dividing the field into `diffuse methods' and `sharp interface methods'. 
Diffuse methods are characterised by a smearing of the immersed boundary over a small number of computational cells. Sharp 
interface methods explicitly account for the interface, thereby retaining the immersed boundary as a sharp interface 
throughout the computation with no spatial smearing.}

\SecondRevision{Many diffuse immersed boundary methods used in recent literature employ artificial volume forces distributed
over a zone of cells close to the interface to effectively represent the pressure force exerted by 
the immersed boundary. This stress can be distributed to the surrounding fluid using, for example,  a dirac delta 
function, a distributed function or a forcing term. Early examples can be 
found in~\cite{peskin1972flow, peskin2002immersed, goldstein1993modeling, tyagi2005large}.
The advantage of this approach is that the wall geometry appears only in the distributed momentum 
forces, while the surface geometry does not otherwise affect the scheme. In particular, intersections
of the wall geometry with computational cells do not have to be accounted for explicitly. 
Recent developments using this approach include particle flow applications~\cite{Wang2008283, luo2017direct} 
and heat transfer applications~\cite{Ren2013694, Xia2014302}.}

\SecondRevision{Another popular class of diffuse boundary methods are the ``fictitious domain'' methods. In this approach, 
both fluid and solid regions are treated mathematically as a fluid, with either a material-dependent rigidity constraint, as in 
Glowinski et al.~\cite{Glowinski1994283}, or the fluid and solid regions are treated as a porous medium, using the Navier Stokes/Brinkmann 
equations with a permeability parameter-based forcing term, as in Angot et al.~\cite{Angot1999} and 
Khadra et al.~\cite{khadra2000fictitious}. Recent, notable developments to the fictitious domain 
approach include Randrianarivelo et al.~\cite{randrianarivelo2005numerical}, Angot~\cite{angot2010fictitious}, 
Angot et al.~\cite{angot2012fast} and Ducassou et al.~\cite{ducassou2017fictitious}, for a range of 
flow types, including homogenous, dilatable, non-homogenous, multi-phase and free surface flows.}
 
\SecondRevision{The diffuse methods considered so far can be further categorised as ``Continuous Forcing Approach'' (CFA) methods, 
in which the forcing terms (e.g., dirac delta function or distributed function) are applied to the continuous form of the equations. 
The ``Discrete Forcing Approach'' (DFA), on the other hand, applies forcing terms after discretisation of the governing equations. 
This approach, introduced by Mohd-Yusof~\cite{mohd1997simulations}, and developed by many researchers, 
including Fadlun et al.~\cite{fadlun2000combined}, Uhlmann~\cite{uhlmann2005immersed} and 
Breugem~\cite{breugem2012second} is documented to have better stability characteristics due to the absence of 
user specified forcing parameters.}

\SecondRevision{Although recent work by Qiu et al.~\cite{qiu2016boundary} has sought to extend the range of diffuse immersed rigid body methods 
into the compressible flow regime (in addition to a recent review of general diffuse-interface capturing methods for compressible two-phase flows by Saurel \& Pantano~\cite{saurel2018diffuse}), 
the majority of diffuse methods presented in the available literature for capturing rigid body flows are applicable to incompressible flow. 
For the compressible moving rigid body flows considered in the current work, an established alternative to the diffuse 
CFA and DFA methods is the ``sharp interface'' approach. This class of methods is therefore considered in the remainder of this section.}

``Ghost-fluid methods'' \cite{fedkiw1999non, fedkiw2002coupling} belong to the class
of sharp interface schemes. That is, the sub-cell geometries of Cartesian grid cells cut by the
boundary are explicitly accounted for, and no artificial smearing of the solution occurs near the
boundaries. These schemes work by first extrapolating the solution at a given time step beyond 
the actual computational domain into a halo of ``ghost cells'' in the immediate vicinity of the
boundary, and then employing standard flux balances for complete cells to advance the solution. 
The wall geometry is re-inserted into the updated full grid cells after a completed time step. 
A disadvantage of ghost fluid methods is that they are generally not conservative with respect to 
individual cut-cells, such that mass, momentum, and energy losses across a rigid wall may occur. 
\SecondRevision{Subsequent modifications to the ghost-fluid method of Fedkiw et al.~\cite{fedkiw1999non} and 
Fedkiw~\cite{fedkiw2002coupling} have sought to improve accuracy and 
robustness~\cite{terashima2009front, liu2003ghost} and to extend the multi-material properties 
of this approach~\cite{kaboudian2015ghost, kaboudian2014ghost, feng2017simulation}.}

Cut-cell methods provide a conservative alternative, where Cartesian cells underlying the solid 
boundary are each divided into constituent components. For flows involving a solid geometry in a 
surrounding fluid, the Cartesian cell underlying the boundary is divided into fluid and solid 
components. The boundary is therefore simply `cut out' of the surrounding Cartesian mesh. Boundary 
conditions are subsequently imposed along the intersecting surface. Dividing the Cartesian cell in 
this manner does, however, generate small sub-cells at the boundary which  may feature 
arbitrarily small volume and highly distorted geometry. These `small cells', which may be orders of 
magnitude smaller than the surrounding Cartesian cells, create issues of stability and efficiency for 
explicit time integration methods. This is termed the ``small cell problem''. 
A brief outline of existing conservative cut-cell approaches can now be described.

Flux redistribution schemes \cite{Pember95, Colella06, HuEtAl2006, schneiders2013accurate} 
are conservative and obtain stable updates for small cut-cells at time steps comparable 
to those achievable on a regular grid in two steps. First, cut-cell updates are determined from 
standard flux balances that are locally computed and disregard the small sub-cell 
problem. In the second step, a suitable fraction of these updates is redistributed to a stencil of 
neighbouring cells. Colella~\cite{Colella1990} argues that handling the time updates for cut-cells 
emerging from the intersection of a regular Cartesian grid with a complex geometry would necessitate 
``unsplit'' numerical schemes, \ie, schemes that avoid directional operator splitting, and accordingly 
introduced an unsplit Godunov-type higher-order scheme in this paper. Yet, in \cite{Klein09} we 
have demonstrated that stable individual updates for small cut-cells can be formulated in an operator 
splitting scheme, and we show here that the same methodology can be extended to moving walls.


``Cell merging'' methods~\cite{Clarke85, Quirk94, Berger03, Ingram03, Xu97, yang1997cartesian, barton2011conservative}, avoid the small sub-cell problem by
creating larger effective control volumes of size comparable to that of a regular cell through 
merging cut-cells with neighbouring larger cells and then computing common updates for the 
merged cells. There is some arbitrariness in these schemes as to which cells and sub-cells are to
be merged to create the enlarged effective control volumes, and a potential loss of resolution 
of the near-wall flow states, both of which we intend to avoid. 
\revision{Notably, Falcovitz et al.~\cite{falcovitz1997two} designed a technique based on directional 
operator splitting, and this is conceptually closest to the present approach.  
It uses a cell merging for cut-cells with volume fractions less than a user defined cell `partial length' to maintain stability in small cells. 
\revisionB{The cell merging approach has been shown to evoke spurious oscillations when applied to moving boundaries, as shown in for example, Schneiders et al.~\cite{schneiders2013accurate}. 
This motivates the use of alternative approaches, such as the one discussed in this article.}}

An approach that extends the merged-cell technique to obtain individual updates for small
cut-cells has been proposed by Hartmann et al.~\cite{HartmannEtAl2011}. In their scheme, a stable 
update for a merged cell is computed first, and then a multi-dimensional reconstruction algorithm 
is invoked to calculate a conservative improved local update for the cut-cell portion of the merged 
cell. It does not appear obvious how their scheme could be formulated in the context of directional 
operator splitting as is the goal of the present work.

The ``H-Box method'' by Berger, Helzel et al.~\cite{Berger03, Helzel05} is a locally second order 
accurate method that achieves stability for small cut-cells by resorting to LeVeque's wave propagation scheme, 
\cite{LeVequeBook2002}, which avoids flux balancing for control volumes but rather obtains
updates by superimposing wave-type increments that emerge from (linearized) Riemann problems 
solved at grid cell interfaces. Using Roe's approach for the formulation of an approximate
Riemann problem, \cite{Roe1981,LeVequeBook2002}, such wave superposition schemes can be formulated 
such that the resulting scheme conserves the primary conserved quantities, \ie, mass, momentum, and 
energy for gas dynamics. Through this construction, the proposed embedded boundary scheme is inherently 
tied to the wave propagation approach, however, and is not flexible with respect to the use of a 
user's flux approximation of choice. Also, the distribution of wave portions in the vicinity of cut 
cells involves complex geometrical computations. Thus, even though a substantially simpler 
variant of the original scheme was proposed in \cite{BergerHelzel2012}, implementation of this 
approach in three space dimensions remains rather tedious. 
\revision{Recently, Mudalidharan \& Menon~\cite{muralidharan2016high} describe a high order 
(third order inviscid fluxes and fourth order viscous fluxes) conservative finite 
volume cut-cell method that uses a high order reconstruction using a cell-centre piece-wise polynomial 
approximation. A cell merging technique is used for small cut-cells. Mudalidharan \& Menon~\cite{muralidharan2016high} 
attempt to overcome numerical oscillations at the cut-cell boundary caused by the irregular stencil used in the 
k-exact CENO reconstruction by applying the piece-wise polynomial reconstruction to clusters of cells. 
\SecondRevision{This static method has recently been extended to reactive flows over moving interfaces by Muralidhran \& Menon~\cite{muralidharan2018simulation}.} 
Other recent progress in the development of incompressible higher order cut-cell methods includes Krause \& Kummer~\cite{krause2017incompressible} 
in which a Discontinuous Galerkin (DG) discretization approach is developed based on the extended 
Discontinuous Galerkin (XDG) approach of Kummer~\cite{kummer2017extended}. This method is simply mentioned 
here for completeness, as we limit the scope of this literature review to compressible finite volume methods.}

The challenges involved in introducing a cut-cell technique are exacerbated when a moving boundary is 
required. In addition to problems caused by continuous reduction in the volume fraction of a single 
cut-cell over time, the characteristics of a cut-cell (and the neighbouring regular cells) generally 
alter over time. This can result in an individual cell changing from a cut-cell to a solid cell over a 
single time step, a regular cell changing to a cut-cell, or a large cut-cell (volume fraction $>0.5$) 
changing to a small cut-cell (or the opposite of all three scenarios). This has significant 
implications for any of the embedded boundary techniques summarized above. In addition, any cut-cell 
method development must satisfy the consistency condition, whereby a solid body moving at the same 
relative velocity as the surrounding fluid should produce an exact solution. Published literature on 
cut-cell methods for moving boundary problems include 
cell merging~\cite{yang1997cartesian, barton2011conservative} methods, implicit 
time-stepping~\cite{aftosmis2000parallel, murman2003implicit} and flux redistribution 
methods~\cite{MeinkeEtAl2013, schneiders2013accurate}, 
\SecondRevision{as well as recent developments by Muralidhran \& Menon~\cite{muralidharan2018simulation}, Lin et al.~\cite{lin2017simulation}, 
Deng \& li~\cite{deng2018simulating} and Patel \& Lakdawala~\cite{patel2018dual}.}

We describe an alternative technique for the solution of moving boundary flows using a 
dimension-splitting cut-cell approach, derived from the static wall method introduced by 
Klein et al.~\cite{Klein09}. Our scheme calculates fluxes needed for a conservative update of the 
near-wall cut-cells as linear combinations of ``standard fluxes'' from a one-dimensional extended 
stencil. Here the standard fluxes are those obtained without regard to the small sub-cell problem, and 
the linear combination weights involve detailed information regarding the cut-cell geometry. This 
linear combination of standard fluxes stabilizes the updates such that the time-step yielding marginal 
stability for arbitrarily small cut-cells is of the same order as that for regular cells. Moreover, it 
renders the approach compatible with a wide range of existing numerical flux-approximation methods. 
The scheme is extended here to time dependent rigid boundaries by reformulating the linear combination 
weights of the stabilizing flux stencil to account for the time dependence of cut-cell volume and 
interface area fractions. 

The basic scheme from \cite{Klein09} and the extension to moving walls are described in 
section~\ref{sec:NumericalMethod}, while results for a number of smooth and discontinuous 
flow test cases are presented in section~\ref{sec:NumericalResults}. Section~\ref{sec:Conclusions}
provides conclusions and an outlook to future work.


\section{NUMERICAL METHOD}
\label{sec:NumericalMethod}


\subsection{Finite Volume Discretisation}
\label{sec:FiniteVolume}

\begin{figure}
\begin{centering}
(a) \includegraphics[width=6cm, clip]{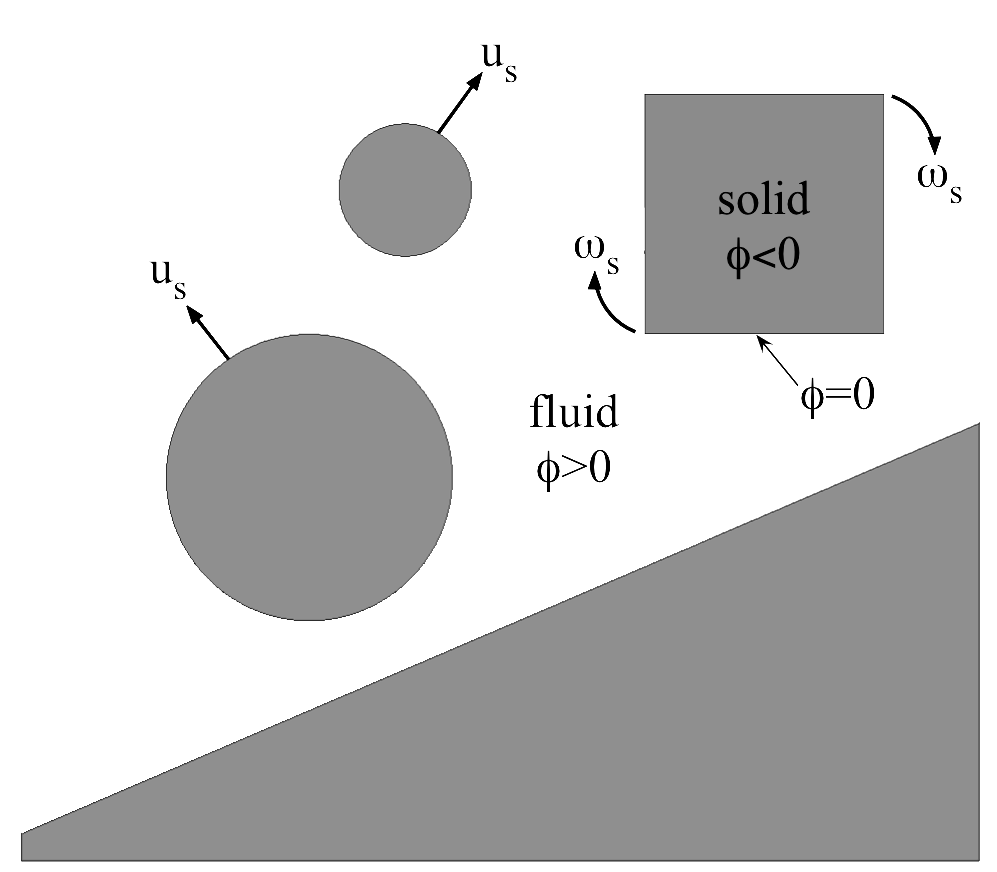}
(b) \includegraphics[width=6cm, clip]{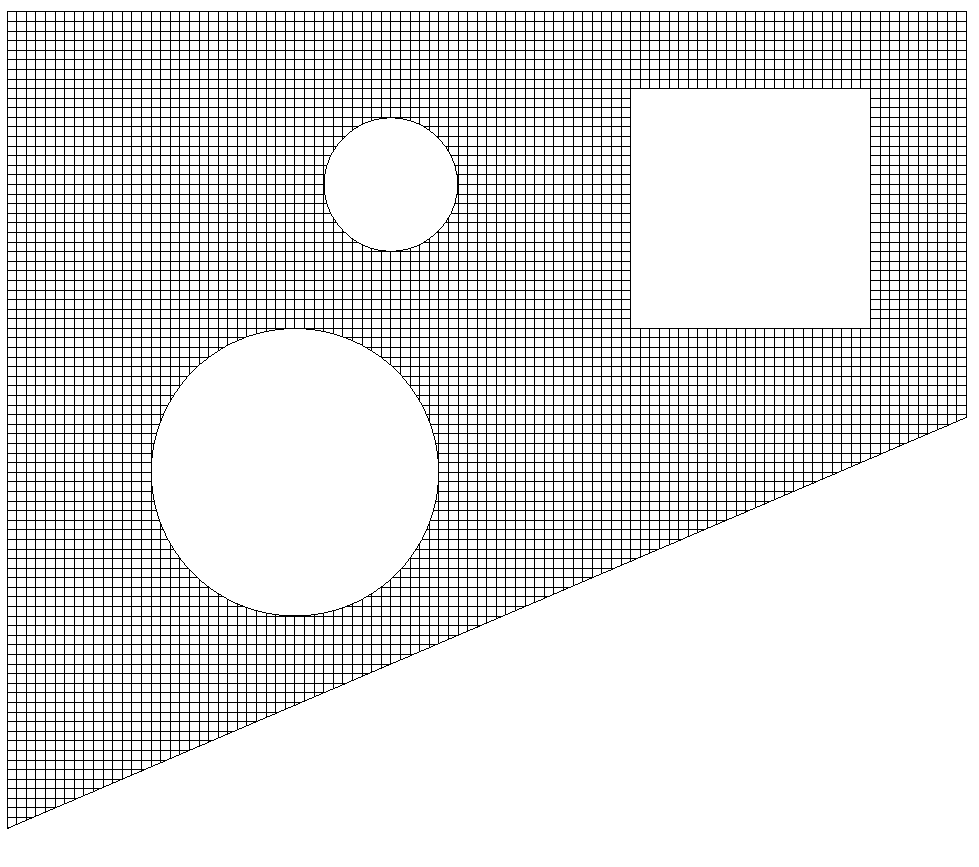}
\caption{Demonstrative low resolution Cartesian mesh containing embedded solid regions moving with velocity $\mathbf{u}_{s}$ and rotation rate $\omega_s$. (a) Physical geometry with indicative signed distance function, $\phi$. (b) Computational domain discretization using a Cartesian mesh. \label{fig:CutCellExample}}
\end{centering}
\end{figure}

We consider a computational domain containing a fluid region, with one or more embedded solid regions of essentially arbitrary shape. 
Each solid region can move freely through the fluid in any direction with velocity, $\mathbf{u}_{s}$, and rotation rate, $\omega_s$, as shown in Figure~\ref{fig:CutCellExample}~(a).
Following standard finite volume methodology, the computational domain is divided into $N_l$ discrete cells in the $l$th spatial direction.
The solid region is then cut out of this domain, leaving just the fluid domain discretised by the Cartesian mesh, as shown in Figure~\ref{fig:CutCellExample}~(b). The current cut-cell implementation represents a curved solid interface using a planar representation in each computational cell. The resultant curvature is therefore mesh dependent. The solid-fluid interface is imposed within each cut-cell using an appropriate boundary condition.

In this paper, we consider an inviscid, single-phase gas for the fluid region. The compressible Euler equations comprise of the mass, momentum and energy conservation equations. These are defined in compact form as:

\begin{equation}
\frac{\partial  \mathbf{Q}}{\partial t} + \nabla \cdot \mathbf{F\left(Q\right)} = 0,
\label{GeneralEulerEqn}
\end{equation} 

\noindent where the vector of conserved variables is given by $\mathbf{Q}=\left[\rho, \rho\mathbf{u}, \rho\left(e+\mathbf{u}^2/2\right)\right]$ and the flux vector is given by $\mathbf{F} = \left[\rho\mathbf{u}, \rho\mathbf{u}\mathbf{u}+p\mathbf{I}, \rho\revision{u}\left(e+\mathbf{u}^2/2  + p/ \rho\right)\right]$. These equations are closed using the perfect gas equation of state, $\revision{e = p/\rho\left(\gamma-1\right)}$, with a constant isentropic exponent, $\gamma$. 

Following integration of Equation~\ref{GeneralEulerEqn} over an arbitrary control volume, $V$, and after applying the Gauss divergence theorem, the discrete form of the Euler equations can be obtained from the resultant integral form by replacing the flux integral around the control volume by a summation of the surface normal fluxes at each computational cell interface. For regular Cartesian cells, away from the cut-cell surface, this is given by:

\begin{equation}
\left(\mathbf{Q}^{n+1}\right)_{l} = \left(\mathbf{Q}^{n}\right)_{l} - \frac{\Delta t}{V} \sum_{m=1}^{N_{f}} \left( \mathbf{F}\left(\mathbf{Q}\right)_{l,m} \cdot \mathbf{n}_{l,m}\right) \mathcal{S}_{l,m},
\label{EulerDiscreteRegular}
\end{equation} 

\noindent where the shorthand cell index, $l=i,j,k$ and $V$ is the regular cell volume. Cut-cells are defined as having volume, $\check{\alpha}^{n}V$, where the cut-cell volume fraction, $\check{\alpha}^{n}$, is defined as the ratio of the fluid volume in the cut-cell to the fluid volume of a regular Cartesian cell, at time step $n$. In the current moving boundary formulation, the solid region is permitted to move over the course of each time step. This movement results in a change to the fraction of the Cartesian cell covered by the solid, altering the fluid volume fraction, $\check{\alpha}$, between time steps $n$ and $n+1$. The discrete form of the Euler equations applied to a moving boundary cut-cell is then given by:

\begin{equation}
\left(\check{\alpha}^{n+1}V\mathbf{Q}^{n+1}\right)_{l} = \left(\check{\alpha}^{n}V\mathbf{Q}^{n}\right)_{l} - \Delta t \sum_{m=1}^{N_{f}} \left( \left( \mathbf{F}\left(\mathbf{Q}\right) - \delta\mathbf{F}^{wall} \right)_{l,m} \cdot \mathbf{n}_{l,m} \right ) \mathcal{S}_{l,m},
\label{EulerDiscreteCutCell}
\end{equation}
In this formulation, $\mathbf{F}^{wall}$ is an additional contribution due to the solid wall movement, denoted here as a wall `flux' for convenience. The kronecker delta, $\delta$, is therefore unity at solid boundaries and zero otherwise. The conservative flux is required normal to both inter-cell boundaries and solid-fluid boundaries, $\mathbf{F}\left(\mathbf{Q}\right) \cdot \mathbf{n}$. The total number of  cell faces in a Cartesian cut-cell may be more than, less than, or equal to the number of faces in a regular Cartesian cell. The total number of cut-cell faces is dependent on the local discretisation of the solid geometry and the geometry curvature. \revision{The moving boundary cut-cell approach described in this paper is a dimension-splitting approach, which isolates the flux component in each spatial direction, considering each as a separate Strang-splitting stage~\cite{strang1968construction, LeVequeBook2002}, in which the operator sequence is chosen to maintain second-order accuracy.}

\subsection{Static Boundary Method Overview}
\label{sec:StaticKleinOverviewSubSection}

\begin{figure}
\begin{centering}
(a) \includegraphics[width=10.5cm]{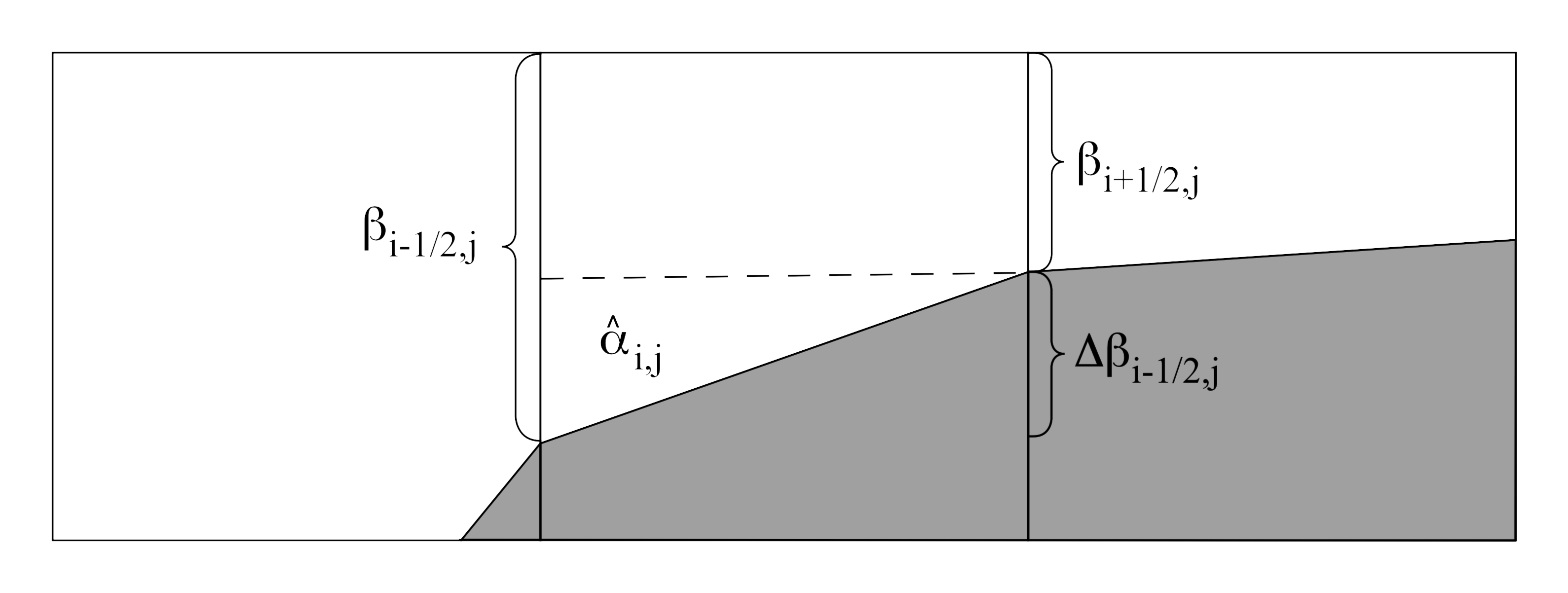}\\
(b) \includegraphics[width=10.5cm]{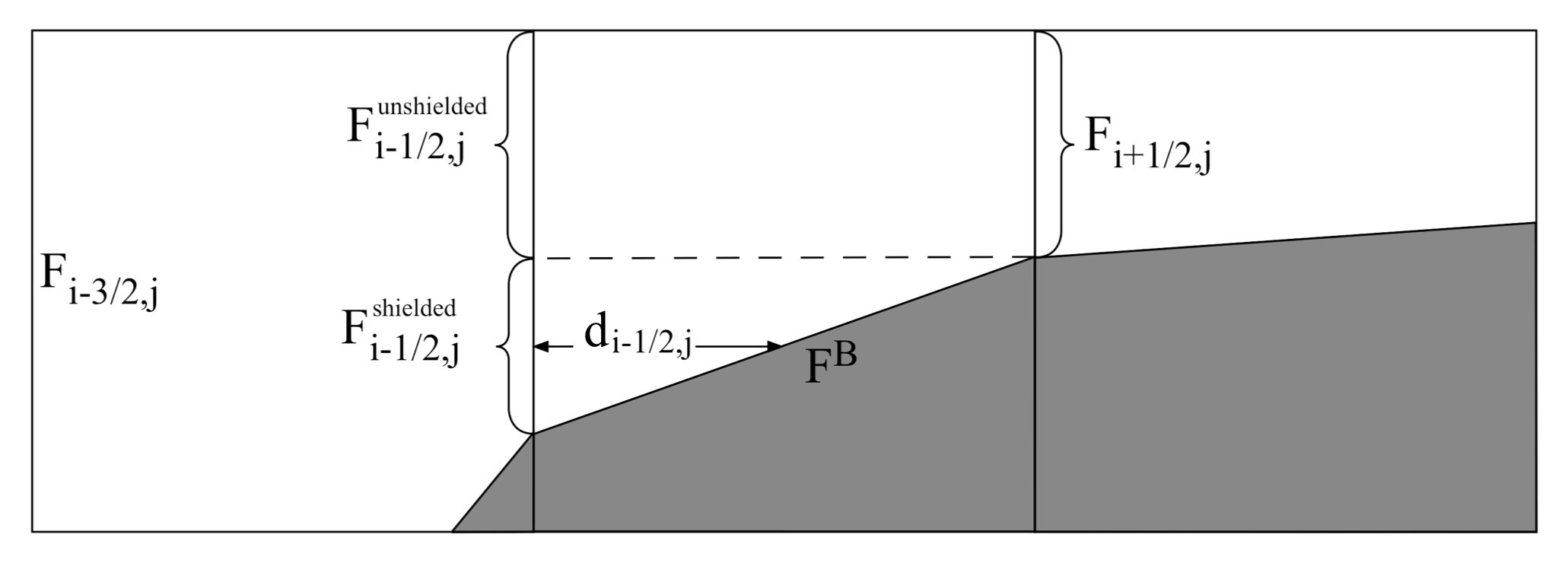}
\caption{Demonstrative solid-fluid interface in three cut-cells. (a) Interface and volume fractions for the two-dimensional cut-cell method of Klein et al.~\cite{Klein09}. (b) Flux components for the two-dimensional cut-cell method of Klein et al.~\cite{Klein09}. \label{fig:Klein2Da}}
\end{centering}
\end{figure}

The conservative, dimension-splitting cut-cell method of Klein et al.~\cite{Klein09} is extended in this paper to flows involving one or more moving solid boundaries. This section starts with a brief overview of the static boundary method of Klein et al.~\cite{Klein09} to provide context before describing the current extension to moving boundary flows.
Figure~\ref{fig:Klein2Da} shows a two-dimensional solid-fluid interface extending across three cut-cells, with indices $\left(i-1,j\right)$, $\left(i,j\right)$ and $\left(i+1,j\right)$. Considering the $x$-sweep \emph{i.e.}, the horizontal sweep, in isolation, the cut-cell face area fluid fractions at $\left(i-\frac{1}{2},j\right)$ and $\left(i+\frac{1}{2},j\right)$ are given by $\beta_{i-\frac{1}{2},j}$ and $\beta_{i+\frac{1}{2},j}$ respectively. These are labelled in Figure~\ref{fig:Klein2Da}~(a). The difference between these cut-cell face area fractions, for cell $\left(i,j\right)$, is given by $\Delta \beta_{i-\frac{1}{2},j}$. This represents the projection of the solid-fluid interface onto the cut-cell face, $\left(i-\frac{1}{2},j\right)$.

The flux stabilization method of Klein et al.~\cite{Klein09} permits a global time step to be imposed in both regular Cartesian cells and cut-cells. This is achieved by modifying the regular cell flux at inter-cell boundaries bordering a cut-cell. To describe this flux `stabilization', we consider the cut-cell face $\Delta \beta_{i-\frac{1}{2},j}$ in Figure~\ref{fig:Klein2Da}~(a).
The projection of the solid-fluid interface onto the cell face $\left(i-\frac{1}{2},j\right)$, given by $\Delta \beta_{i-\frac{1}{2},j}$, is termed in Klein et al.~\cite{Klein09} the `shielded' region. Specifically, this region is directly influenced in the x-direction by the presence of the solid-fluid interface. The remaining fluid portion of the cut-cell face area, i.e., $\beta_{i-\frac{1}{2},j}-\Delta \beta_{i-\frac{1}{2},j}$, is the interface portion not directly influenced by the solid-fluid interface. This is termed the `unshielded' region. 

Figure~\ref{fig:Klein2Da}~(b) shows the shielded and unshielded portions of the inter-cell flux for the x-direction sweep along the interface $\left(i-\frac{1}{2},j\right)$. These fluxes are defined as $\mathbf{F}_{i-\frac{1}{2},j}^{\textrm{shielded}}$ and $\mathbf{F}_{i-\frac{1}{2},j}^{\textrm{unshielded}}$ respectively. The total flux along this cell interface, $F^{2D}_{i-\frac{1}{2},j}$, is defined by a weighted combination of the shielded and unshielded fluxes, as:

\begin{equation}
\mathbf{F}_{i-\frac{1}{2},j}^{2D} = \frac{1}{\beta_{i-\frac{1}{2},j}}\left[ \beta_{i+\frac{1}{2},j}\mathbf{F}_{i-\frac{1}{2},j}^{\textrm{unshielded}} + \left( \beta_{i-\frac{1}{2},j} - \beta_{i+\frac{1}{2},j}\right )\mathbf{F}^{\textrm{shielded}}_{i+\frac{1}{2},j} \right ]. \label{Klein2Dtotalfluxa}
\end{equation}

\noindent For the convex geometry shown in Figure~\ref{fig:Klein2Da}, the unshielded flux component, $\mathbf{F}_{i-\frac{1}{2},j}^{\textrm{unshielded}}$, is simply the regular Cartesian cell flux:

\begin{equation}
\mathbf{F}_{i-\frac{1}{2},j}^{\textrm{unshielded}} = \mathbf{F}_{i-\frac{1}{2},j}. \label{Klein2Dunshieldeda}
\end{equation}

\noindent The shielded flux creates a stable flux by combining the boundary flux, $\mathbf{F}_{i-\frac{1}{2},j}^{\textrm{B}}$, with the regular cell flux, $\mathbf{F}_{i,j}$, as:

\begin{equation}
\mathbf{F}_{i-\frac{1}{2},j}^{\textrm{shielded}} = \mathbf{F}_{i,j}^{B} + \revision{d_{i-\frac{1}{2},j}} \left(\mathbf{F}_{i-\frac{1}{2},j} - \mathbf{F}_{i,j}^{B} \right), \label{Klein2shieldedFlux}
\end{equation}

\noindent where $\revision{d_{i-\frac{1}{2},j}}$ is the average distance between the interface at $\left(i-\frac{1}{2},j\right)$ and the solid-fluid interface, as shown in Figure~\ref{fig:Klein2Da}~(b). The average distance, $\revision{d_{i-\frac{1}{2},j}}$, is computed from the shielded volume fraction, $\hat{\alpha}_{i,j}$, shown in Figure~\ref{fig:Klein2Da}~(a), as:

\begin{equation}
\revision{d_{i-\frac{1}{2},j}} = \frac{\hat{\alpha}_{i,j}}{\beta_{i-\frac{1}{2},j}-\beta_{i+\frac{1}{2},j}}.\label{Klein2DalphaHat}
\end{equation}

\noindent For clarity, the notation of the average distance from the cut-cell interface to the stabilized cell interface, $\revision{d_{i-\frac{1}{2},j}}$, is distinct from the cut-cell shielded volume fraction, $\hat{\alpha}_{i,j}$, and the total cut-cell volume fraction, $\check{\alpha}_{i,j}$.  The total cut-cell volume fraction is defined as the combination of the shielded and unshielded cut-cell volume fractions.
The shielded flux in Equation~\ref{Klein2shieldedFlux} has the property that, as the cut-cell volume fraction is reduced, and therefore $\revision{d_{i-\frac{1}{2},j} \rightarrow 0}$, then $\mathbf{F}_{i-\frac{1}{2},j}^{\textrm{shielded}} \rightarrow \mathbf{F}_{i,j}^{B}$. As the cut-cell tends towards a regular cell, $\revision{d_{i-\frac{1}{2},j} \rightarrow 1}$, the shielded flux tends towards the regular cell flux, $\mathbf{F}_{i-\frac{1}{2},j}^{\textrm{shielded}} \rightarrow \mathbf{F}_{i-\frac{1}{2},j}$.

The boundary flux, $\mathbf{F}_{i,j}^{\textrm{B}}$, is shown in Figure~\ref{fig:Klein2Da}~(b) along the solid-fluid interface. The flux at this interface is computed using a reflective wall boundary condition. Caution is required here, as in order that the dimension splitting remains conservative, the advective part of the boundary flux in all directions must be computed for a common solid-fluid interface state. That is, we can separate the full flux at the boundary into an advective flux and a pressure flux:

\begin{equation}
\mathbf{F}\left(\mathbf{Q}^{\textrm{B}}_{i,j}\right) = \mathbf{F}^{\textrm{ad}}\left(\mathbf{Q}^{\textrm{B}}_{i,j}\right) + \mathbf{F}^{\textrm{pr}}\left(\mathbf{Q}^{\textrm{B}}_{i,j}\right),\label{KleinAdvectivePressureSplitA}
\end{equation}

\noindent where, the x-direction (horizontal) advective and pressure fluxes for the Euler equations in two-dimensions are respectively $\mathbf{F}^{\textrm{ad}}\left(\mathbf{Q}^{\textrm{B}}_{i,j}\right) = \left[\rho^{*}u^{*}, \rho^{*} u^{*}u^{*}, \rho^{*} u^{*}v^{*}, \rho^{*}\revision{u^{*}}\left(e+\mathbf{u^{*}}^2/2\right)\right]$ and $\mathbf{F}^{\textrm{pr}}\left(\mathbf{Q}^{\textrm{B}}_{i,j}\right) = \left[0,p^{*},0,u^{*}p^{*}\right]$ based on the boundary state, $\mathbf{Q}^{\textrm{B}}$. 
Similarly, the advective flux and pressure flux in the y-direction are given by $\mathbf{G}^{\textrm{ad}}\left(\mathbf{Q}^{\textrm{B}}_{i,j}\right)$ and $\mathbf{G}^{\textrm{pr}}\left(\mathbf{Q}^{\textrm{B}}_{i,j}\right)$ respectively. The advective flux is computed before the onset of the first dimensional sweep, at the start of each time-step. The pressure flux in the momentum equation is, however, computed at each dimensional splitting stage using the updated boundary state.

The neighbouring fluxes, at $\mathbf{F}_{i-\frac{3}{2},j}$ and $\mathbf{F}_{i+\frac{1}{2},j}$, are also shown in Figure~\ref{fig:Klein2Da}~(b). These fluxes are, in turn, a combination of the local shielded and unshielded flux contributions from the solid-fluid interface in cells $\left(i-1,j\right)$ and $\left(i+1,j\right)$ respectively for this example.

The complete two-dimensional update for cell $\left(i,j\right)$ is therefore given by:

\begin{eqnarray}
\mathbf{Q}_{i,j}^{x} = \mathbf{Q}_{i,j}^{n} + \frac{\Delta t}{\check{\alpha}_{i,j} \Delta x} \left[\beta_{i-\frac{1}{2},j} \mathbf{F}_{i-\frac{1}{2},j}^{2D} - \beta_{i+\frac{1}{2},j} \mathbf{F}_{i+\frac{1}{2},j}^{2D} + \left(\beta_{i+\frac{1}{2},j}-\beta_{i-\frac{1}{2},j}\right) \mathbf{F}_{i,j}^{\textrm{B}} \right],\label{Klein2DcompleteA}\\
\mathbf{Q}_{i,j}^{n+1} = \mathbf{Q}_{i,j}^{x} + \frac{\Delta t}{\check{\alpha}_{i,j} \Delta y} \left[\beta_{i,j-\frac{1}{2}} \mathbf{G}_{i,j-\frac{1}{2}}^{2D} - \beta_{i,j+\frac{1}{2}} \mathbf{G}_{i,j+\frac{1}{2}}^{2D} + \left(\beta_{i,j+\frac{1}{2}}-\beta_{i,j-\frac{1}{2}}\right) \mathbf{G}_{i,j}^{\textrm{B}} \right].
\label{Klein2DcompleteB}
\end{eqnarray}

Klein et al.~\cite{Klein09} show that, using the flux stabilization from Equation~\ref{Klein2shieldedFlux}, this update can be rewritten for cut-cells in such a way that the division by $\check{\alpha}_{i,j}$ is eliminated and singular behaviour for vanishing cut-cell volume fractions is avoided.

\subsection{Moving Boundary Extension}
\label{sec:MovingBoundarySubSection}

\subsubsection{Moving Boundary Issues}
\label{sec:MovingBoundaryIssues}

\begin{figure}
\begin{centering}
(a) \includegraphics[width=4cm]{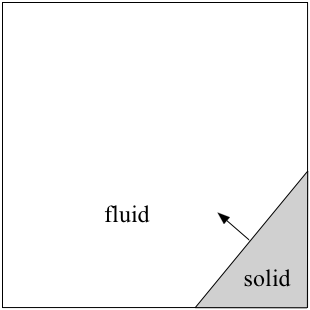}
(b) \includegraphics[width=4cm]{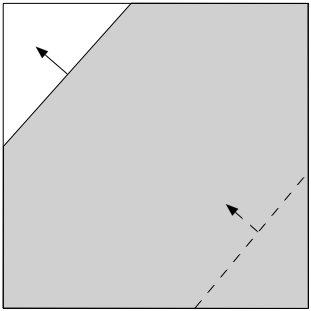}
(c) \includegraphics[width=4cm]{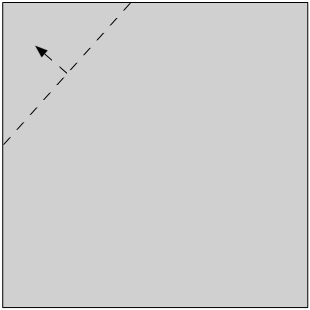}
\caption{Interaction of a moving solid-fluid interface with a cut-cell. (a) The solid first appears in the subject Cartesian cell, turning the regular cell into a cut-cell. (b) The cut-cell volume fraction gradually decreases as the solid covers more of the cut-cell. (c) The solid covers the Cartesian cell completely, turning the cut-cell into a solid cell. Solid line: Solid-fluid interface at the current time step. Dashed line: Solid-fluid interface at the previous time step. The arrow indicates the direction of the solid-fluid interface movement.\label{fig:CutCellCoverUp}}
\end{centering}
\end{figure}

A number of additional challenges arise in the extension of the static cut-cell method of Klein et al.~\cite{Klein09} to accommodate moving solids. Whereas the status of the cut-cells and regular Cartesian fluid cells are unchanging over the course of a static boundary computation, a transient cut-cell status now exists due to the motion of the solid-fluid interface across each Cartesian cell. As the solid-fluid interface first crosses one of the regular cell boundaries,  the regular Cartesian cell changes to a cut-cell which requires stabilization of the boundary fluxes, Figure~\ref{fig:CutCellCoverUp}~(a). As the solid-fluid interface moves further into the cut-cell, the cut-cell volume fraction becomes a smaller fraction of the regular cell fluid volume, Figure~\ref{fig:CutCellCoverUp}~(b), tending towards a zero volume fraction when the solid-fluid interface passes through the final cut-cell fluid boundary. At this point the cut-cell changes to a solid cell, which is no longer used in the fluid computation, Figure~\ref{fig:CutCellCoverUp}~(c).

Conversely, a cut-cell can be created from a solid cell when the solid-fluid interface passes through one of the inter-cell boundaries of a solid cell. As the solid-fluid interface moves across the new cut-cell, the fluid volume fraction becomes an increasing fraction of the regular cell volume, until the solid-fluid interface passes out of the cut-cell and the cut-cell becomes a regular Cartesian fluid cell. 

The problem caused by the decreasing cut-cell volume fraction, and the related changes of the cut-cell face area fractions, can be identified by rearranging the discrete Euler equation, Equation~\ref{EulerDiscreteCutCell}, to make the state variable the subject, i.e:

\begin{equation}
\mathbf{Q}_{l}^{n+1} = \frac{\check{\alpha}_{l}^{n}}{\check{\alpha}_{l}^{n+1}}\left(\mathbf{Q}^{n}\right)_{l} - \frac{\Delta t}{\check{\alpha}_{l}^{n+1}V} \sum_{m=1}^{N_{f}} \left( \left( \mathbf{F}\left(\mathbf{Q}\right) - \delta_{m}\mathbf{F}^{wall} \right)_{l,m} \cdot \mathbf{n}_{l,m} \right ) \mathcal{S}_{l,m},
\label{EulerDiscrete2}
\end{equation}
where the shorthand index, $l = \left(i,j,k\right)$. As the cut-cell volume fraction tends towards zero over time, the reciprocal of the cut-cell volume fraction at the updated time, $\check{\alpha}^{n+1}$, causes the state variables at the new time, $\mathbf{Q}^{n+1}$, to increase to large values. Merging cut-cell methods, as used for moving boundaries by Xu et al.~\cite{Xu97}, Yang et al.~\cite{yang1997cartesian} and Barton et al.~\cite{barton2011conservative}, for example, deal with this issue by grouping cut-cells of small volume fraction with an adjacent regular cell (or large cut-cell) to form one cell with a combined volume. In contrast, Schneiders et al.~\cite{schneiders2013accurate, schneiders2016efficient} deal with this issue through a redistribution of the numerical error from the small volume cut-cell to the surrounding cells, following a time integration of all cells using regular cell fluxes.
The vanishing volume fraction issue is dealt with in the current method through an augmentation of the cut-cell shielded flux. 

For a solid-fluid interface moving at a velocity that matches the local velocity, the work done on the fluid by the solid-fluid interface reduces to zero and a constant wall pressure is recovered, $p_{wall} = p_{l}$. This is described in, for example, Toro~\cite{Toro}. In terms of the present cut-cell method, this constraint dictates that the mass, momentum or energy displaced by the solid-fluid boundary movement within a cut-cell should equal the total advective mass, momentum or energy flux through the computational cell boundaries, for each spatial direction. This provides a sound test of any new moving boundary method, as any resultant pressure fluctuations at the solid-fluid interface are therefore due to errors in matching the boundary movement to the advective flux, or due to fluctuations in the solid geometry through inadequate geometry discretisation and movement across the Cartesian mesh over time.\\

\subsubsection{Moving Boundary Geometry Approximation}
\label{sec:MovingBoundaryGeometryApproximation}

The moving boundary cut-cell method described in this paper is a straightforward extension to the static boundary method of Klein et al.~\cite{Klein09} described in Section~\ref{sec:StaticKleinOverviewSubSection}. The moving boundary extension maintains the principle of dividing each cut-cell interface flux into a `shielded' and `unshielded' contribution, based on whether the inter-cell boundary directly faces the solid-fluid interface during each dimensional sweep.

\begin{figure}
\begin{centering}
\includegraphics[width=10.5cm]{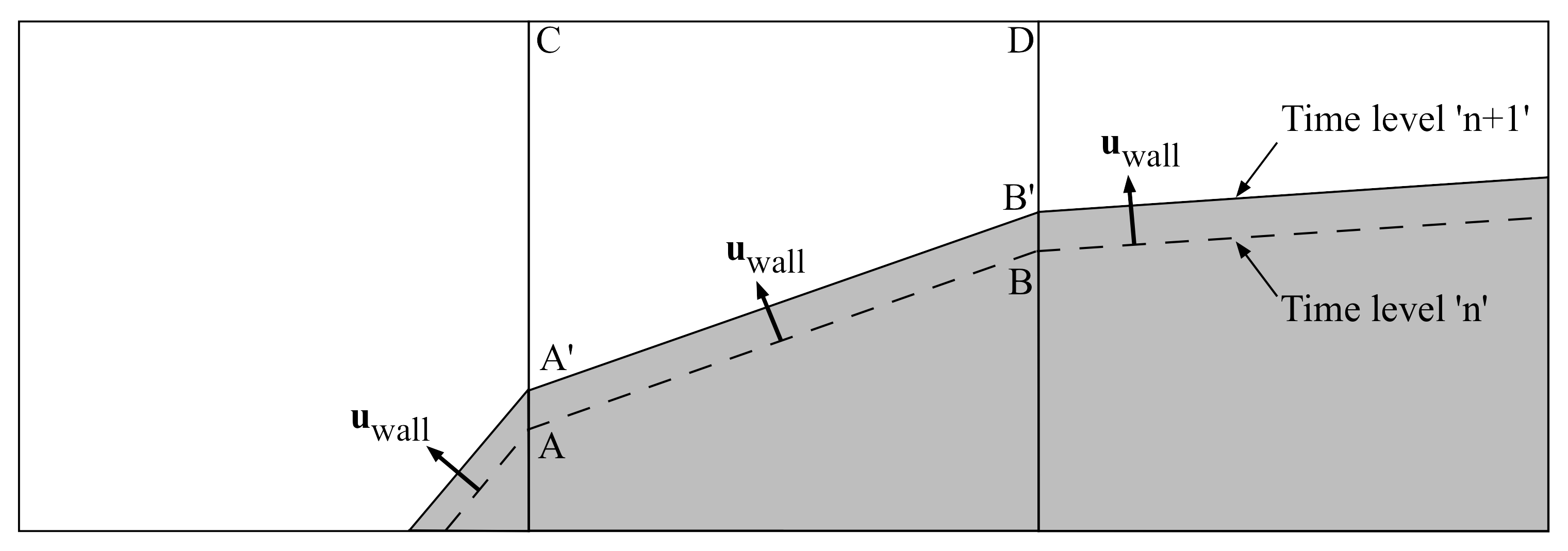}
\caption{Moving solid interface across three cut-cells. The solid-fluid interface in cell $\left(i,j\right)$ moves from A-B to A'-B' between time level $n$ (dashed line) and $n+1$ (solid line). \label{fig:Klein2DgeomMovingA}}
\end{centering}
\end{figure}

For the purpose of illustration, we can consider the same solid-fluid interface as in the static boundary section, intersecting three cut-cells, $\left(i-1,j\right)$, $\left(i,j\right)$ and $\left(i+1,j\right)$, this time with movement of the solid-fluid interface over a typical time step, $\Delta t$, as shown in Figure~\ref{fig:Klein2DgeomMovingA}. The direction of the solid-fluid interface movement is indicated by arrows. Over the time interval, $\Delta t$, the cut-cell interface in cell $\left(i,j\right)$ moves such that the intersection points with the Cartesian cell boundaries, $\left(x,y\right)_{A}$ and $\left(x,y\right)_{B}$, at time $n$ become $\left(x,y\right)_{A'}$ and $\left(x,y\right)_{B'}$ at time $n+1$. The cut-cell volume correspondingly changes from $V_{ABCD}$ to $V_{A'B'CD}$.

Murman et al.~\cite{murman2003implicit} classify the numerical representation of the space-time geometry development in increasing order of fidelity as either `sequential-static', `staircase-in-time', `linear-in-time' or `full space-time approximation'. In terms of this classification, the current method would be considered a `linear-in-time' approach, in which the geometry movement, and hence the cut-cell area and volume fractions, are assumed to change linearly  between time steps $n$ and $n+1$. The current moving boundary method is introduced by initially considering a simpler `moving wall' boundary condition approach applied to the method of Klein et al.~\cite{Klein09} in Section~\ref{sec:Method1MovingWall}.

\subsubsection{The Moving Wall Boundary Condition Approach}
\label{sec:Method1MovingWall}

\begin{figure}
\begin{centering}
\includegraphics[width=10cm]{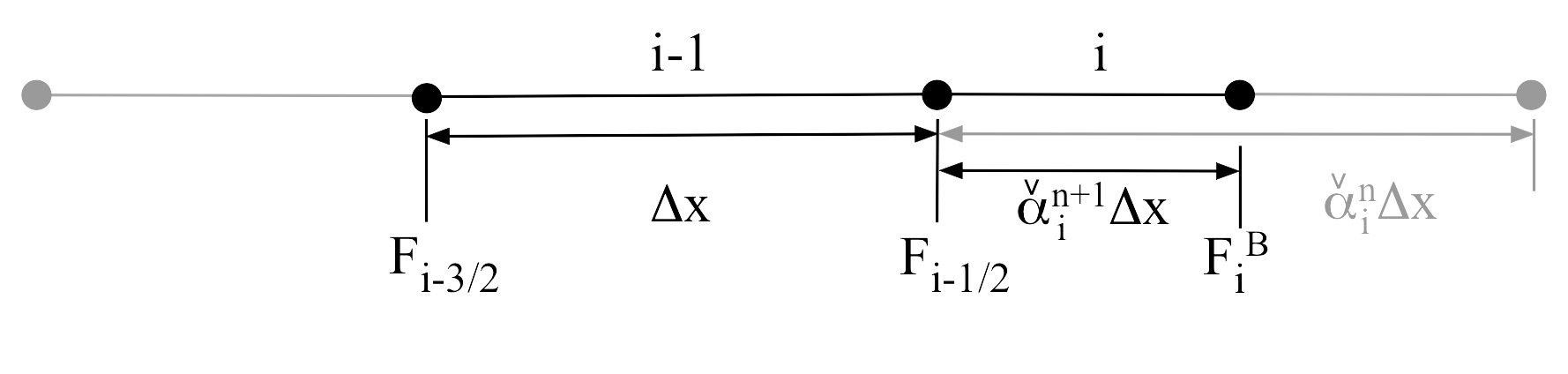}
\caption{One-dimensional moving boundary example.  The cut-cell reduces in size from $\check{\alpha}^{n}_{i}\Delta x$ to $\check{\alpha}^{n+1}_{i}\Delta x$ between times $n$ and $n+1$, where $0 \leq \check{\alpha}_{i} \leq 1$. As the total cut-cell volume fraction, $\check{\alpha}_{i}$, the shielded cut-cell volume fraction, $\hat{\alpha}_{i}$, and the average distance from the stabilized cell interface to the cut-cell interface, $\revision{d_{i-\frac{1}{2}}}$, are equivalent in a one-dimensional problem, we use the notation $\check{\alpha}_{i}$ here. \label{fig:Klein1Dgeom}}
\end{centering}
\end{figure}

The simplest extension to the method of Klein et al.~\cite{Klein09} involves fixing the geometry at each time level and changing the boundary flux to a `moving wall' boundary flux, as described for a compressible Riemann-based approach in Toro~\cite{Toro}.  At the moving solid-fluid interface, the boundary flux incorporates the local boundary velocity by imposing a ghost-cell velocity, $u_{g}$, according to:

\begin{equation}
\textbf{u}_{g} = \textbf{u}_{i,j} - 2\left(\textbf{u}_{i,j}\cdot\textbf{n}\right)\textbf{n} + 2\left(\textbf{u}_{wall}\cdot\textbf{n}\right)\textbf{n},
\label{riemann_wallvel}
\end{equation}

\noindent where, $\textbf{u}_{wall}$ is the local solid-fluid interface velocity vector and $\textbf{u}_{i,j}$ is the fluid velocity vector in the cut-cell $\left(i,j\right)$, as shown in Figure~\ref{fig:Klein2DgeomMovingA}. The ghost-cell density and pressure can be imposed for an inviscid reflective boundary condition according to $\rho_{g}=\rho_{i,j}$ and $p_{g}=p_{i,j}$. Finding the resultant boundary state in terms of pressure, density and velocity results in a boundary velocity equal to the wall velocity, $\textbf{u}^{*}=\textbf{u}_{wall}$. For an advancing wall, the wall pressure, $p^{*} > p_{i,j}$, and density, $\rho^{*} > \rho_{i,j}$. Conversely, for a retreating wall, $p^{*} < p_{i,j}$ and $\rho^{*} < \rho_{i,j}$. For a simple one-dimensional problem, approximating the non-penetration reflective boundary condition using this `moving wall' boundary state results in a mass flux $\rho^{*}u^{*}$ `through' the solid-fluid interface which matches the fluid mass displaced by the movement of the solid-fluid interface over this time step. Using an example of a one-dimensional cut-cell reducing in length from $\check{\alpha}^{n}_{i}\Delta x$ to $\check{\alpha}^{n+1}_{i}\Delta x$ over the time interval $\Delta t$, as shown in Figure~\ref{fig:Klein1Dgeom}, the cut-cell update is given by:

\begin{equation}
\check{\alpha}_{i}^{n+1} \textbf{Q}_{i}^{n+1} = \check{\alpha}_{i}^{n}\textbf{Q}_{i}^{n} + \frac{\Delta t}{\Delta x} \left[\textbf{F}_{i-\frac{1}{2}} - \textbf{F}_{i}^{\textrm{B}} \right] - \frac{\Delta t}{\Delta x}u_{wall}\textbf{Q}^{B}_{i},
\label{Klein1DcompleteMovingMethod1}
\end{equation}

\begin{figure}
\begin{centering}
(a) \includegraphics[width=10.5cm]{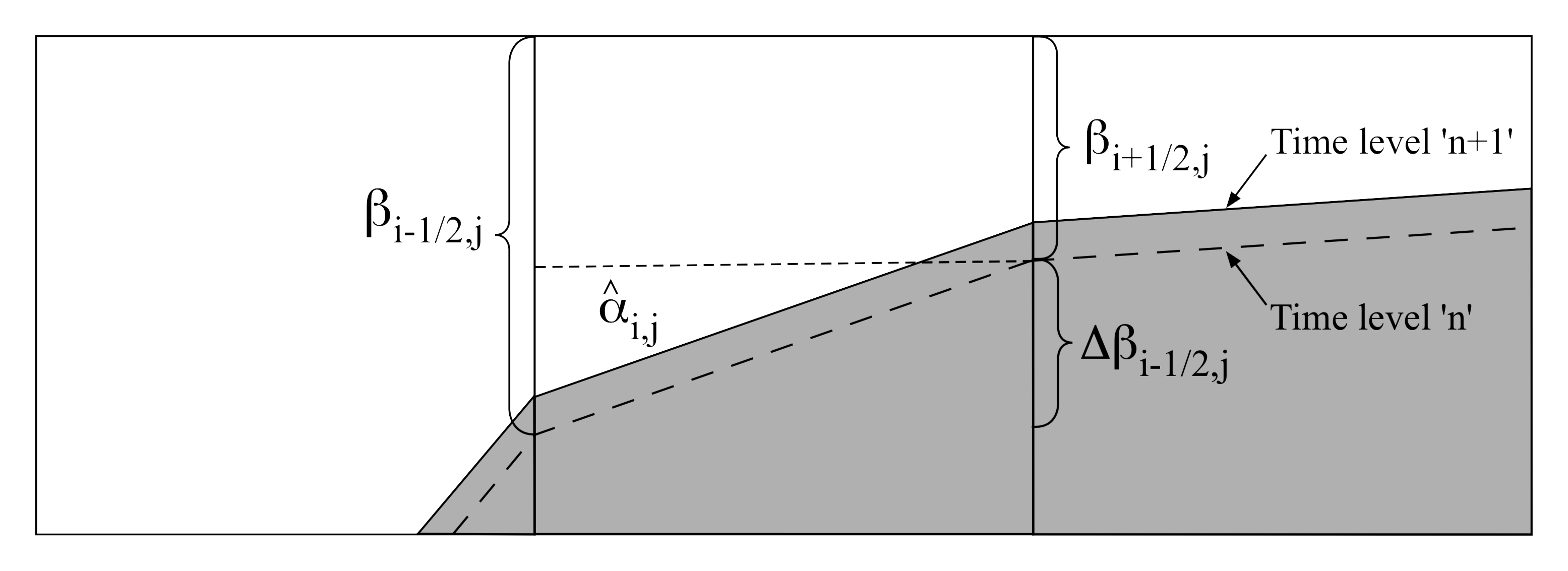}\\
(b) \includegraphics[width=10.5cm]{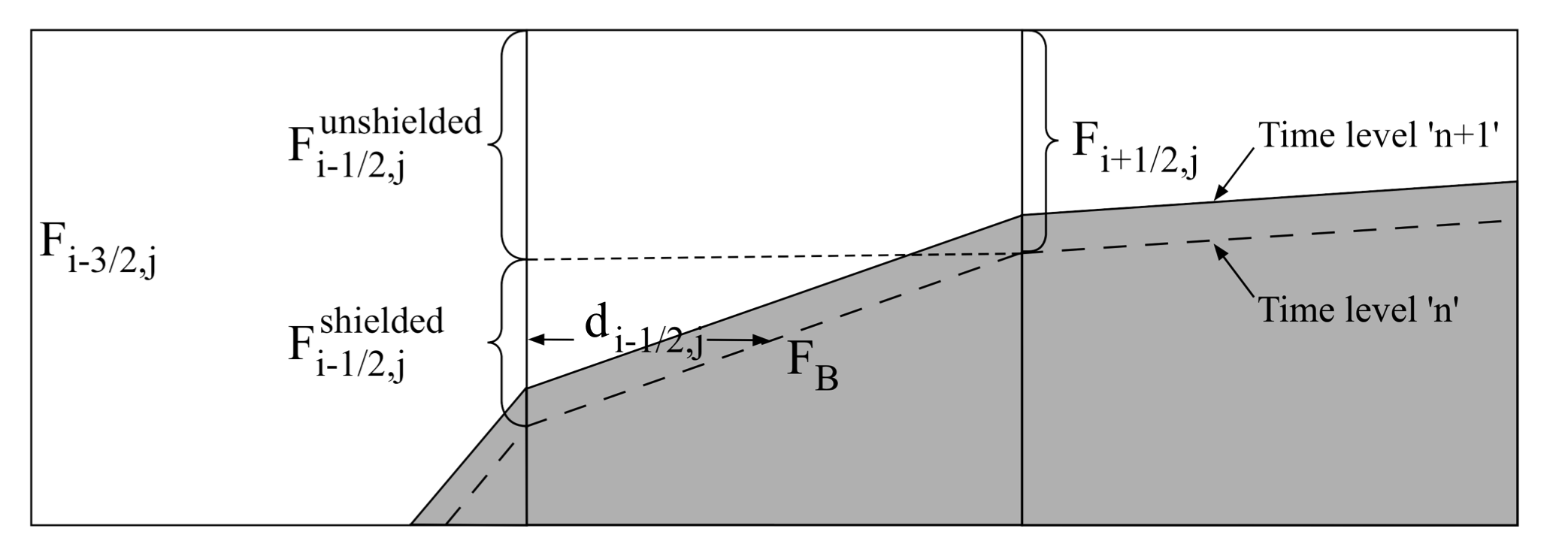}
\caption{Moving solid interface in three cut-cells. (a) Interface and volume fractions for the two-dimensional cut-cell method of Klein et al.~\cite{Klein09} based on the solid-fluid interface at time, $n$. (b) Interface fluxes based on the solid fluid interface at time $n$. \label{fig:Klein2DgeomMovingB}}
\end{centering}
\end{figure}

\noindent where the boundary flux, $\mathbf{F}^{B}_{i}$, and boundary state, $\textbf{Q}^{B}_{i}$, are non zero in mass, momentum and energy. We explicitly note that, in the one-dimensional case, the total cut-cell volume fraction, $\check{\alpha}_{i}$, the shielded cut-cell volume fraction, $\hat{\alpha}_{i}$, and the average distance from the stabilized cell interface to the cut-cell interface, $\revision{d_{i-\frac{1}{2}}}$, are equivalent. For clarity, we therefore use the total cut-cell volume fraction notation, $\check{\alpha}_{i}$, in Equation~\ref{Klein1DcompleteMovingMethod1} and Figure~\ref{fig:Klein1Dgeom}. \revisionB{It is noted that the moving boundary time integration/accuracy is explained further in Section~\ref{sec:StrictConservationMethod2}.}

The corresponding two-dimensional cut-cell update, for the cell $\left(i,j\right)$ shown in Figure~\ref{fig:Klein2DgeomMovingB}, would then be given by:

\begin{eqnarray}
\check{\alpha}_{i,j}^{x}\mathbf{Q}_{i,j}^{x} = \check{\alpha}_{i,j}^{n}\mathbf{Q}_{i,j}^{n} + \frac{\Delta t}{\Delta x} \left[\beta_{i-\frac{1}{2},j} \mathbf{F}_{i-\frac{1}{2},j}^{2D} - \beta_{i+\frac{1}{2},j} \mathbf{F}_{i+\frac{1}{2},j}^{2D} + \left(\beta_{i+\frac{1}{2},j}-\beta_{i-\frac{1}{2},j}\right) \mathbf{F}_{i,j}^{\textrm{B}} - \left(\beta_{i+\frac{1}{2},j}-\beta_{i-\frac{1}{2},j}\right)\mathbf{u}^{x}_{wall}\mathbf{Q}_{i,j}^{\textrm{B}}\right],\label{Klein2DcompleteMovingMethod1a}\\
\check{\alpha}_{i,j}^{n+1}\mathbf{Q}_{i,j}^{n+1} = \check{\alpha}_{i,j}^{x}\mathbf{Q}_{i,j}^{x} + \frac{\Delta t}{\Delta y} \left[\beta_{i,j-\frac{1}{2}} \mathbf{G}_{i,j-\frac{1}{2}}^{2D} - \beta_{i,j+\frac{1}{2}} \mathbf{G}_{i,j+\frac{1}{2}}^{2D} + \left(\beta_{i,j+\frac{1}{2}}-\beta_{i,j-\frac{1}{2}}\right) \mathbf{G}_{i,j}^{\textrm{B}} - \left(\beta_{i,j+\frac{1}{2}}-\beta_{i,j-\frac{1}{2}}\right)\mathbf{u}^{y}_{wall}\mathbf{Q}_{i,j}^{\textrm{B}}\right].
\label{Klein2DcompleteMovingMethod1b}
\end{eqnarray}

where $\mathbf{u}^{x}_{wall}$ and $\mathbf{u}^{y}_{wall}$ are the wall velocity $x$ and $y$ direction components. $\mathbf{Q}^{B}_{i,j}$ is the boundary state. As shown in Ben-Artzi \& Falcovitz~\cite{ben2003generalized} and Falcovitz et al.~\cite{falcovitz1997two}, the resultant contribution of the moving boundary in the $x$ and $y$ directions is then:

\begin{eqnarray}
\mathbf{F}^{B}_{res} = \left[0,p^{*},0,u^{*}p^{*}\right] = \mathbf{F}^{B}_{i,j} - \mathbf{u}_{wall}^{x}\mathbf{Q}_{i,j}^{B},\\
\mathbf{Q}^{B}_{res} = \left[0,0,p^{*},v^{*}p^{*}\right] = \mathbf{G}^{B}_{i,j} - \mathbf{u}_{wall}^{y}\mathbf{Q}_{i,j}^{B}.
\label{Klein2DMovingResultantFluxes}
\end{eqnarray}

Thus, the contribution from the movement of the non-penetrative solid wall is through the boundary pressure in the momentum and energy equations. The boundary interface area that this pressure acts upon is the time averaged boundary area, as outlined in Section~\ref{sec:StrictConservationMethod2}.

\begin{figure}
\begin{centering}
(a) \includegraphics[width=5cm]{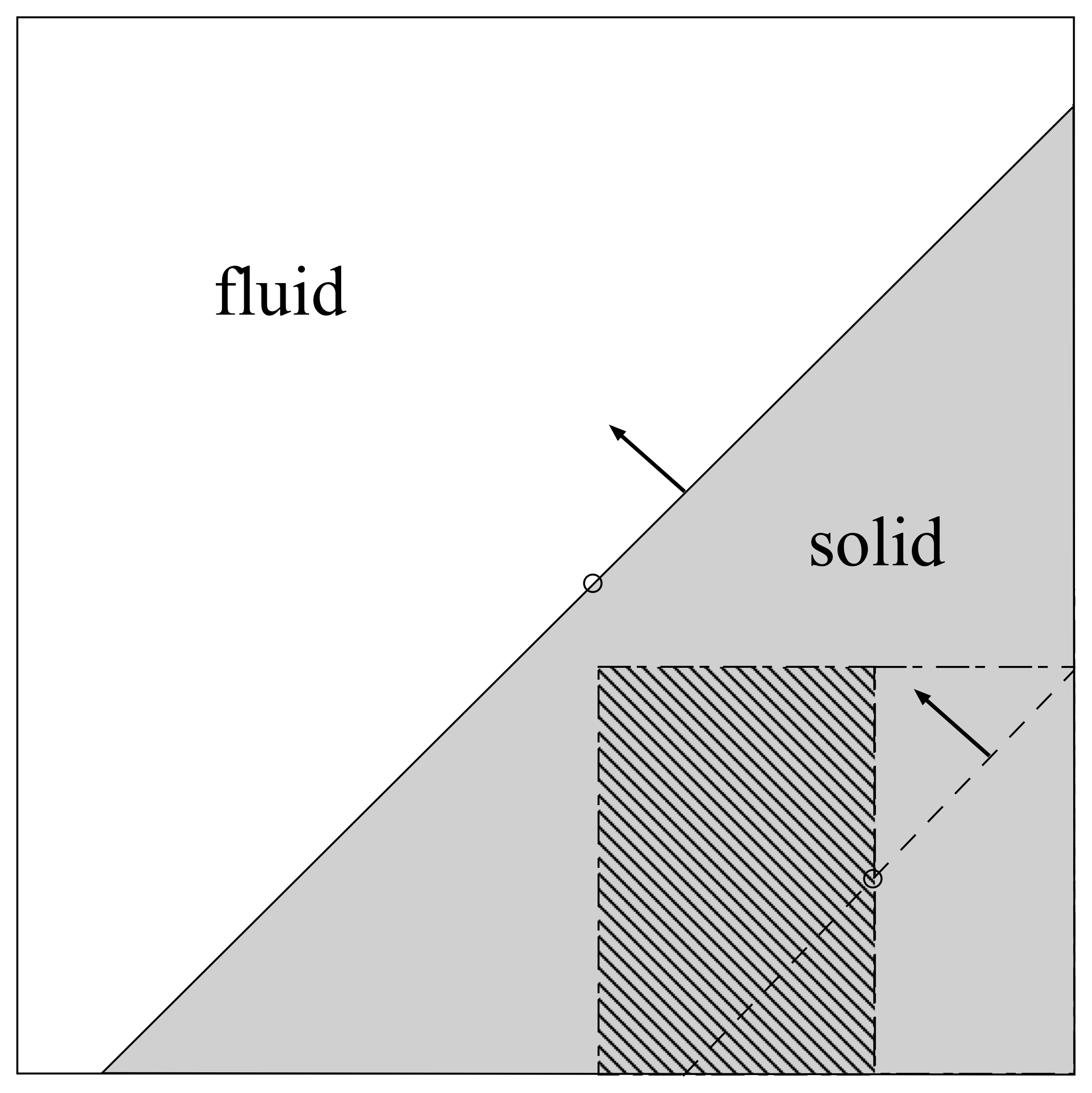}
(b) \includegraphics[width=5cm]{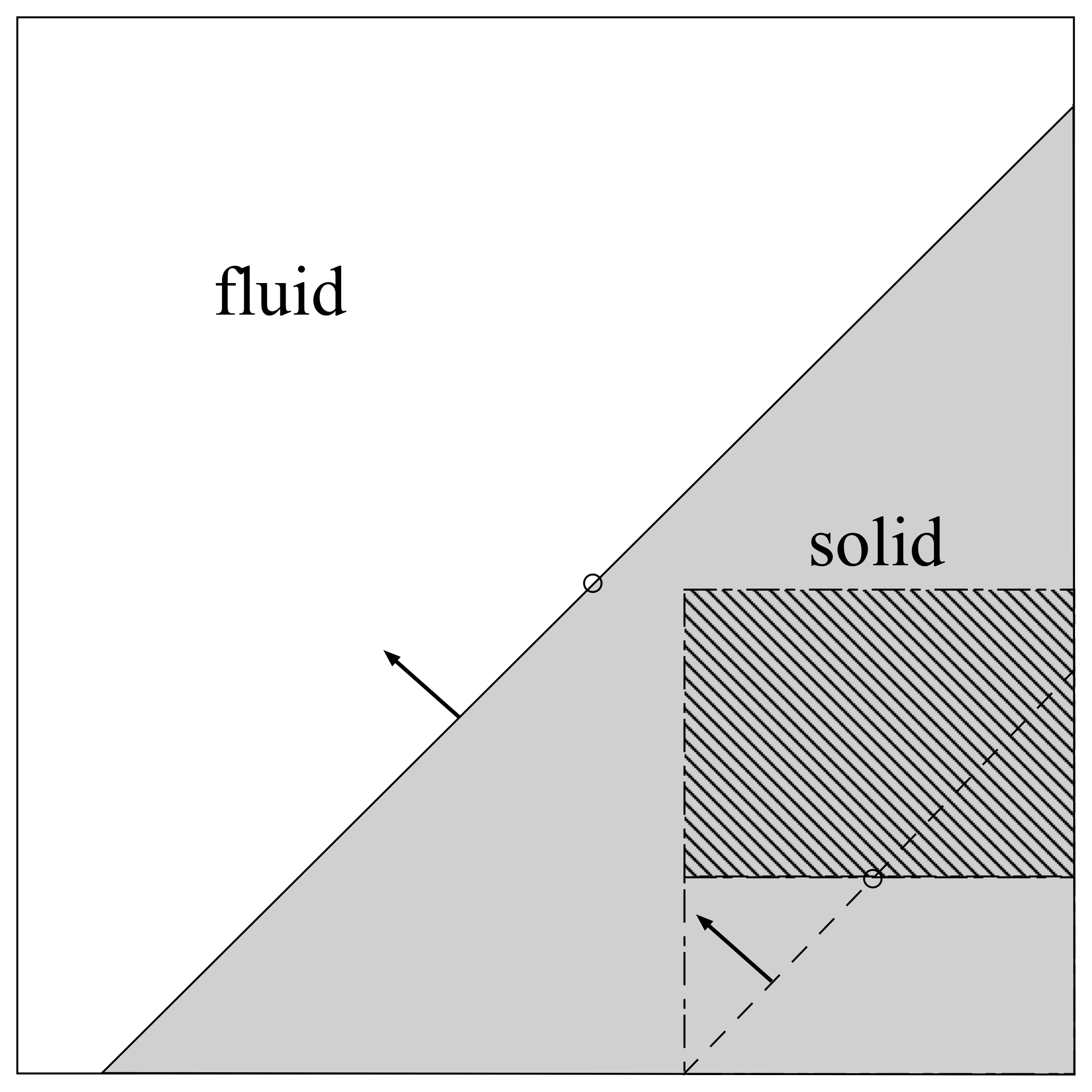}
\caption{Comparison of the volume fraction change as a diagonal solid-fluid interface diagonally traverses a Cartesian cell. (a) Volume change after the x-direction sweep. (b) Volume change after the y-direction sweep. Dashed line: Solid-fluid interface at time level $n$. Solid line: Solid-fluid interface at time level $n+1$. Striped region: Volume change in each direction between time level $n$ and $n+1$. Dash-dot line: Solid-fluid interface projection in the x-direction or y-direction.\label{fig:SplitMassNonConservation}}
\end{centering}
\end{figure}

The necessity of precisely accounting for the movement of the solid boundary, in this case through a time averaged boundary in the conservation equation, is highlighted by Gokhale~\cite{Gokhale14} through consideration of the loss in conservation for a dimension-splitting method applied to a diagonal piston problem. This is presented here by considering the movement of the solid-fluid interface in the $x$ and $y$ directions independently, as shown in Figure~\ref{fig:SplitMassNonConservation}. The resultant fluid volume displacement arising from the independent x-direction and y-direction movement does not necessarily equal the two-dimensional volume displacement.

As the rigid body interface moves across the Cartesian cell in Figure~\ref{fig:SplitMassNonConservation}, the overall loss of conservation is accompanied by an under-prediction of the boundary work done by the solid-fluid interface. We attempt to preserve conservation for a moving rigid body by defining time averaged boundary properties in the conservation equations, together with a conservation check for newly covered cells. These are outlined in Section~\ref{sec:StrictConservationMethod2} and Section~\ref{sec:CutCellInitialisationSubmergence} respectively.

\subsubsection{Time Averaged Boundary Modification}
\label{sec:StrictConservationMethod2}

The moving boundary method developed here uses the time-averaged solid-fluid interface for defining the shielded and unshielded cell area fractions, $\beta$, and the average shielded distance from the solid, $\revision{d_{i-\frac{1}{2},j}}$. These can be formally defined as: 

\begin{equation}
\tilde{\beta}_{i-\frac{1}{2},j}=\frac{1}{\Delta t}\int_{t^{n}}^{t^{n+1}}\beta_{i-\frac{1}{2},j}\left(t\right) dt,
\label{timeAveragedBeta}
\end{equation}

\begin{equation}
\revision{\tilde{d}_{i-\frac{1}{2},j}=\frac{1}{\Delta t}\int_{t^{n}}^{t^{n+1}}d_{i-\frac{1}{2},j}\left(t\right) dt,}
\label{timeAveragedAlpha}
\end{equation}

These are used for calculating $\beta_{\textrm{shielded}}$, $\beta_{\textrm{unshielded}}$ and $\revision{d_{i-\frac{1}{2},j}}$. By using the time averaged cell areas, together with the conservative state after each dimensional sweep, the resultant flux through each cut-cell interface provides a better approximation than shown in Figure~\ref{fig:SplitMassNonConservation}. Similarly, the solid-fluid interface unit normal is also computed as the time averaged surface normal.

\begin{figure}
\begin{centering}
(a) \includegraphics[width=5cm]{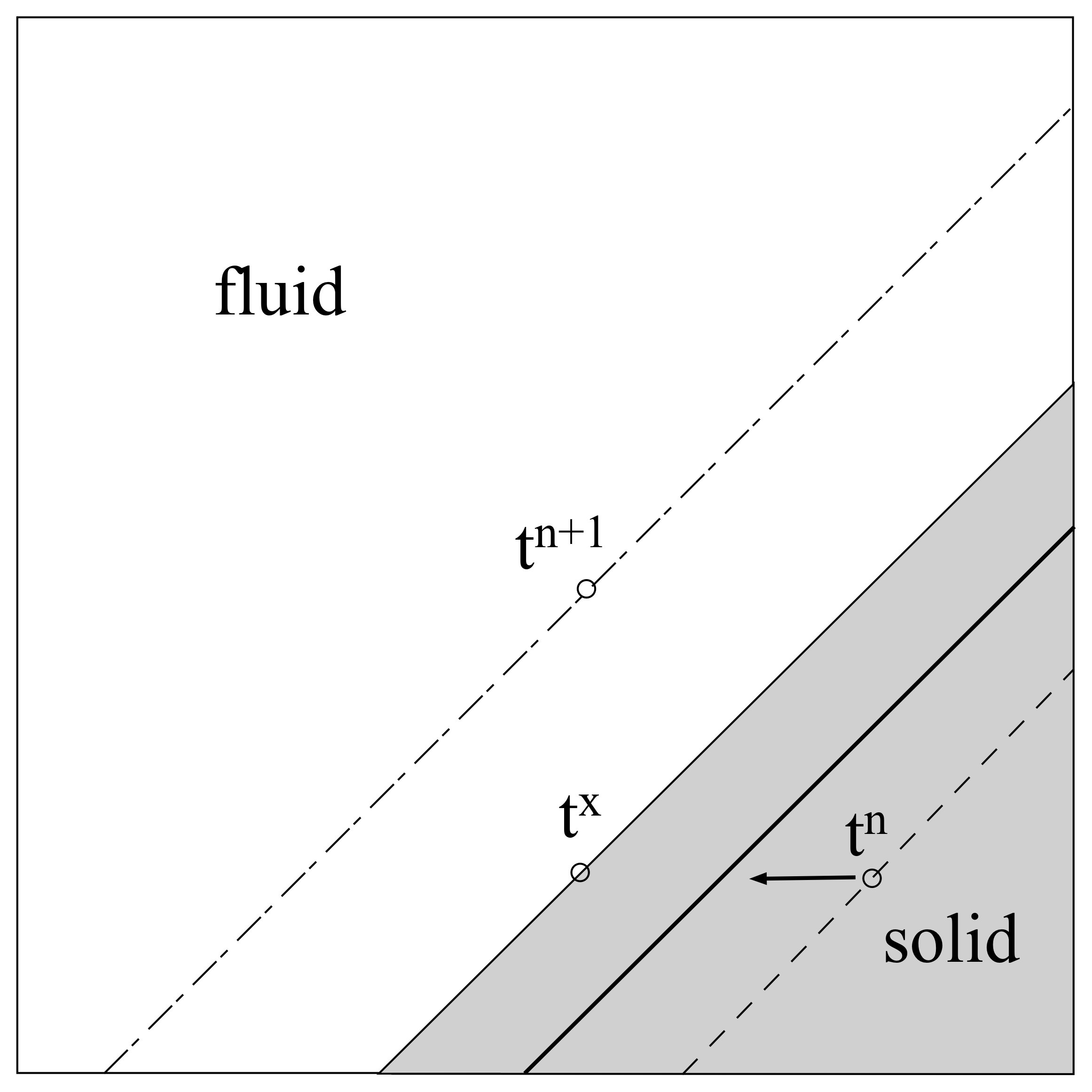}
(b) \includegraphics[width=5cm]{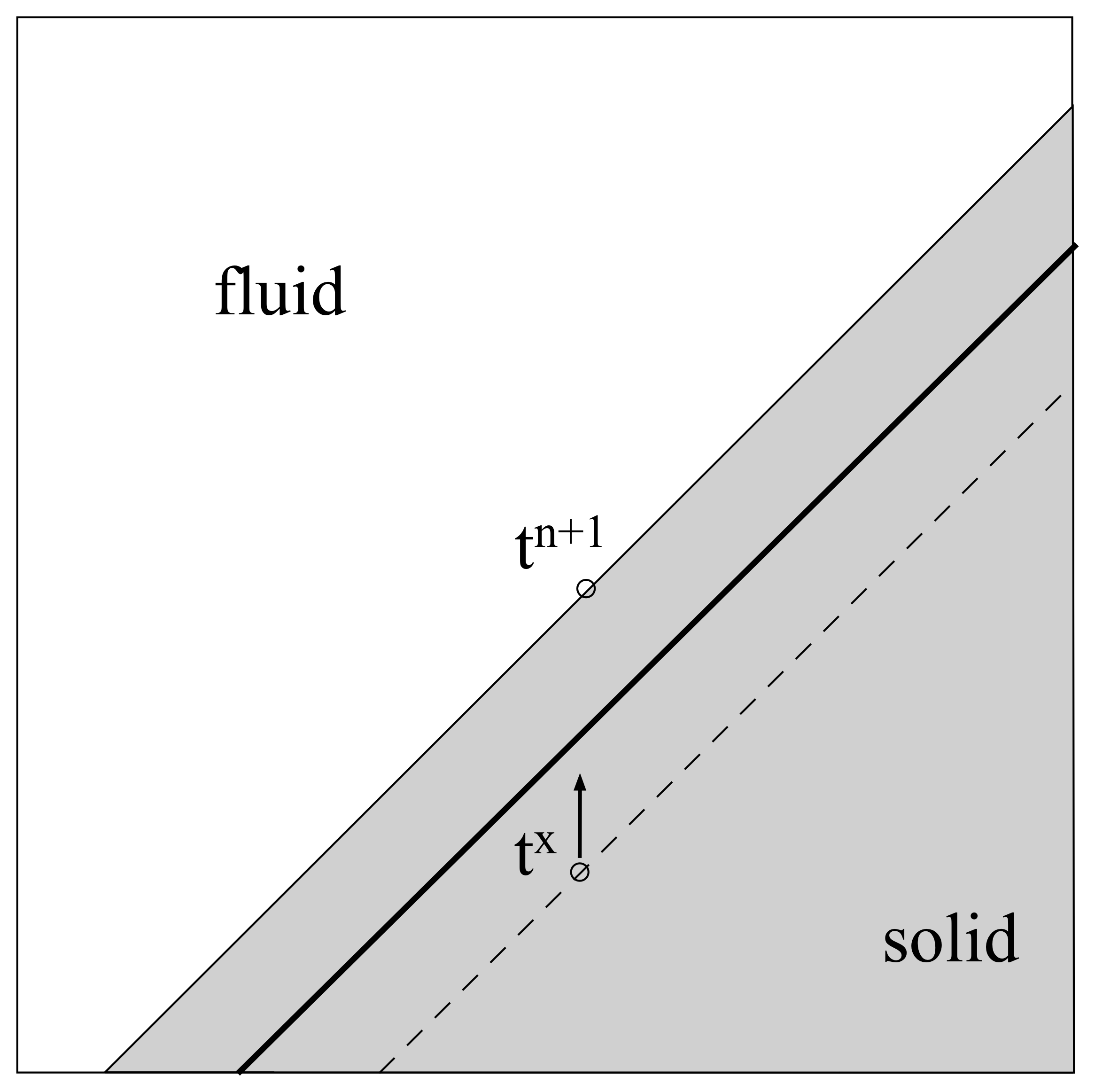}
\caption{x-sweep and y-sweep for the diagonal moving piston. (a) x-sweep, starting from the dashed line, $t^{n}$, and ending at the solid line, $t^{x}$. Bold solid line: Time averaged boundary for the x-sweep. (b) y-sweep, starting from the dashed line, $t^{x}$, and ending at the solid line, $t^{n+1}$. Bold solid line: Time averaged boundary for the y-sweep.\label{fig:SplitMassNonConservation_XandYsweeps}}
\end{centering}
\end{figure}

The x-sweep and y-sweep for a single time step, between times $n$ and $n+1$, is shown in Figure~\ref{fig:SplitMassNonConservation_XandYsweeps} for the same diagonal piston problem of a solid moving diagonally across a static fluid. Figure~\ref{fig:SplitMassNonConservation_XandYsweeps}~(a) shows the motion of the solid during the x-sweep. The initial position of the solid-fluid interface is given by a dashed line. The final position of the solid-fluid interface, at the end of the x-sweep, is given by the solid line, with the solid shaded in grey. The final position of the solid-fluid interface, at the end of the time step $n+1$, is given as a dash-dot line for reference. The time-averaged location of the solid-fluid interface during the x-sweep is given as a bold solid line, located between the dashed line and the thin solid line. Similarly, the y-sweep is shown in Figure~\ref{fig:SplitMassNonConservation_XandYsweeps}~(b). The spatial location of the solid-fluid interface at the end of the x-sweep, $t^{x}$, is given as a dashed line in this figure. This is the start position for the y-sweep. The location of the solid-fluid interface at the end of the y-sweep, at time $t^{n+1}$, is given as a solid line with the solid at this time shaded in grey. The time-averaged location of the solid-fluid interface during the y-sweep is given by a bold solid line, located between the dashed line and the thin solid line.
Following Ben-Artzi \& Falcovitz~\cite{ben2003generalized}, splitting the motion into individual x and y-sweeps (i.e. horizontal and vertical motion) simplifies the determination of the time averaged cut-cell face fraction, $\tilde{\beta}$, the cut-cell volume fraction and hence the average cut-cell distance, $\tilde{d}_{i-\frac{1}{2},j}$ compared to non-dimensional-splitting alternatives. For example, in Figure~\ref{fig:SplitMassNonConservation_XandYsweeps}~(a) during an x-sweep, the boundary movement is isolated to the horizontal direction, maintaining an invariant projection of the solid surface (between adjacent geometric vertices) onto the vertical cell interfaces, i.e., $\left(i+\frac{1}{2},j\right)$. The same argument holds for the projection of the same solid-fluid interface section on the cell interface $\left(i,j+\frac{1}{2}\right)$ during the y-sweep in Figure~\ref{fig:SplitMassNonConservation_XandYsweeps}~(b). The change in the cut-cell face area fractions, $\beta_{i+\frac{1}{2},j}$, and the average cut-cell distance, $\revision{d_{i-\frac{1}{2},j}}$, over a time step then becomes a function of the angle between the cell interface and the local solid-fluid interface (i.e., the surface slope in the current piece-wise planar surface representation), together with the local wall velocity component in the direction of the current sweep. 

For the numerical examples shown in this paper, in which a non-deforming rigid body is defined, we have made the assumption that the time step, $\Delta t$, is small enough such that the time averaged quantities can be approximated using a linear movement of $\frac{1}{2}\Delta t$ in both $x$ and $y$ sweeps. It is recognised that this is not a good approximation in general, and an improvement to this assumption to define an efficient general method for describing the time averaged cut-cell area fractions and length fractions is under development. Falcovitz et al.~\cite{falcovitz1997two} describes one such method.

Using the time averaged $\tilde{\beta}$ and $\tilde{d}$, the modified version of the shielded flux is therefore given by:

\begin{equation}
\tilde{\mathbf{F}}_{i-\frac{1}{2},j}^{\textrm{shielded}} = \tilde{\mathbf{F}}_{i,j}^{B} + \tilde{d}_{i-\frac{1}{2},j} \left(\mathbf{F}_{i-\frac{1}{2},j} - \tilde{\mathbf{F}}_{i,j}^{B} \right), \label{Klein2shieldedFluxMethod2}
\end{equation}

\noindent where, 

\begin{equation}
\tilde{\mathbf{F}}_{i,j}^{B} = \left[\rho^{*}\mathbf{u}^{*}, \rho^{*}\mathbf{u}^{*}\mathbf{u}^{*}+p^{*}\mathbf{I}, \rho^{*}\left(e+\frac{1}{2}\left(\mathbf{u}^{*}\right)^{2} + \frac{p^{*}}{\rho^{*}}\right)\right]. 
\label{ModifiedBoundaryFluxMethod2}
\end{equation}

\noindent The total flux at the boundary, $\left(i-\frac{1}{2},j\right)$, is then given by:

\begin{equation}
\tilde{\mathbf{F}}_{i-\frac{1}{2},j}^{2D} = \frac{1}{\tilde{\beta}_{i-\frac{1}{2},j}}\left[ \tilde{\beta}_{i+\frac{1}{2},j}\mathbf{F}_{i-\frac{1}{2},j}^{\textrm{unshielded}} + \left( \tilde{\beta}_{i-\frac{1}{2},j} - \tilde{\beta}_{i+\frac{1}{2},j}\right )\tilde{\mathbf{F}}^{\textrm{shielded}}_{i+\frac{1}{2},j} \right ], \label{Klein2DtotalfluxaMethod2}
\end{equation}

\noindent where the unshielded flux, $\mathbf{F}_{i-\frac{1}{2},j}^{\textrm{unshielded}}$ is the regular Cartesian cell flux given by Equation~\ref{Klein2Dunshieldeda}. 

\begin{figure}
\begin{centering}
\includegraphics[width=9cm]{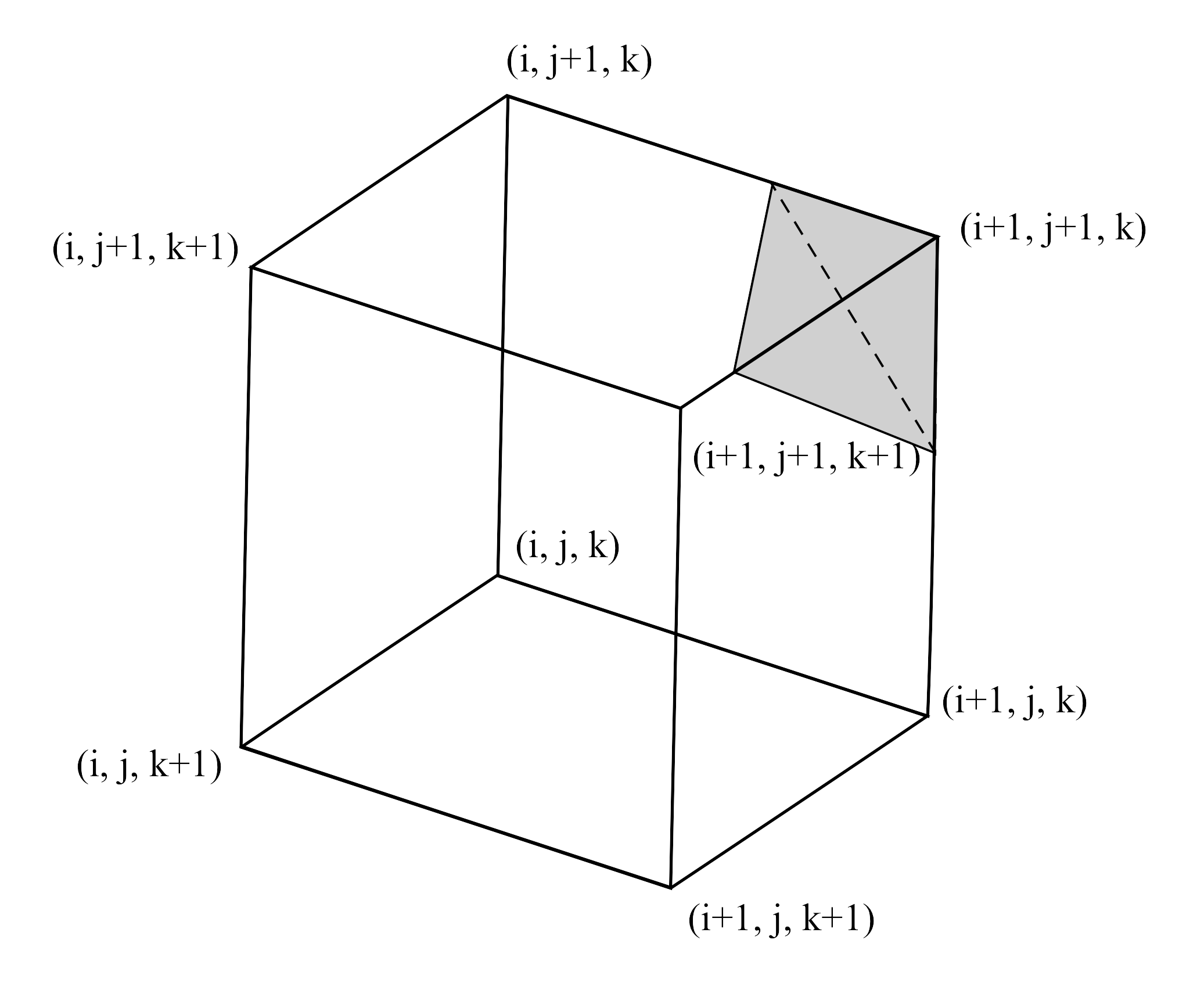}
\caption{Three-dimensional computational cell orientation, with the solid region highlighted in grey. The remaining (white) space contains a single phase fluid. \label{fig:ThreeDimensionalCube}}
\end{centering}
\end{figure}

In the current dimensional-split formulation, the time-averaged  boundary contribution in the $x$, $y$ and $z$ directions are applied independently. The final form can therefore be divided into a boundary movement in the $x$ direction, followed by a boundary movement in the $y$ direction, followed by a boundary movement in the $z$ direction, with the corresponding component of the time-averaged boundary movement applied at each stage. The three-dimensional form of the moving boundary conservation equations are therefore:

\begin{eqnarray}
\check{\alpha}_{l}^{x}\mathbf{Q}_{l}^{x} &=& \check{\alpha}_{l}^{n}\mathbf{Q}_{l}^{n} + \frac{\Delta t}{\Delta x} \left[\tilde{\beta}_{i-\frac{1}{2},j,k} \mathbf{F}_{i-\frac{1}{2},j,k}^{2D} - \tilde{\beta}_{i+\frac{1}{2},j,k} \mathbf{F}_{i+\frac{1}{2},j,k}^{2D} + \Delta\tilde{\beta}_{i-\frac{1}{2},j,k}\mathbf{F}_{l}^{\textrm{B}} - \Delta\tilde{\beta}_{i-\frac{1}{2},j,k}\mathbf{u}^{x}_{wall}\mathbf{Q}_{l}^{\textrm{B}}\right],\label{Klein2DcompleteMovingMethod2a}\\
\check{\alpha}_{l}^{xy}\mathbf{Q}_{l}^{xy} &=& \check{\alpha}_{l}^{x}\mathbf{Q}_{l}^{x} + \frac{\Delta t}{\Delta y} \left[\tilde{\beta}_{i,j-\frac{1}{2},k} \mathbf{G}_{i,j-\frac{1}{2},k}^{2D} - \tilde{\beta}_{i,j+\frac{1}{2},k} \mathbf{G}_{i,j+\frac{1}{2},k}^{2D} + \Delta\tilde{\beta}_{i,j-\frac{1}{2},k}\mathbf{G}_{l}^{\textrm{B}} - \Delta\tilde{\beta}_{i,j-\frac{1}{2},k}\mathbf{u}^{y}_{wall}\mathbf{Q}_{l}^{\textrm{B}}\right],\label{Klein2DcompleteMovingMethod2b}\\
\check{\alpha}_{l}^{n+1}\mathbf{Q}_{l}^{n+1} &=& \check{\alpha}_{l}^{xy}\mathbf{Q}_{l}^{xy} + \frac{\Delta t}{\Delta z} \left[\tilde{\beta}_{i,j,k-\frac{1}{2}} \mathbf{H}_{i,j,k-\frac{1}{2}}^{2D} - \tilde{\beta}_{i,j,k+\frac{1}{2}} \mathbf{H}_{i,j,k+\frac{1}{2}}^{2D} + \Delta\tilde{\beta}_{i,j,k-\frac{1}{2}}\mathbf{H}_{l}^{\textrm{B}} - \Delta\tilde{\beta}_{i,j,k-\frac{1}{2}}\mathbf{u}^{z}_{wall}\mathbf{Q}_{l}^{\textrm{B}}\right].\label{Klein2DcompleteMovingMethod2c}
\end{eqnarray}

\noindent where, $l=\left(i,j,k\right)$ and $\Delta\tilde{\beta}$ are used for compactness. The three-dimensional computational cell orientation and solid region used in this definition are shown in Figure~\ref{fig:ThreeDimensionalCube}.

\revision{To maintain a second-order accurate scheme, the x-y-z sequence is reversed from Equations~\ref{Klein2DcompleteMovingMethod2a}-\ref{Klein2DcompleteMovingMethod2c} in the next sweep via Strang-splitting, away from the solid surface. A detailed introduction to Strang-splitting for second-order accurate multi-dimensional schemes is given by Leveque~\cite{LeVequeBook2002}. As explained in Leveque~\cite{LeVequeBook2002}, choosing to reverse the x-y-z sequence in the Strang-splitting approach can be a more efficient alternative than taking half time-steps.}

Finally, it is recognised that in the limit as $\textbf{u}_{s} \rightarrow 0$, i.e., for a stationary solid, these equations reduce to the static boundary version of Klein et al.~\cite{Klein09},  i.e., the two-dimensional form of Equations~\ref{Klein2DcompleteMovingMethod1a}-\ref{Klein2DcompleteMovingMethod1b} reduce to Equations~\ref{Klein2DcompleteA}-\ref{Klein2DcompleteB} respectively.

\subsubsection{Covering and Uncovering Cut-Cells}
\label{sec:CutCellInitialisationSubmergence}

During the transit of the solid-fluid interface through the computational domain, computational cells can change over the period of a time step from regular Cartesian cells to cut-cells, from cut-cells to solid cells, or the reverse of these two. The change from a regular cell to a cut-cell is dealt with implicitly by the moving boundary method outlined in the preceding sections, as summarised in Equations~\ref{Klein2shieldedFluxMethod2}-\ref{Klein2DcompleteMovingMethod2c}. The uncovering of a cut-cell (transition from a solid cell to a cut-cell) and the `covering' of a cut-cell (transition from a cut-cell to a solid cell), however, require additional measures to maintain strict conservation.

\begin{figure}
\begin{centering}
(a) \includegraphics[width=8cm]{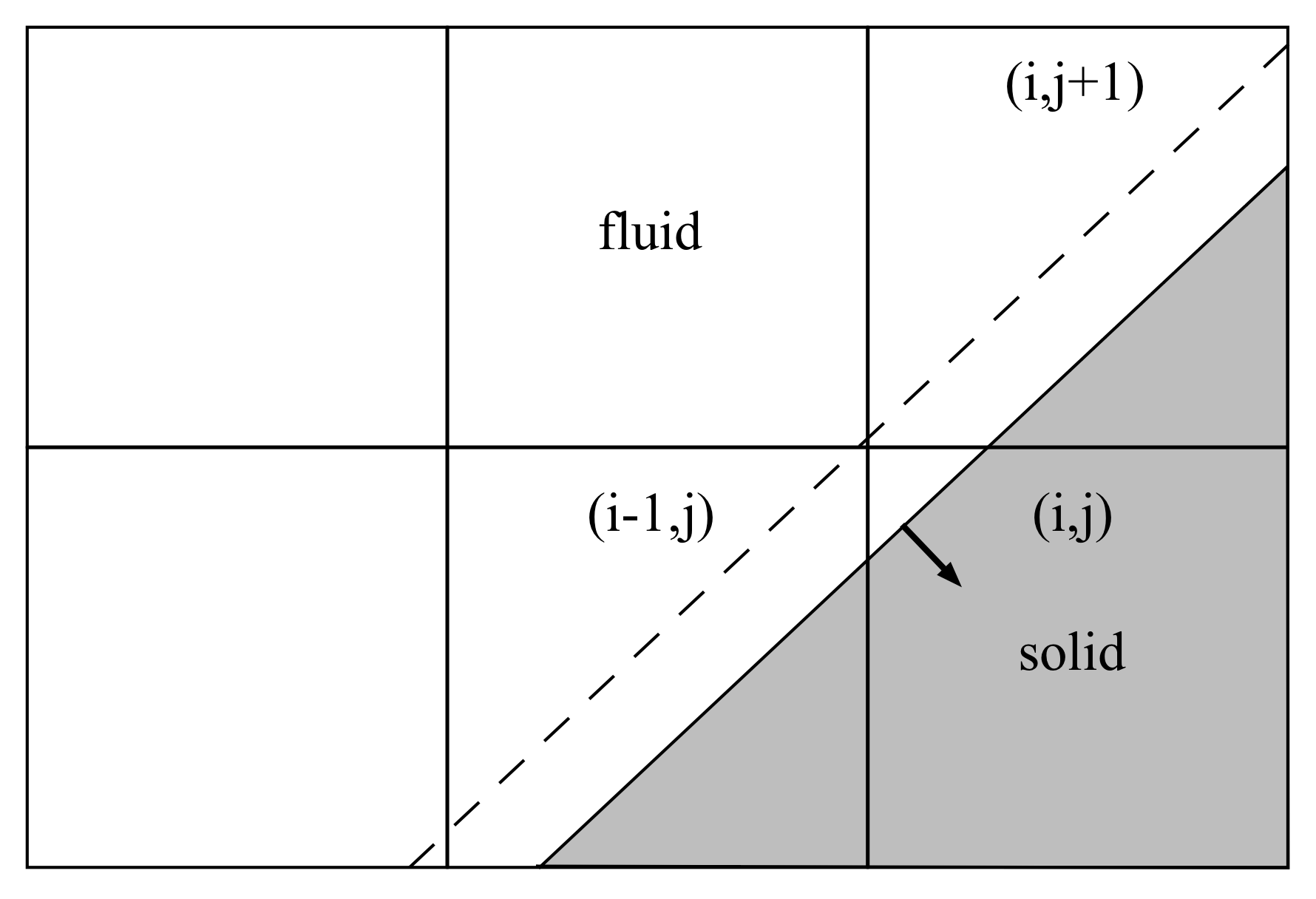}
(b) \includegraphics[width=8cm]{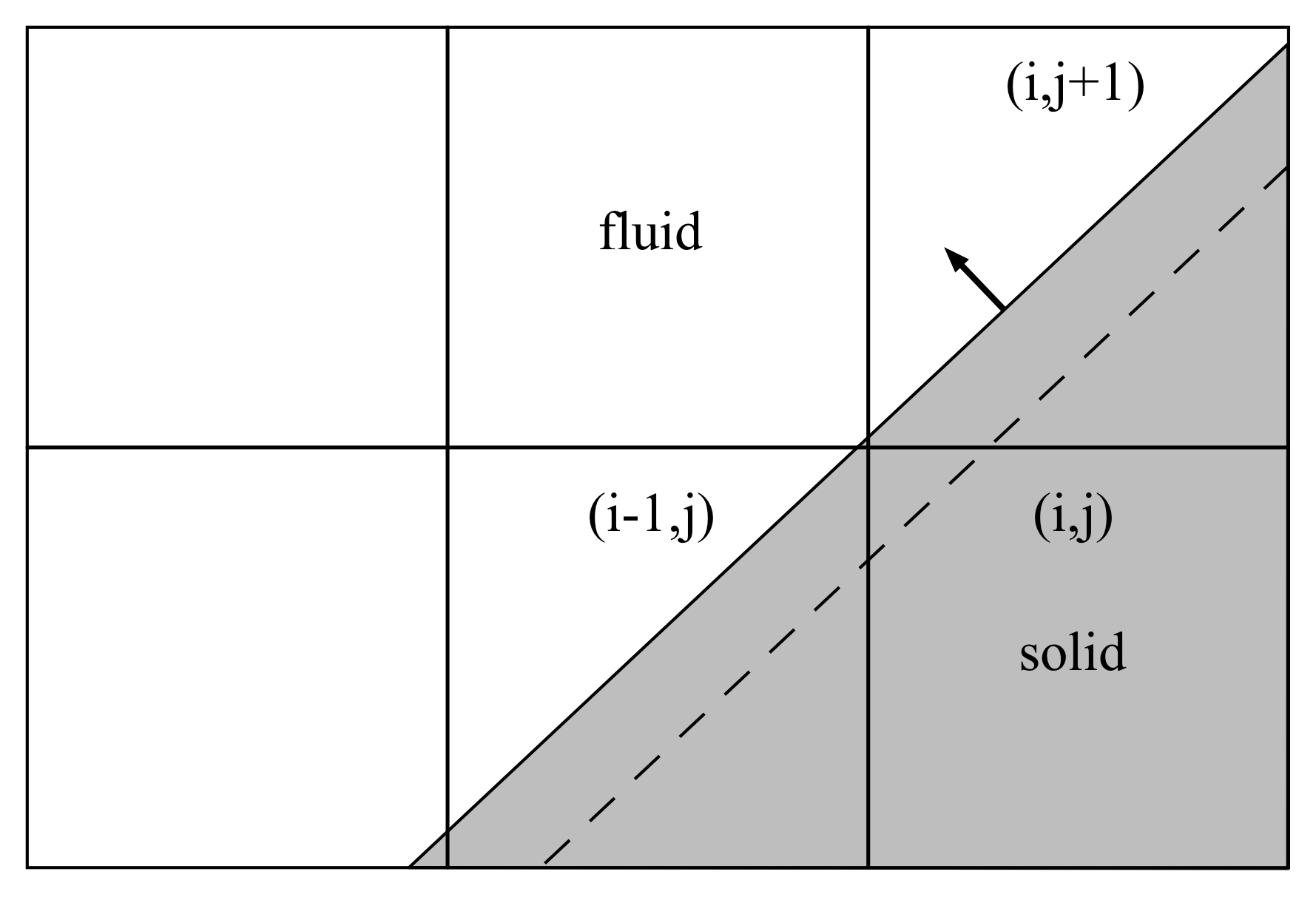}
\caption{The covering and uncovering of a cut-cell due to the solid-fluid interface moving from the dashed line at time level $n$ to the solid line at time level $n+1$. (a) An uncovered cut-cell at $\left(i,j\right)$. (b) A covered cut-cell at $\left(i,j\right)$. \label{fig:EmergingSubmergingCell}}
\end{centering}
\end{figure}

The change from a solid cell to a cut-cell (uncovering) is shown diagrammatically in Figure~\ref{fig:EmergingSubmergingCell}~(a). The reverse situation, of a cut-cell changing to a solid cell, is shown in Figure~\ref{fig:EmergingSubmergingCell}~(b).

Solid cells in the context of the current method are not updated and therefore have no valid conservative variable state. The change over time from a solid cell to a cut-cell therefore requires initialisation of the cut-cell before continuing the cut-cell update procedure. 

To maintain strict global conservation, an initial state is computed from a volume weighted average of the neighbouring cells, with the contribution from each neighbouring cell being reduced from its own conserved state.  Following the method outlined in Schneiders et al~\cite{schneiders2013accurate}, we define a volume weighting, $\sigma$, for a cell with shorthand index, $l=\left(i,j,k\right)$, as:

\begin{equation}
\sigma_{l} = \frac{\left(\check{\alpha} V\right)_{l}}{\sum\limits_{k\in M_{c}} \left(\check{\alpha} V\right)_{k}}, \label{volumeWeighting}
\end{equation}

\noindent where $M_{c}$ is the total number of neighbouring cells used as donors for the newly uncovered cell. This includes the uncovered cell itself. In this work, donor cells are defined as those neighbouring cells horizontal or vertical to the uncovered cell. For a two-dimensional computation this may include donor cells at $\left(i+1,j\right)$, $\left(i-1,j\right)$, $\left(i,j+1\right)$, $\left(i,j-1\right)$ and the uncovered cell, $\left(i,j\right)$. The uncovered cell, $\left(i,j\right)$ in Figure~\ref{fig:EmergingSubmergingCell}, is then initialised using a conserved variable state of:

\begin{equation}
\left(\check{\alpha} V\mathbf{Q}^{\textrm{new}}\right)_{i,j} = \sigma_{i,j} \sum\limits_{k\in M_{d}} \left(\check{\alpha} V\mathbf{Q}\right)_{k}
\label{cutCellInitialisation}
\end{equation}

\noindent where $M_{d}$ is the total number of donor cells (excluding the uncovered cell). Note that $\check{\alpha}=0$ for a solid cell and $0 < \check{\alpha} \leq 1$ for fluid cells. This ensures only cut-cells or regular fluid cells contribute to the new cell initialisation. Each of the donor cells is then updated with a modified state according to:

\begin{equation}
\left(\check{\alpha} V\mathbf{Q}\right)^{\textrm{modified}}_{d} = \left(\check{\alpha} V\mathbf{Q}\right)_{d} - \sigma_{l} \left(\check{\alpha} V\mathbf{Q}\right)_{d}. \label{cutCellInitialisationDonorCellUpdate}
\end{equation}

\noindent where subscripts $d$ and $l$ distinguish the donor and uncovered cell respectively for the general case. The penalty of maintaining conservation in this way is a local reduction in the primitive variables for those donor cells surrounding the uncovered cell. \revisionAB{The influence of this penalty can be reduced through increased use of Adaptive Mesh Refinement (AMR), a method of locally refining the computational mesh around flow features of interest using a hierarchical mesh configuration based on a user defined mesh refinement criterion. The density is used as the refinement criterion in all the test cases described in this paper that use AMR, employing a user-defined refinement threshold which is individually tailored for each test case to maintain a tight confinement of the refined mesh around the compressible flow features, while maintaining efficiency in preventing unnecessary sub-division of the refined region into excessively small AMR blocks.}

To attempt to maintain global conservation, the reverse situation of a cut-cell being covered completely by the solid, requires the conserved variables in the cut-cell, $\left(i,j\right)$, to be completely expelled through the fluid portion of the inter-cell boundary. The cut-cell conserved variables are gradually passed to the neighbouring cells through the shielded flux as the cut-cell volume fraction $\check{\alpha} \rightarrow 0$. To ensure strict conservation is maintained after the cut-cell is completely covered, any residual fluid remaining in the cut-cell is distributed to the neighbouring cells. The residual conserved state remaining in the covered cut-cell is computed as:

\begin{equation}
\left(\check{\alpha}^{n}V \mathbf{Q}\right)^{res} = \left(\check{\alpha} V\mathbf{Q}\right)^{n}_{l} - \Delta t \sum_{m=1}^{N_{f}} \left( \mathbf{F}\left(\mathbf{Q}\right)_{m} \cdot \mathbf{n}_{m}\right)_{l} \left(\mathcal{S}_{m}\right)_{l}.
\label{massDeficitEquation}
\end{equation}

\noindent This is distributed to the surrounding fluid cells as:

\begin{equation}
\left(\check{\alpha} V\mathbf{Q}\right)^{\textrm{modified}}_{r} = \left(\check{\alpha} V\mathbf{Q}\right)_{r} + \frac{\left(\check{\alpha} V\right)_{r}}{\sum\limits_{k\in M_{r}} \left(\check{\alpha} V\right)_{k}} \left(\check{\alpha}^{n}V \mathbf{Q}\right)^{res}, \label{cutCellCoverDonation}
\end{equation}

\noindent where, $M_{r}$ is the total number of recipient cells surrounding the covered cut-cell and subscript $r$ denotes the recipient cell index.

In the next section, the moving boundary cut-cell method described here is applied to a range of compressible flows involving the time dependent interaction of a moving rigid body with a single phase fluid. \revisionB{To compute the regular cell fluxes, and those of the cut-cell unshielded flux component, the MUSCL-Hancock solver (see, for example, Toro~\cite{Toro}) is used in combination with the HLLC Riemann solver and van-Leer slope limiter.} \revision{Currently, complex geometries can be imported from the standard STL file format, allowing simulation of Computer Aided Design (CAD) models. However, as the test case geometries in this paper are simple geometries, each geometry was explicitly specified using simple polygons, or the NACA profile.}


\section{NUMERICAL RESULTS}
\label{sec:NumericalResults}


\subsection{The Sloped Channel Flow - Moving Boundary Validation}

\begin{figure}
\begin{center}
\includegraphics[width=14cm]{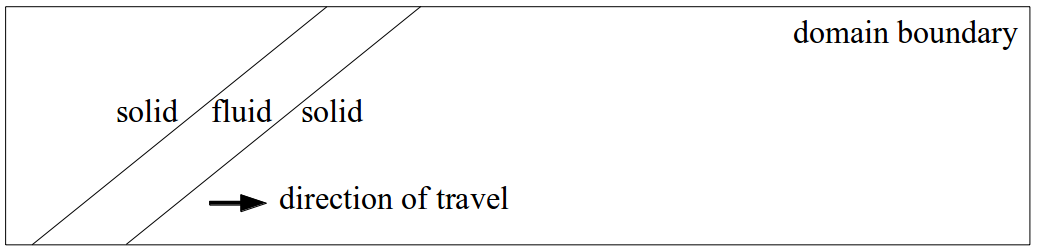}
\caption{Sloping channel flow configuration. The test geometry consists of a channel 
oriented an angle of $30^{\circ}$ to the horizontal boundary. 
The channel moves in the positive horizontal direction over time.}
\label{ChannelDomain_pic}
\end{center}
\end{figure}

\begin{figure}
\begin{center}
\includegraphics[width=10cm]{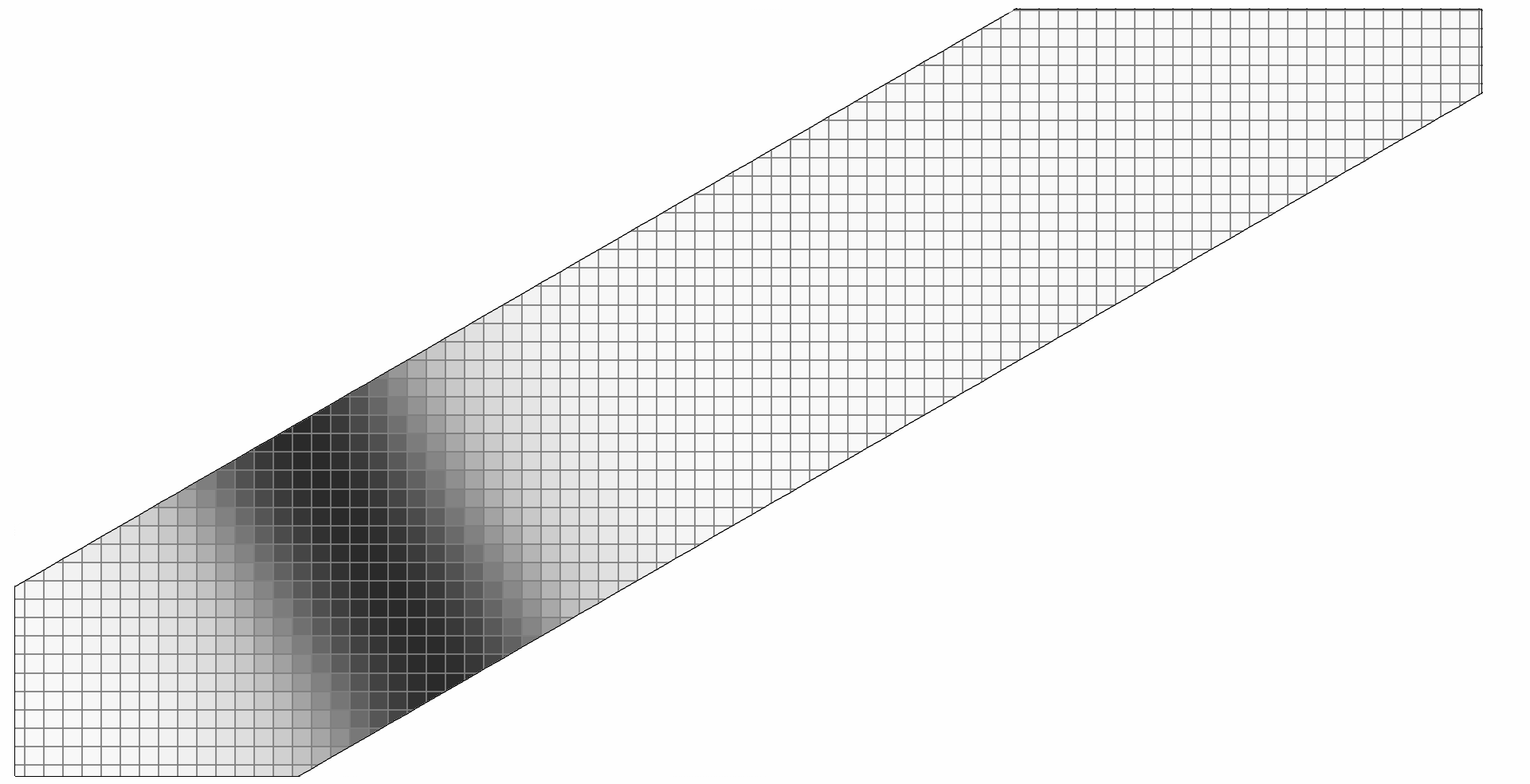}
\caption{Sloping channel case mesh showing the relative orientation of 
the Cartesian mesh, the channel boundaries and the initial density perturbation. 
The density perturbation is shown using a grey scale pseudo-colour plot.}
\label{ChannelMesh}
\end{center}
\end{figure}

We consider as our first test case the propagation of a simple perturbation in density within an inclined channel. 
The simulation configuration for this problem is shown in Figure \ref{ChannelDomain_pic}. 
Two parallel interfaces at an angle of $30^{\circ}$ to the horizontal plane form 
an embedded boundary, creating cut-cells with a large range of volume
fractions. The initial condition consists of a simple density wave, 
centred at $x=0, y=0$ and travelling in a direction parallel 
to the channel walls. The channel itself moves in the positive x-direction over time.
This test case is intended to assess how the moving boundary approach handles reducing and 
increasing cut-cell volume fractions, as well as the covering and uncovering of cut-cells. By eliminating the 
force of the cut-cell boundary movement on the fluid, we therefore effectively simulate a static channel test 
case resolved by a mesh that is moving in the negative horizontal direction. 
Conservation is maintained through an identical movement of the two channel walls. This is confirmed during the computation.
The solid wall boundary condition is assessed for the moving boundary cut-cell method in subsequent test cases. The CFL number used in this test case is 0.5. Figure~\ref{ChannelMesh} shows the relative orientation of 
the Cartesian mesh, the channel boundaries and the initial density perturbation, which is shown as a grey scale pseudo-colour plot.

The simulation is performed on a domain $0.4$m by $0.08$m, with
the width of the channel set to $0.02$m. 
The final time for each simulation is chosen as $t=1$ms. This allows the density pulse to propagate 
across the majority of the domain, and the entire cut-cell surface to experience the full range of cut-cell volume fractions at each point.

\begin{figure}
\begin{center}
\includegraphics[width=12cm, clip]{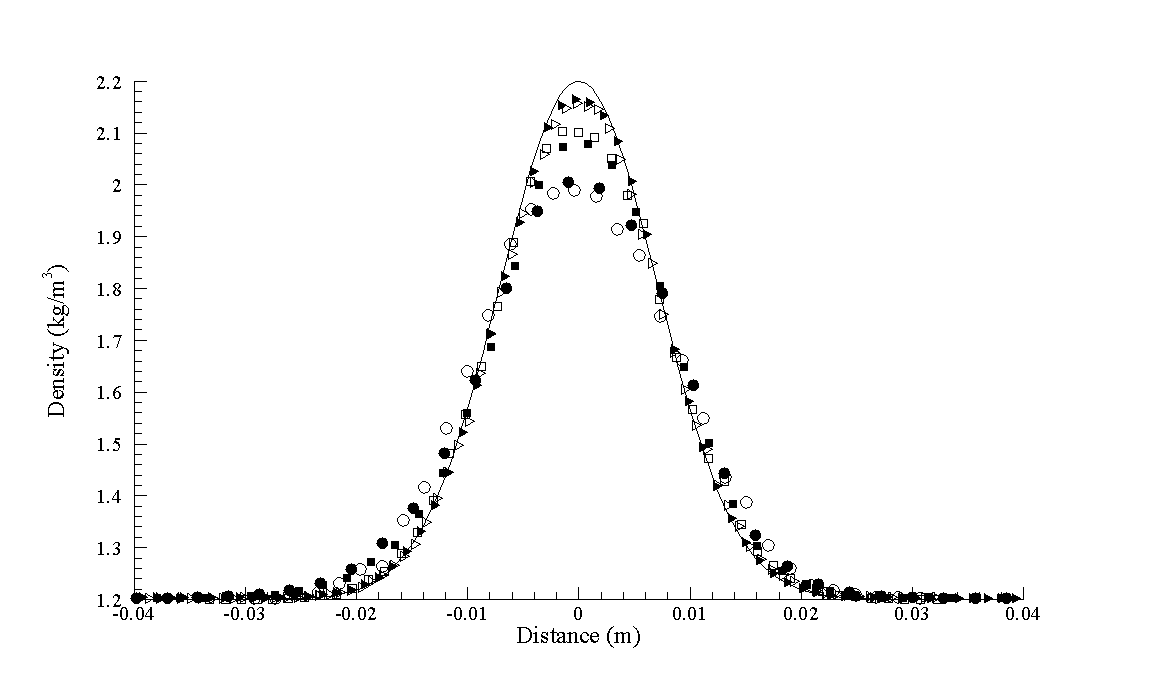}
\caption{Density profiles along the lower channel surface at $t=1$ms. Comparison of the moving boundary method with a static boundary method for a static channel. 
Comparison with an exact analytical density profile. $640 \times 128$ (triangular symbols), $320 \times 64$ (square symbols), $160 \times 32$ (circles).
Static cut-cell case (filled symbols), Moving cut-cell case (hollow symbols). The black line shows the exact analytical density profile at $t=1$ms.}
\label{channeldensityprofile}
\end{center}
\end{figure}

The density along the lower channel surface at a time of $t=1$ms is given in figure \ref{channeldensityprofile} for mesh resolutions of $160\times 32$, $320\times 64$ and $640\times 128$ cells. 
This figure shows a comparison of the moving boundary results against the exact analytical density profile and against an equivalently 
resolved static boundary simulation using the current solver, applied to a static channel.

A satisfactory comparison between the moving boundary and static boundary results is shown for the three mesh resolutions given. Numerical diffusion of the cut-cell method is shown at the 
outer edges of the density pulse. This, of course, varies with mesh resolution, with the lowest diffusion evident in the $640 \times 128$ results. 
Numerical dissipation is also evident at the centre of the pulse resulting in a reduction in the density pulse peak value.

The cut-cell error with increasing mesh resolution is shown for the static boundary results at mesh resolutions of 
$160 \times 32$, $320 \times 64$, $640 \times 128$, $1280 \times 256$ and $2560 \times 512$ in table~\ref{TableSimpleWaveChannelErrorsStatic}. 
This is compared with the moving boundary cut-cell error in table~\ref{TableSimpleWaveChannelErrorsMoving}.

\begin{table}
\centering
\begin{tabular}{|c|c|c|c|c|c|c|c|}
\hline
Resolution & Cell size $\left(\Delta x=\Delta y\right)$ & $L_1$ norm & $L_1$ order & $L_2$ norm & $L_2$ order & $L_{\inf}$ norm & $L_{\inf}$ order \\
\hline
$160 \times 32$ & 2.5(-3)m  & 4.74(-2) & - & 1.03(-1) & - & 5.99(-1) & - \\
$320 \times 64$ & 1.25(-3)m & 1.79(-2) & 1.40 & 5.32(-2) & 9.53(-1) & 4.19(-1) & 5.16(-1) \\
$640 \times 128$ & 6.25(-4)m & 5.61(-3) & 1.67 & 2.64(-2) & 1.01 & 2.53(-1) & 7.28(-1) \\
$1280 \times 256$ & 3.12(-4)m & 1.69(-3) & 1.73 & 1.01(-2) & 1.39 & 1.21(-1) & 1.06 \\
$2560 \times 512$ & 1.56(-4)m & 4.61(-4) & 1.87 & 3.82(-3) & 1.40 & 5.87(-2) & 1.07 \\
\hline
\end{tabular}
\caption{Static boundary results. Global $L_1$, $L_2$ and $L_{\inf}$ error norms  for the propagation of a density
  perturbation in a static boundary sloped channel at $t=1$ms.
  Values given in compact form as $A(B)$ to represent $A\times 10^B$.}
\label{TableSimpleWaveChannelErrorsStatic}
\end{table}

\begin{table}
\centering
\begin{tabular}{|c|c|c|c|c|c|c|c|}
\hline
Resolution & Cell size $\left(\Delta x=\Delta y\right)$ & $L_1$ norm & $L_1$ order & $L_2$ norm & $L_2$ order & $L_{\inf}$ norm & $L_{\inf}$ order \\
\hline
$160 \times 32$ & 2.5(-3)m  & 6.84(-2) & - & 2.11(-1) & - & 8.82(-1) & - \\
$320 \times 64$ & 1.25(-3)m & 3.01(-2) & 1.18 & 1.07(-1) & 9.80(-1) & 7.11(-1) & 3.11(-1) \\
$640 \times 128$ & 6.25(-4)m & 1.07(-2) & 1.49 & 5.37(-2) & 9.95(-1) & 5.38(-1) & 4.02(-1) \\
$1280 \times 256$ & 3.12(-4)m & 3.32(-3) & 1.69 & 2.51(-2) & 1.10 & 2.99(-1) & 8.47(-1) \\
$2560 \times 512$ & 1.56(-4)m & 9.50(-4) & 1.81 & 9.91(-3) & 1.34 & 1.51(-1) & 9.86(-1) \\
\hline
\end{tabular}
\caption{Moving boundary results. Global $L_1$, $L_2$ and $L_{\inf}$ error norms  for the propagation of a density
  perturbation in a moving boundary sloped channel at $t=1$ms. 
  Values given in compact form as $A(B)$ to represent $A\times 10^B$.}
\label{TableSimpleWaveChannelErrorsMoving}
\end{table}

The $L_1$ and $L_2$ norm error order indicate that the global order of accuracy increases towards second order as the mesh resolution increases. This is evident for the static boundary test case in table~\ref{TableSimpleWaveChannelErrorsStatic} and the moving boundary test case in table~\ref{TableSimpleWaveChannelErrorsMoving}. The cut-cell method is locally first order accurate. The global error increases above unity as the influence of the first order boundary region diminishes with an increasing proportion of second order accurate regular cells at higher mesh resolutions. The $L_{\inf}$ norm error shows a smaller order of accuracy, increasing to around unity for both the static and moving boundary cases. The $L_{\inf}$ norm error uses the maximum error in the domain. This is therefore limited by the first order accurate cut-cell boundary region. The increase in the global order of accuracy with increasing mesh resolution justifies the introduction of Adaptive Mesh Refinement (AMR) to resolve flow features of large density gradient and regions close to the cut-cell boundary, restricting the spatial influence of the first order accurate cut-cell method with the current second-order accurate numerical solver. \revisionAB{Refinement of the mesh using AMR is test case specific and can be chosen to refine across, for example, shock waves, contact surfaces or vortices through adjustment of a refinement criterion. Computational cells are flagged for refinement based on a user defined state variable threshold value. Adjacent computational cells with a difference in the chosen state variable are flagged for refinement. In the current computation, the refinement variable is density. If desired, in the current implementation, the rigid body surface can also be independently flagged for refinement.}\newline
\revision{The analysis presented so far in this section studies the convergence properties of the current explicit scheme at fixed CFL number, Tables~\ref{TableSimpleWaveChannelErrorsStatic} and~\ref{TableSimpleWaveChannelErrorsMoving}, yielding the global accuracy of the method in both time and space as the mesh resolution increases. In an extension to this analysis, we isolate the space and time resolution contributions by adjusting the CFL number as the mesh is refined. Suppose we have a scheme that is formally second order accurate in both space and time. A measure of the spatial accuracy in isolation can then be computed by allowing the CFL number to vary between mesh resolutions, according to:}

\begin{equation}
\left(CFL\right)_{n} = 2^{-n} \left(CFL\right)_{0}
\label{CFLvarying}
\end{equation}

\noindent \revision{where, $n$ is a positive integer representing the current refinement level. This value is defined as $0$ for the least refined mesh case and increases by one as the mesh resolution is doubled in all spatial directions. $CFL_{0}$ in Equation~\ref{CFLvarying} therefore denotes the CFL number used in the least refined case. With the CFL number linked to the mesh refinement in this manner, the time step approaches zero more rapidly than the mesh resolution, therefore the spatial error will dominate in this limit. Using this strategy, the global error for the moving channel case is given in Table~\ref{TableSimpleWaveChannelErrorsMovingSpace}. The most refined three mesh resolutions from Table~\ref{TableSimpleWaveChannelErrorsMoving} have been tested here. Table~\ref{TableSimpleWaveChannelErrorsMovingSpace} documents the CFL number as the mesh resolution increases.}

\begin{table}
\centering
\begin{tabular}{|c|c|c|c|c|c|c|c|c|}
\hline
Resolution & Cell size $\left(\Delta x=\Delta y\right)$ & CFL & $L_1$ norm & $L_1$ order & $L_2$ norm & $L_2$ order & $L_{\inf}$ norm & $L_{\inf}$ order \\
\hline
$640 \times 128$ & 6.25(-4)m & 0.5 & 9.27(-3) & -  & 3.92(-2) & -  & 4.62(-1) & - \\
$1280 \times 256$ & 3.12(-4)m & 0.25 & 3.08(-3) & 1.59 & 1.92(-2) & 1.03 & 2.64(-1) & 8.07(-1) \\
$2560 \times 512$ & 1.56(-4)m & 0.125 & 8.70(-4) & 1.82 & 7.45(-3) & 1.37 & 1.38(-1) & 9.36(-1) \\
\hline
\end{tabular}
\caption{\revision{Moving boundary results. Isolation of the spatial component of the global $L_1$, $L_2$ and $L_{\inf}$ error norms for the propagation of a density
  perturbation in a moving boundary sloped channel at $t=1$ms. 
  Values given in compact form as $A(B)$ to represent $A\times 10^B$.}}
\label{TableSimpleWaveChannelErrorsMovingSpace}
\end{table}

\revision{Table~\ref{TableSimpleWaveChannelErrorsMovingSpace} indicates that the influence of the first order cut-cell method is restricted to a small region close to the surface in terms of the global order of accuracy when the mesh is sufficiently well resolved. When considered together with Table~\ref{TableSimpleWaveChannelErrorsMoving}, which demonstrated a similar convergence to second order accuracy, the implication is that the accuracy of the second-order method is globally maintained in both time and space for the most refined mesh resolution cases.}


\subsection{Circular Cylinder Interaction with a Stationary Shock Wave}

An assessment of the current moving boundary cut-cell method for computing the interaction of a curved surface with a moderately strong shock wave is presented in the second test case. A stationary shock wave initially separates two regions of fluid with a discontinuous rise in pressure and density, accompanied by a decrease in fluid velocity, in the downstream direction. The circular cylinder moves downstream with the same velocity as the initial low pressure region, eventually passing through the stationary shock wave. This models the experimental situation of a stationary cylinder and moving shock wave with the same rise in pressure and density across the shock wave. An ideal gas is used, with specific heat capacity ratio, $\gamma=1.4$, and specific gas constant, $R=287.058$J/kgK. The CFL number is 0.5.

\begin{figure}
\begin{center}
(a) \includegraphics[width=5cm, clip]{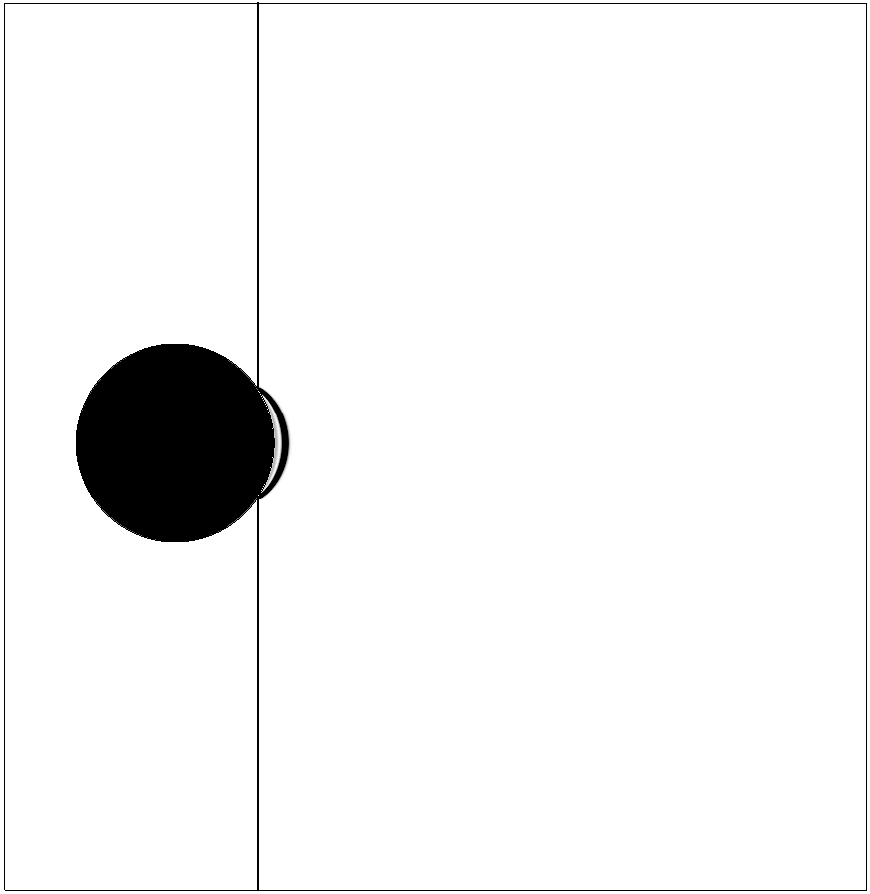}
(b) \includegraphics[width=5cm, clip]{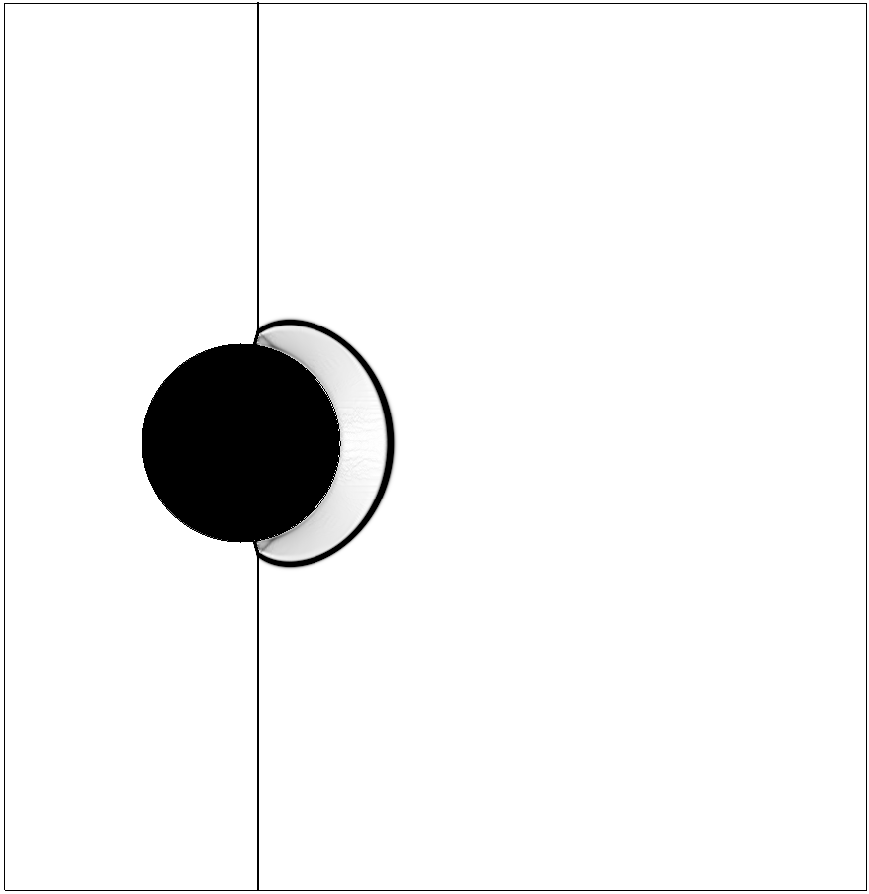}\\
(c) \includegraphics[width=5cm, clip]{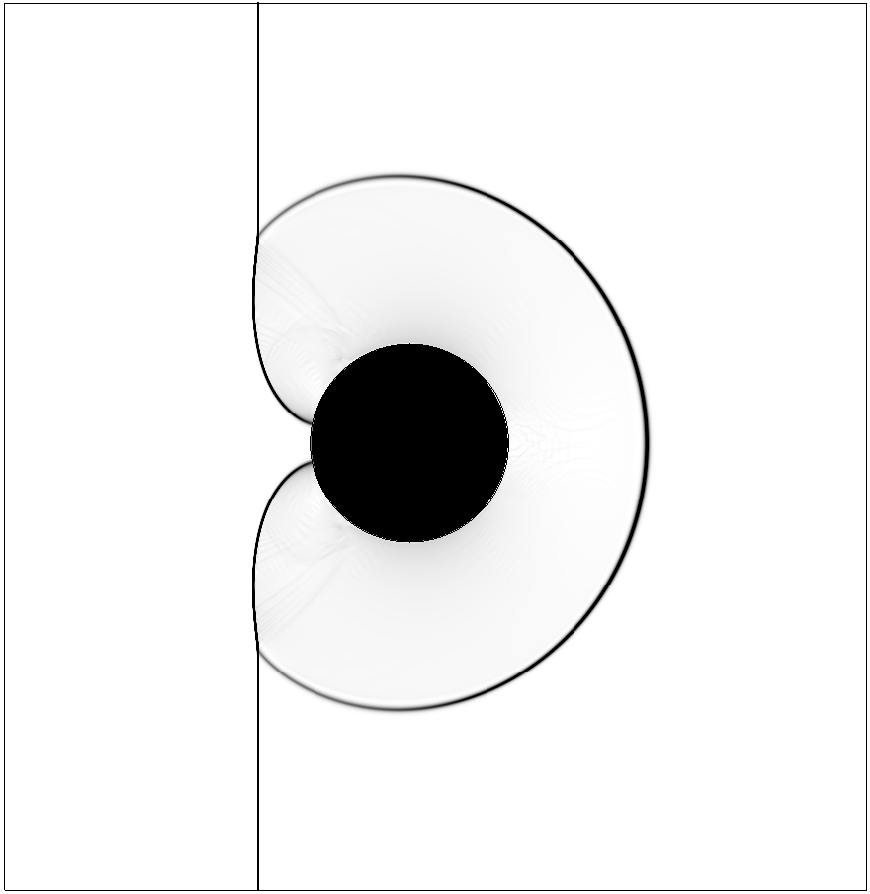}
(d) \includegraphics[width=5cm, clip]{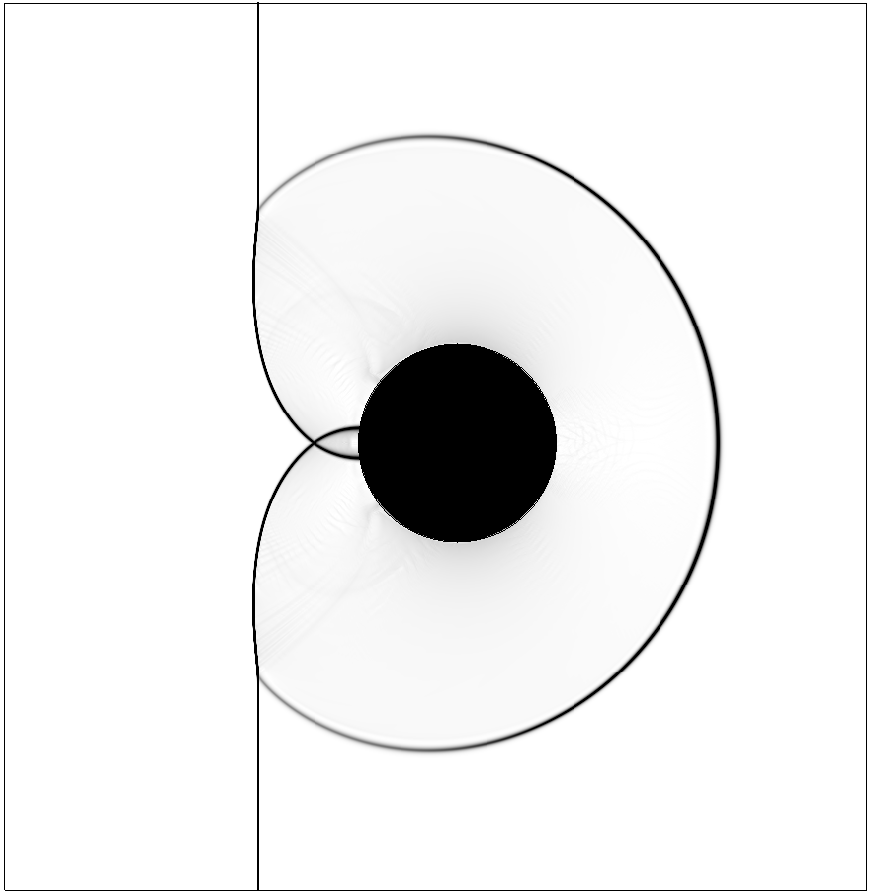}
\caption{Numerical Schlieren plot series showing a circular cylinder passing through a stationary shock wave. (a) 6$\mu$s, (b) 35$\mu$s, (c)106$\mu$s, (d) 126$\mu$s. Times taken with respect to the initial contact of the cylinder and shock wave.}
\label{cylinderShock}
\end{center}
\end{figure}

Figure~\ref{cylinderShock} shows the interaction of a stationary Mach 1.34 shock wave with a moving cylinder of diameter 0.04m. 
Specifically, numerical Schlieren plots are used to highlight compressible flow features at times 6$\mu$s, 35$\mu$s, 106$\mu$s and 126$\mu$s after the initial contact of the shock wave with the circular cylinder surface.

This Mach 1.34 shock wave test case is initialised by globally imposing the stationary shock wave flow properties throughout the computational domain. Specifically, the flow properties to the left and right-hand sides of the shock wave are $\rho_L=$ 0.595kg/m$^3$, $u_L=$ 459.54m/s, $p_L=$ 0.05MPa and $\rho_R=$0.944kg/m$^3$, $u_R=$289.86m/s, $p_R=$0.096MPa. The circular cylinder velocity is $u_{cyl}=$459.54m/s, matching the fluid velocity on the left-hand side of the shock wave. The cylinder is initially located one diameter upstream of the shock wave. Before contact with the shock wave, therefore, the circular cylinder initially moves with the same velocity as the surrounding fluid. The absence of pressure waves in the low pressure fluid at this time demonstrates that the consistency condition is satisfied by the moving cut-cell method.

\begin{figure}
\begin{center}
\includegraphics[width=8cm]{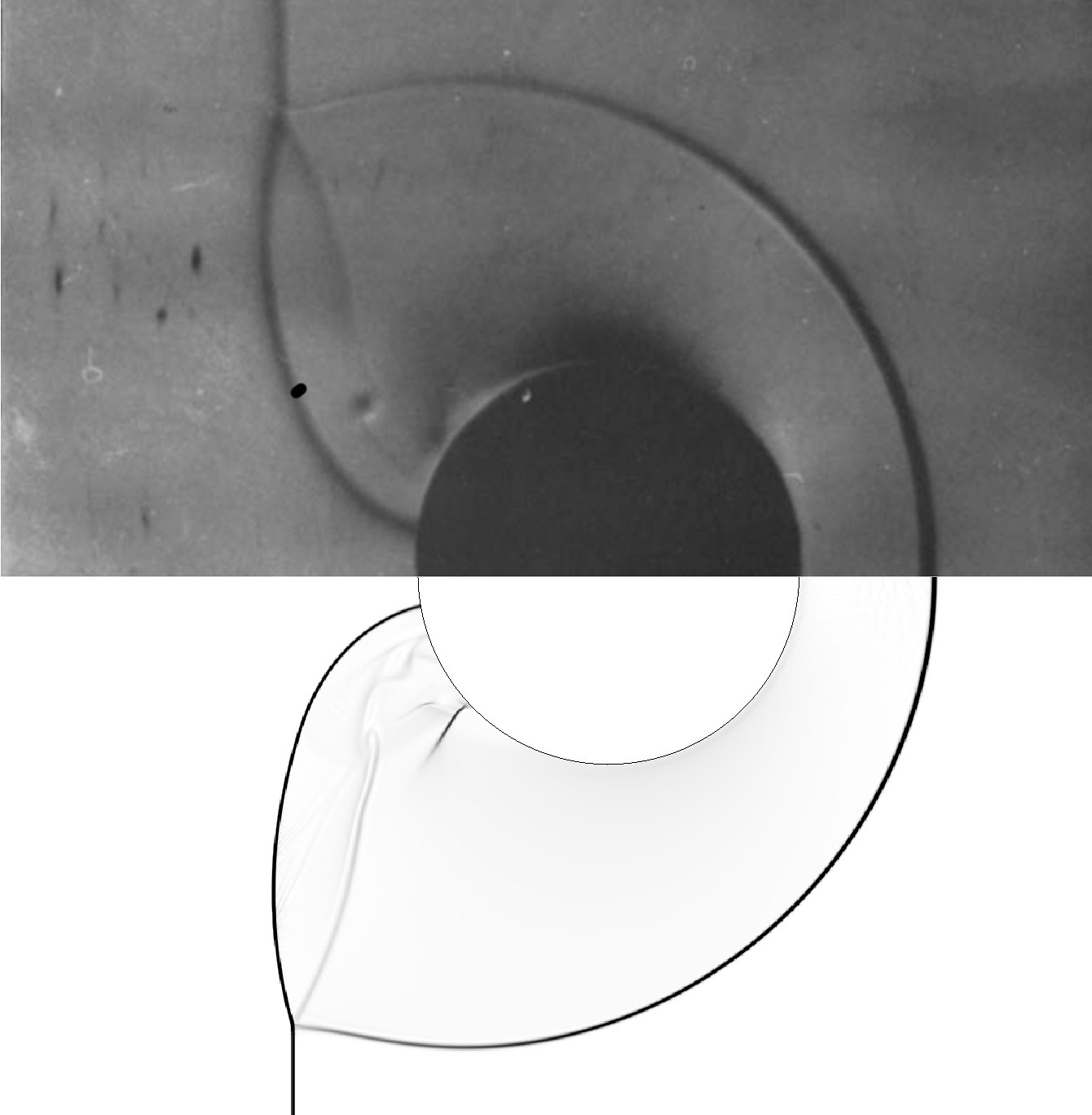}
\caption{Circular cylinder passing through a Mach 2.82 shock wave. Top: Schlieren image from the experiment of Bryson \& Gross~\cite{bryson1961diffraction} in which a moving shock wave passes over a fixed cylinder. Bottom: Numerical Schlieren image using the current moving boundary cut-cell method and modelling a stationary shock wave with a moving cylinder.}
\label{cylinderSchlierenComparison}
\end{center}
\end{figure}

The location of the initial shock wave and the development of additional compressible flow features, such as a bow shock and shock wave triple points are evident in Figure~\ref{cylinderShock}. As outlined in Zdravkovich~\cite{zdravkovich1997flow}, the location, strength and development of these compressible flow features are dependent on the magnitude of the pressure and density rise across the shock wave. Even moderate increases in the shock wave Mach number result in an altered shock wave pattern, providing a sound basis with which to assess the moving boundary cut-cell surface condition. Figure~\ref{cylinderSchlierenComparison} shows a comparison between the current moving boundary cut-cell method and an experiment by Bryson \& Gross~\cite{bryson1961diffraction} at a higher shock wave Mach number of $M = 2.82$. At this Mach number, the pressure and density ratio across the shock wave is $p_{R}/p_{L} = 9.045$Pa and $\rho_{R}/\rho{L} = 3.674$kgm$^{-3}$. The Cartesian mesh resolution used for Figure~\ref{cylinderSchlierenComparison} is $\Delta x = \Delta y = 5 \times 10^{-3}D$, where $D=0.04$m is the circular cylinder diameter.

The distance from the circular cylinder to the bow shock, and the location of the shock triple point, in the numerical Schlieren compare well with the experiment of Bryson \& Gross~\cite{bryson1961diffraction}. The contact discontinuity, which runs between the circular cylinder and the shock triple point is also well located, indicating the moving cut-cell method is able to capture the correct advection dominated flow behaviour regarding the shock wave interaction with the moving circular cylinder. 


\subsection{Schardin Experiment}
\label{sec:SchardinWedge2D}

The interaction of a solid body with an incident shock wave is extended in this test case to the classic experiment of Schardin \cite{Schardin57},
in which a planar shock interacts with a triangular wedge. This interaction results in the generation of a relatively
complex flow pattern, with transonic tip vortices extending downstream of the wedge, as well as Mach stems, triple points, reflected and
diffracted shocks, slip lines and acoustic waves. Schardin's original shadowgraph images provide an initial basis for comparison, with further experimental imaging by Chang and Chang
\cite{chang2000shock} allowing for more detailed verification. As with the circular cylinder test case, we model a stationary shock wave, matching the experimental pressure and density rise across the shock wave. A triangular wedge, moving with the same velocity as the surrounding fluid on one side of the shock wave, is then allowed to pass through the stationary shock wave. This models the experiment, in which a fixed stationary wedge in static fluid is subjected to the impingement of a moving shock wave. An ideal gas is used, with specific heat capacity ratio, $\gamma=1.4$, and specific gas constant, $R=287.058$J/kgK. The CFL number used in this test case is 0.5.

The solid body in this test case takes the form of an equilateral triangle, with base length 20mm. 
The triangular wedge is placed in a fluid with a density and pressure of $\rho=0.595$ kg m$^{-3}$ and $p=0.05$ MPa. The incident shock wave has Mach number 1.34, giving a post-shock state of $\rho=0.944$ kg m$^{-3}$ and $p=0.0964$ MPa.

The numerical results are computed on the domain of size $0.2 \times 0.2$ m, with a regular Cartesian mesh of cell size $\Delta x =
\Delta y = 5 \times 10^{-4}$ m. A transmissive boundary condition is defined at the computational domain extents. 
The simulation is run to a final time of $t=200 \mu$s. 
The start time, $t=0$, is taken as the time at which the initial contact of the triangular wedge and the shock wave occurs. 

\begin{figure}
\begin{center}
(a) \includegraphics[width=5cm]{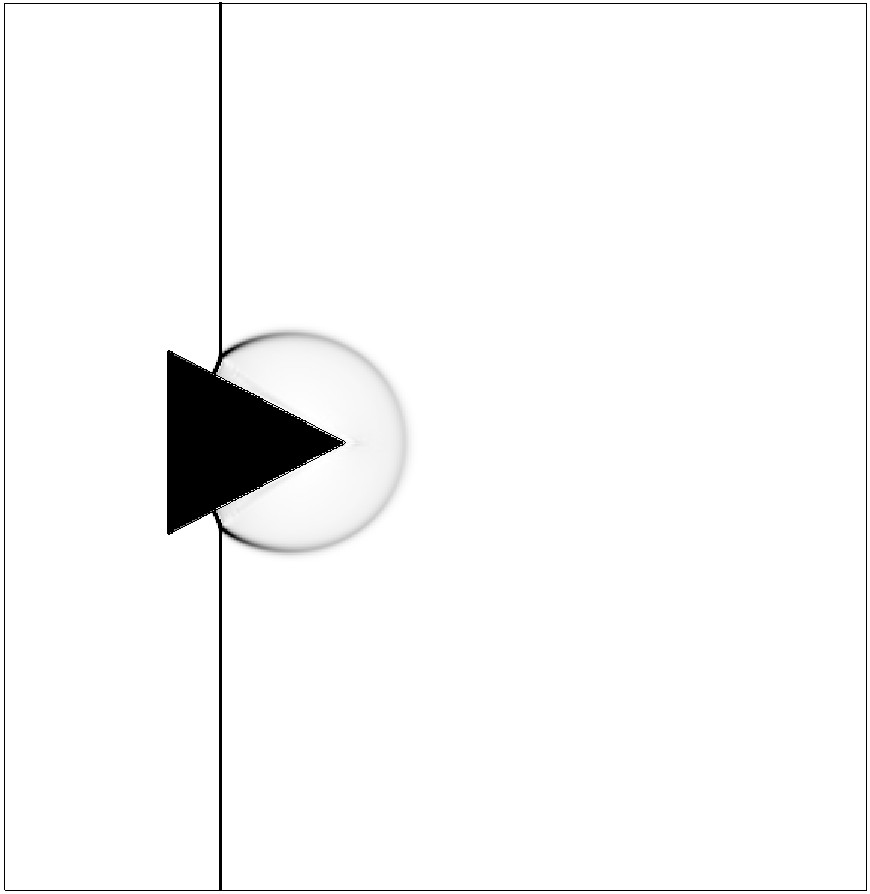}
(b) \includegraphics[width=5cm]{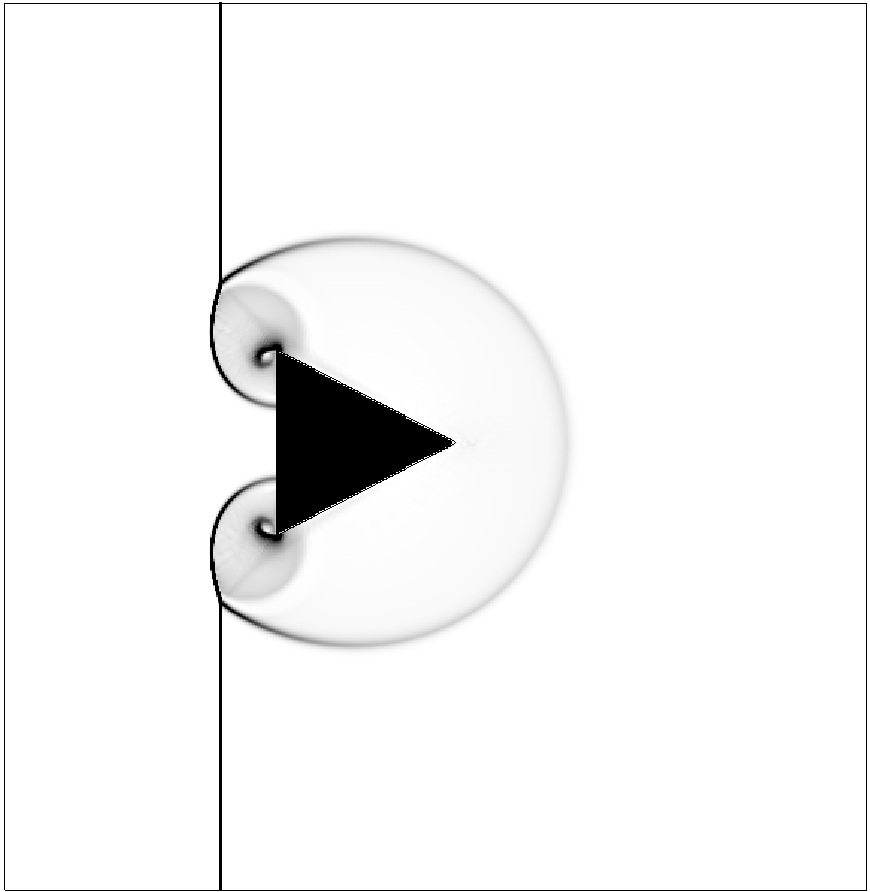} \\
(c) \includegraphics[width=5cm]{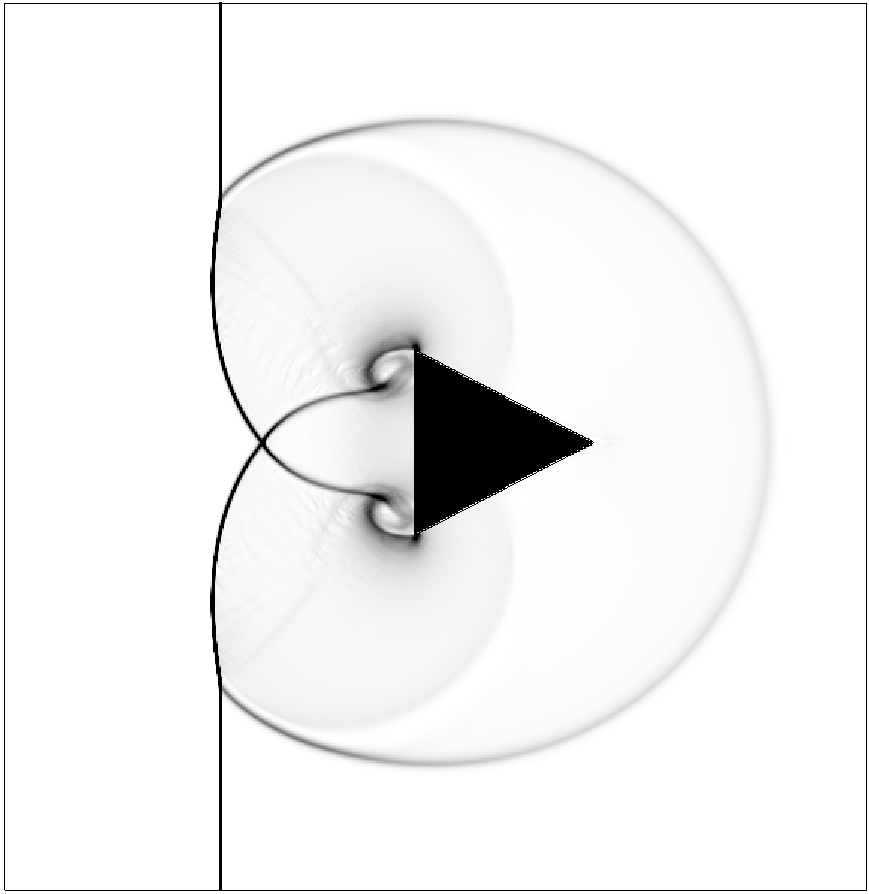}
(d) \includegraphics[width=5cm]{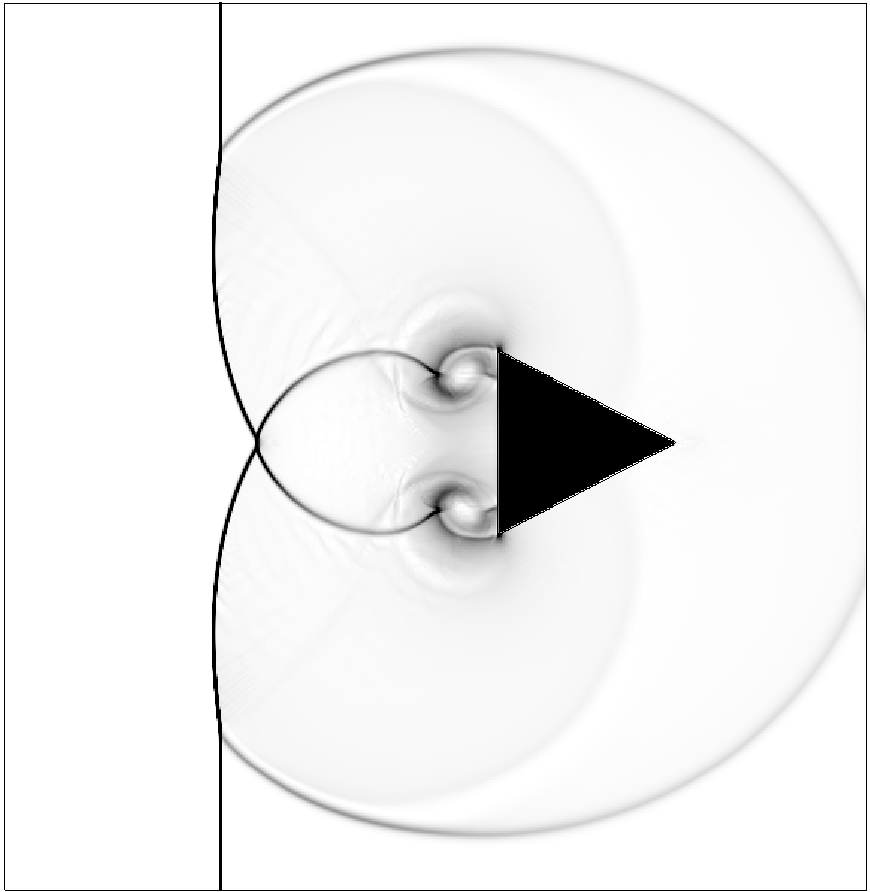} \\
\caption{Numerical Schlieren plots at (a) 28 $\mu$s,
  (b) 53 $\mu$s, (c) 79 $\mu$s and (d) 102
$\mu$s after the instant the shock wave contacts the triangular wedge. 
This corresponds to Figure 5 of \cite{chang2000shock}.}
\label{schardin_snapsnots}
\end{center}
\end{figure}

The development of a compressible flow pattern around the wedge as it passes through the static shock wave is documented at times, $t=28 \mu$s, $53 \mu$s, $79 \mu$s and $102 \mu$s in Figure \ref{schardin_snapsnots}, using numerical Schlieren to highlight density gradient. These correspond to the experimental interferograms and computational results in \cite{chang2000shock}.

\begin{figure}
\begin{center}
\includegraphics[height=8cm,width=9cm]{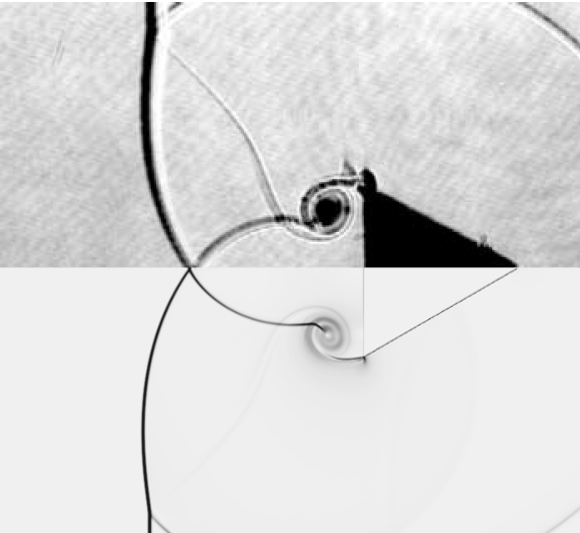}
\caption{Shock wave development past a Schardin Wedge at 91$\mu$s after initial contact. Top: Experimental Schlieren image by Chang $\&$ Chang~\cite{chang2000shock}. Bottom: Numerical Schlieren image using the current moving boundary cut-cell method.}
\label{schardinComparison}
\end{center}
\end{figure}

Numerical Schlieren results from the current moving boundary simulation are compared with experimental Schlieren imaging from Chang $\&$ Chang~\cite{chang2000shock}, at 91$\mu$s after the initial contact of the triangular wedge with the shock wave, in Figure~\ref{schardinComparison}.

\begin{figure}
\begin{center}
\includegraphics[width=9cm]{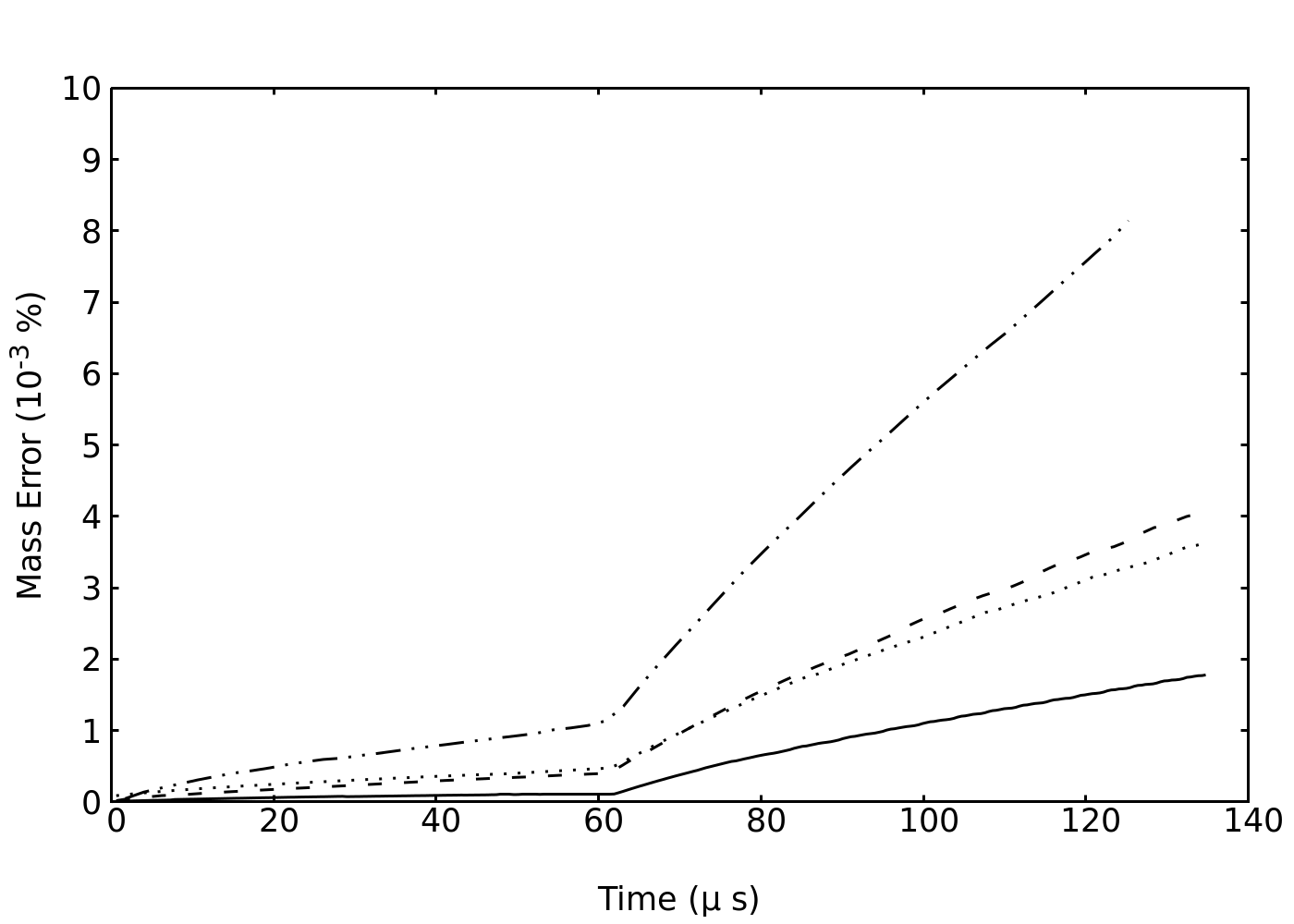}
\caption{\revisionB{Mass conservation error for three mesh resolutions. Wedge apex coincident with the computational cell boundary, solid line: $\Delta x = \Delta y = 2.5 \times 10^{-4}$m. Dashed line: $\Delta x = \Delta y = 5 \times 10^{-4}$m, Dash-dot line: $\Delta x = \Delta y = 10 \times 10^{-4}$m. Dotted line: Wedge apex coincident with the computational cell centre, with $\Delta x = \Delta y = 5 \times 10^{-4}$m.}}
\label{schardinMassError}
\end{center}
\end{figure}

The tip vortices are evident downstream of the triangular wedge in both the numerical Schlieren image and the experiment. The location of the main flow features, the bow shock, tip vortices, downstream shock triple point and the contact surfaces that extend between the tip vortices and the shock triple point are all well located. The size of the tip vortex matches the experimental Schlieren, with an embedded shock wave located in the vortex at the same relative location. The bow shock in Figure~\ref{schardinComparison} extends outside of the photograph extent, with the shock triple point located along the upper edge. These results provide further evidence of the ability for the current moving boundary cut-cell method to accurately reproduce the essential flow physics. \revisionAB{It is noted that in this test case, the apex of the wedge coincides with the computational cell edge. As the current implementation of the cut-cell method represents the rigid body surface as a linear approximation in each computational cell, sharp corners that fall within a single computational cell are therefore not resolved. This represents a limitation to the current implementation. The influence on the global mass conservation of this under-resolution of the region around the three corners is, however, expected to be reduced through judicious use of AMR to better resolve the rigid body surface.}
\revisionB{Whilst capturing discontinuous surface gradients (such as the wedge corners) within the confines of a single cut-cell is the subject of further development, we attempt to quantify the level of error introduced in the current method from this, and other sources of error. These include the accuracy to which slender regions of the rigid body are represented by the level-set method, used in the current method to track the movement of the rigid body. These sources of error, which can alter the local shape of the rigid body as the surface moves across a cut-cell, are resolution dependent. We therefore repeat the same Schardin wedge test case in order to quantify the mass conservation error as the wedge passes through the stationary shock wave. We ensure that the influence of the wedge on the surrounding flow does not extend beyond the computational domain boundary by limiting the end time of the computation and increasing the computational domain size. The results are calculated as the difference in the total mass in the domain with respect to the total mass at the start of the computation. This mass error is presented as a percentage of the total mass at the start of the computation. For these mass conservation tests, a single, uniform mesh is defined across the computational domain in order to eliminate the possibility of the fine-coarse AMR boundary updates negatively influencing the results. The mass conservation error is given in Figure~\ref{schardinMassError}. Three mesh resolutions are compared in this figure, with the leading corner of the sharp wedge coincident with a computational cell boundary. To account for changes in mass conservation resulting from the location of the corners, the mass conservation error from a further test is shown in which the sharp leading corner of the wedge is coincident with the centre of a computational cell with $\Delta x = \Delta y = 5 \times 10^{-4}$m.}

\revisionB{Figure~\ref{schardinMassError} shows that, as expected, the error decreases significantly with increasing resolution as the representation of the rigid body surface improves and changes in the representation of the wedge corners between time steps is confined to a smaller spatial area. The location of the wedge apex is also shown to have an effect on the mass conservation. The slight improvement in mass conservation when the leading corner is located at a computational cell centre is due, in part, to an improvement in the representation of the other two corners in this test. The reduction in the mass conservation error with increasing mesh resolution justifies the judicious use of AMR to improve the representation of the rigid body while minimising the associated increase in computational effort.}

\subsection{Oscillating NACA 0012 Aerofoil}

This fourth test case models a NACA 0012 aerofoil pitching between $2.526^{\circ}$ and $-2.494^{\circ}$ from the oncoming free stream flow direction. The free-stream Mach number is $M_{\infty}=0.755$. This test case has been used to validate a number of numerical methods, including Venkatakrishnan \& Mavriplis~\cite{venkatakrishnan1995implicit}, Krishnan \& Liu~\cite{kirshman2006flutter}, as well as cut-cell methods by Schneiders et al.~\cite{schneiders2013accurate} and Murman et al.~\cite{murman2003implicit}. The free stream density, velocity and pressure are $\rho_{\infty} = 1.225 kg/m^{3}$, $u_{\infty} = 256.9 m/s$ and $p_{\infty} = 101325 Pa$ respectively. This is used to globally initialise the computational domain at the start of the computation. The computation develops initially using a static aerofoil, imposed for the first $t=0.2$ seconds to allow the flow around the aerofoil to develop. The aerofoil then oscillates with incidence angle $\varphi(t) = 0.016 +2.51 \textrm{sin}\left(\omega t \right)$, where the reduced frequency, $\omega c/2u_{\infty}=0.0814$ based on the aerofoil chord length, $c$, and the free-stream velocity. A perfect gas is used, with specific heat capacity ratio, $\gamma=1.4$, and specific gas constant, $R=287.058$J/kgK.

\begin{figure}
\begin{center}
(a)\includegraphics[width=10cm]{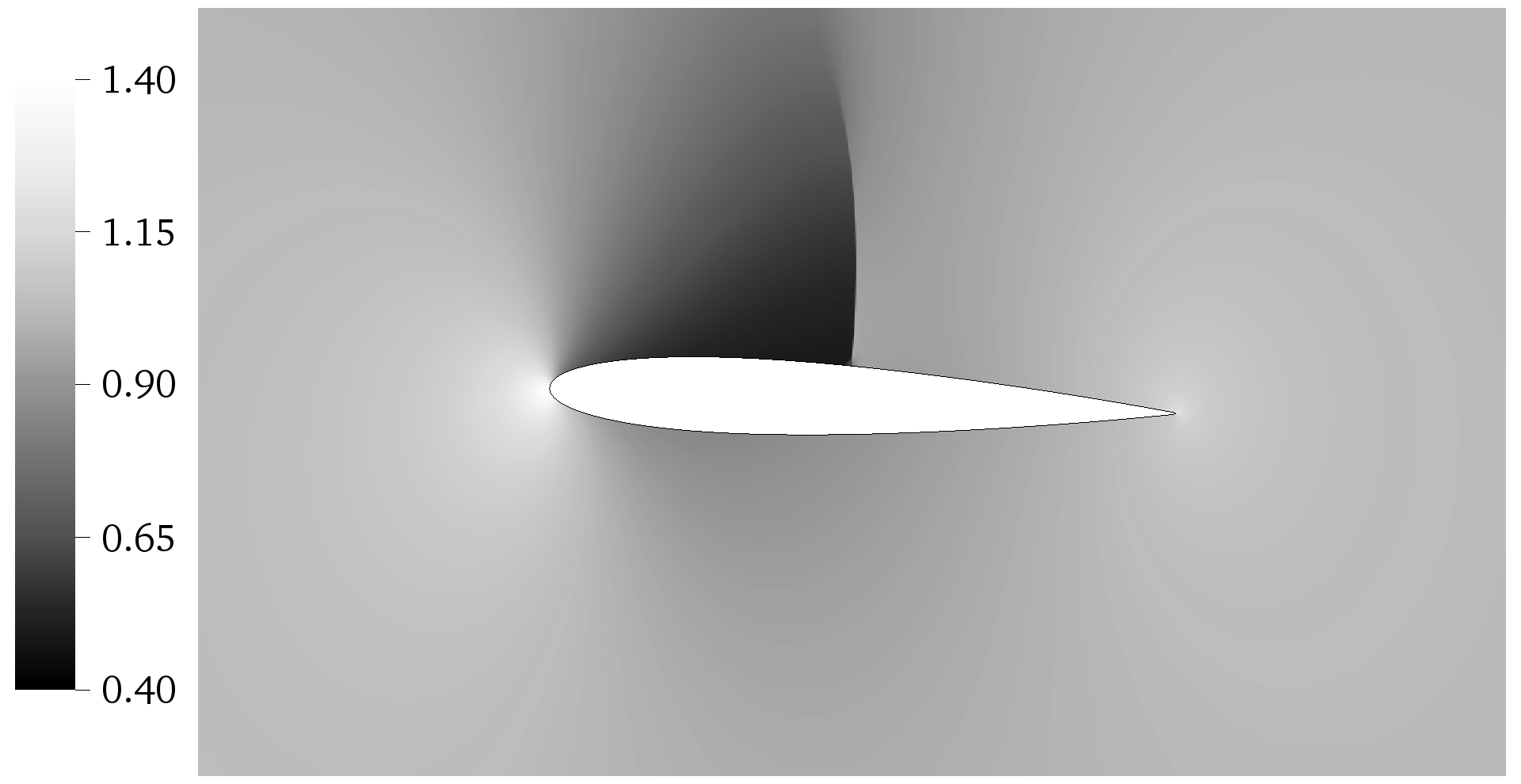} \\
(b)\includegraphics[width=10cm]{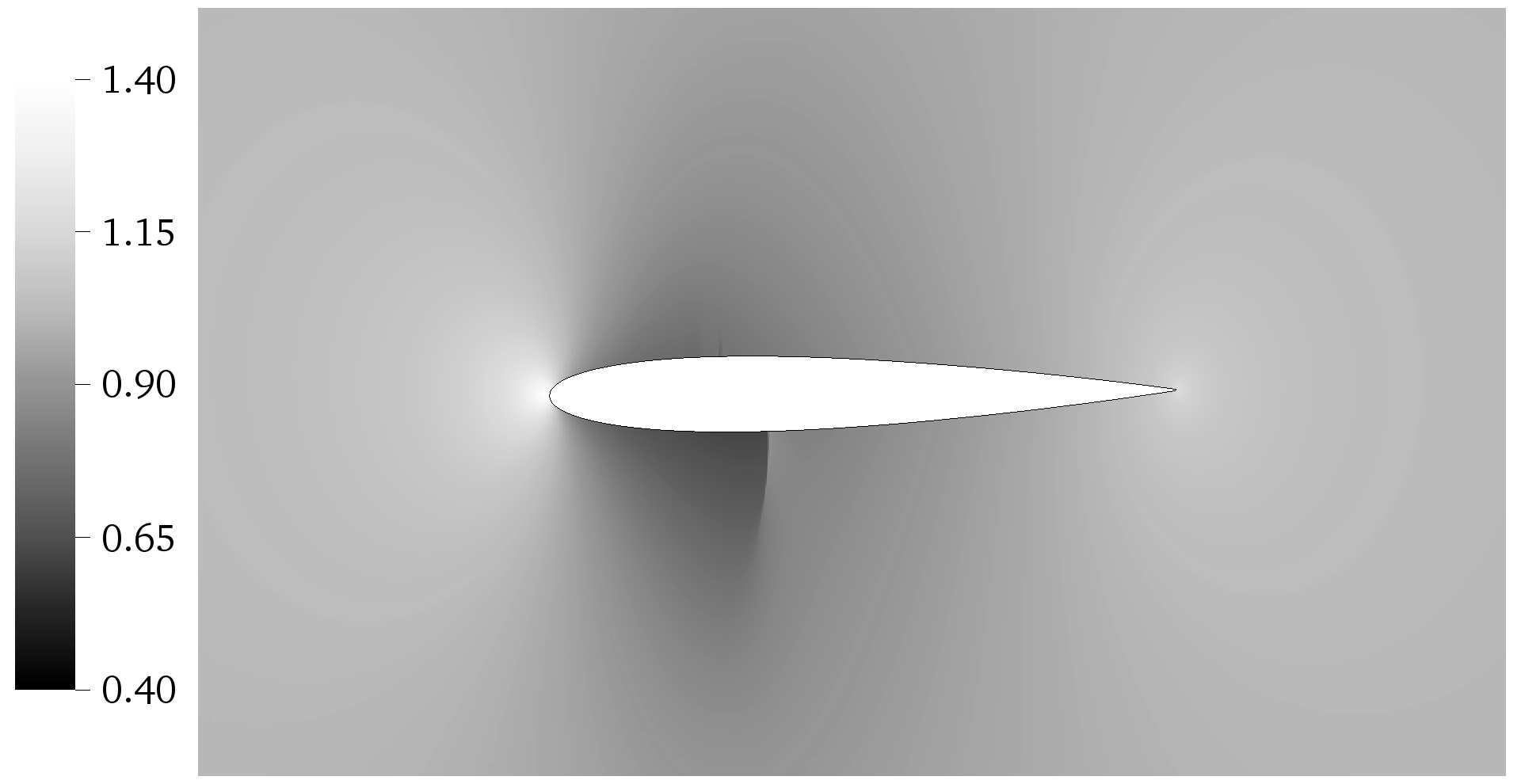}
\caption{Oscillating NACA 0012 aerofoil. Angle of attack, (a) $\varphi(t) = 2.34^{\circ}$, (b) $\varphi(t) = -0.54^{\circ}$. Non-dimensional pressure pseudo-colour, $p/p_{\infty}$.}
  \label{NACA0012contours}
\end{center}
\end{figure}

\begin{figure}
\begin{center}
\includegraphics[width=12cm]{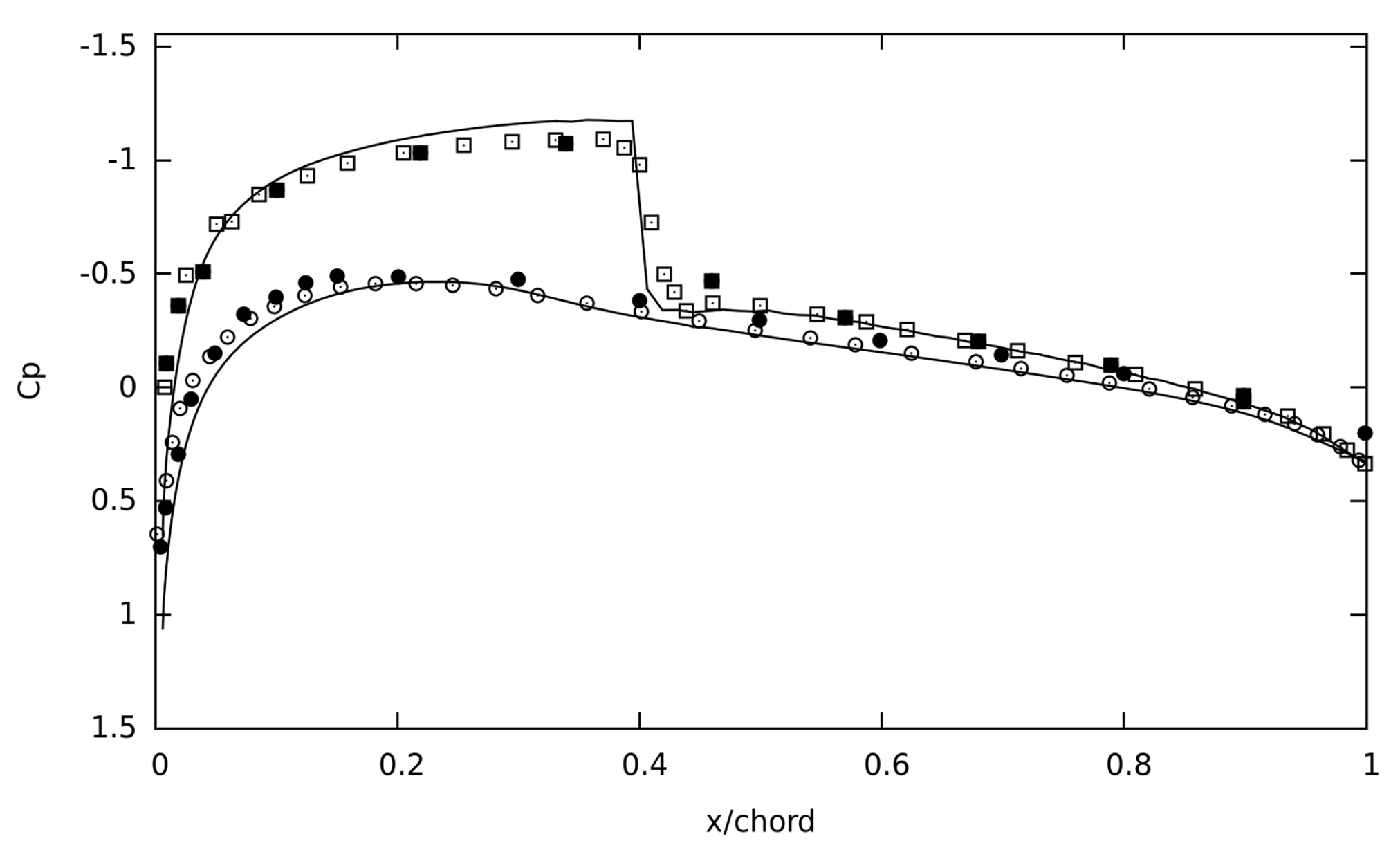}
\caption{Oscillating NACA 0012 aerofoil pressure coefficient at $\varphi(t) = 2.34^{\circ}$. Solid line: Current simulation. Open symbols: Numerical simulation of Kirschman \&  Liu~\cite{kirshman2006flutter}. Solid symbols: Measurements of Landon~\cite{landon1982compendium}. Square symbols denote suction surface, circles denote pressure surface. }
\label{NACA0012Cp}
\end{center}
\end{figure}

As the NACA 0012 aerofoil pitch angle changes over time, an unsteady shock pattern develops close to the surface. Figure~\ref{NACA0012contours} shows this unsteady shock wave pattern at two different times in the pitching cycle, highlighting the change in shock strength and location with time. Figure~\ref{NACA0012contours}~(a) is a pseudo-colour plot of non-dimensional pressure, $p/p_{\infty}$, at a pitching angle of $\varphi(t) = 2.34^{\circ}$ with respect to the horizontally aligned free stream flow. At a later time, the pressure distribution at a pitching angle of $\varphi(t)=-0.54^{\circ}$ is shown in Figure~\ref{NACA0012contours}~(b). An inverted shock pattern is observed between these two times. The results are quantitatively validated by comparison with both numerical and experimental surface pressure profiles along the upper and lower aerofoil surfaces in Figure~\ref{NACA0012Cp} for a pitching angle of $\varphi(t)=2.34^{\circ}$. \revisionAB{Adaptive Mesh Refinement (AMR) is used for this test case based on a density refinement criterion of $\rho = 1 \times 10^{-4}$, with the most refined cells at the aerofoil surface of dimension $\Delta x = \Delta y = 5 \times 10^{-3}c$. The surface of the aerofoil is also flagged for refinement throughout the computation. The CFL number used in this test case is 0.5.}

\begin{figure}
\begin{center}
\includegraphics[width=8cm]{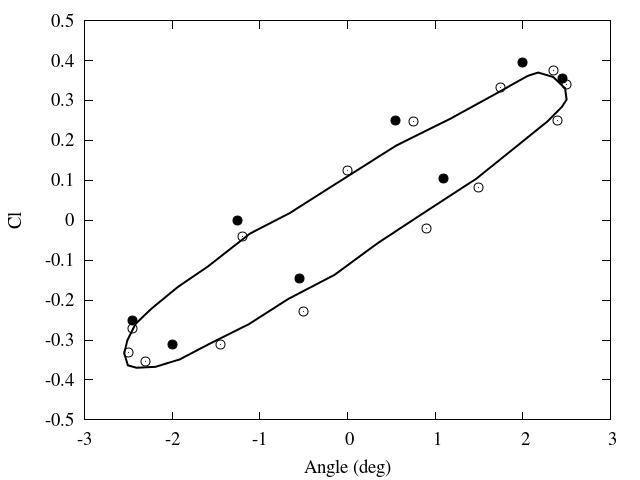}
\caption{Oscillating NACA 0012 aerofoil. Lift coefficient. Solid line: Current simulation. Open symbols: Venkatakrishnan \& Mavriplis~\cite{venkatakrishnan1995implicit} boundary conforming mesh simulation. Solid symbols: Landon experiment~\cite{landon1982compendium}. }
 \label{NACA0012LiftWithAngle}
\end{center}
\end{figure}

The change in the pressure distribution around the aerofoil over each complete pitching cycle results in a cyclical change in the aerofoil lift and drag forces. The cyclical variation in lift, between a positive and negative maximum over the course of each pitching event, is shown in Figure~\ref{NACA0012LiftWithAngle} as a plot of lift coefficient against pitching angle.

Following the onset of aerofoil pitching, the lift coefficient variation undergoes a transient period, characterised by a cycle-to-cycle variation in the lift coefficient maxima and minima. At later times, the lift coefficient settles into a self-sustaining quasi-steady pattern. It is during this quasi-steady period that the lift coefficient for Figure~\ref{NACA0012LiftWithAngle} is recorded.

The current numerical simulation compares well with the numerical simulation of Venkatakrishnan \& Mavriplis~\cite{venkatakrishnan1995implicit} and the measurements of Landon~\cite{landon1982compendium}, indicating the correct time-dependent compressible flow behaviour is captured by the current moving boundary cut-cell method. \revisionB{In addition, it is noted for completeness, that although not directly compared here, the current numerical simulation also compares favourably with the other moving boundary cut-cell methods noted in the test case introduction, specifically Schneiders et al.~\cite{schneiders2013accurate} and Murman et al.~\cite{murman2003implicit}.}


\subsection{\revision{Axisymmetric Open-Ended Shock Tube}}

\begin{figure}
\begin{center}
\includegraphics[width=10cm]{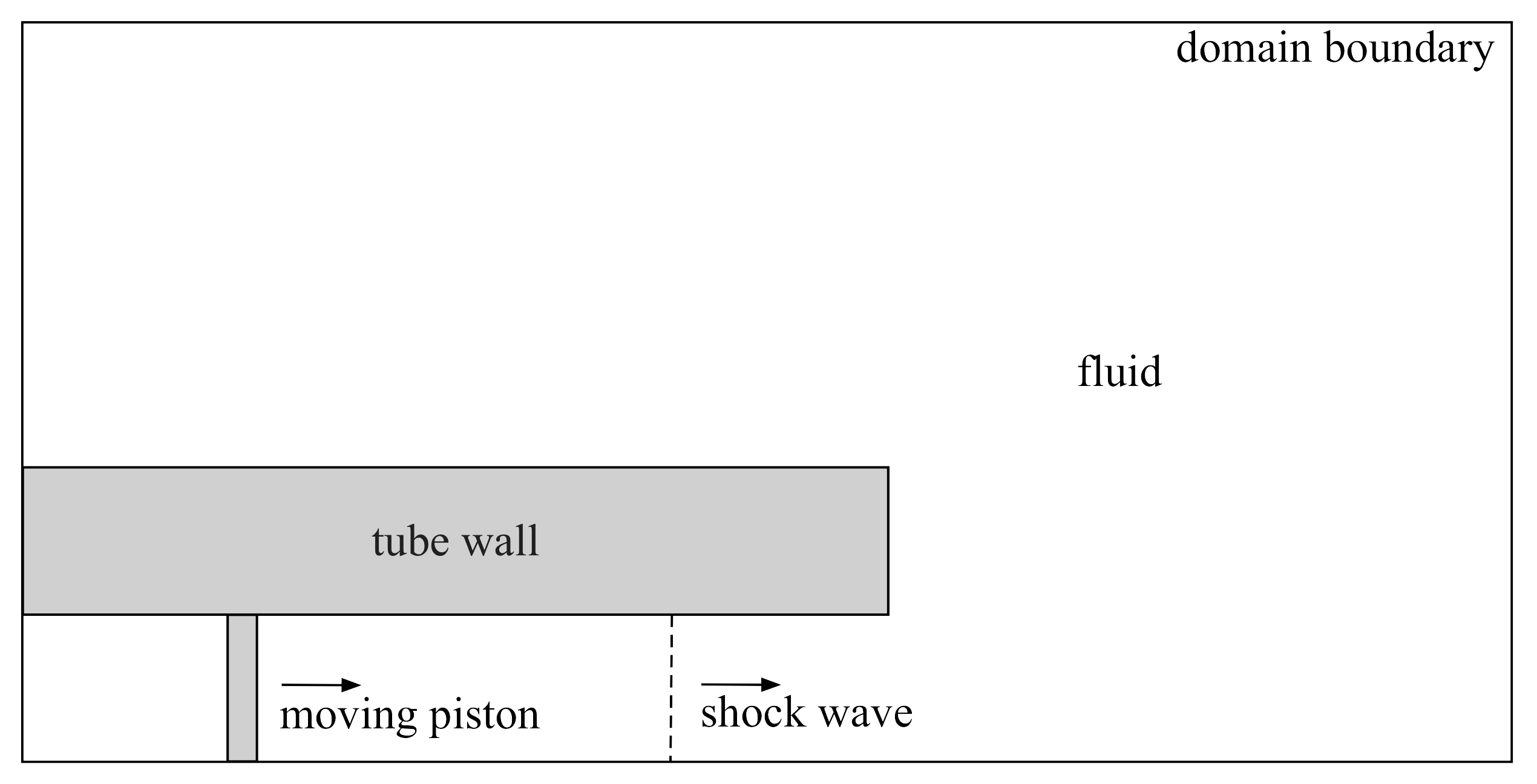}
\caption{\revision{Shock tube test case configuration. A piston moving through a quiescent fluid inside a surrounding open-ended rigid tube induces a shock wave which travels along the tube. Half of the tube diameter is defined, with the full circular tube case approximated through an axisymmetric form of the Euler equations.}}
 \label{PistonSketch}
\end{center}
\end{figure}

\revision{The fifth test case is intended to demonstrate the current cut-cell methods ability to model relative motion between two rigid bodies.
A rigid circular disc moves at a constant imposed velocity through an open-ended cylindrical tube, initiating the onset of a Mach 1.76 shock wave ahead of the disc. As the shock wave exits the shock tube and moves radially outwards through the surrounding quiescent air, a characteristic flow pattern forms close to the shock tube exit. The post-shock fluid in the tube has a resultant velocity in the direction of the tube exit. As this subsonic fluid exits the tube, a circumferential shear layer of fluid forms from the inner lip of the tube, terminating in a ring vortex further from the exit. Over time, the shock wave continues to move radially away from the shock tube exit through the quiescent fluid. The shear layer increases in length as the ring vortex gradually moves further from the shock tube exit. The local acceleration of the fluid around the ring vortex results in local compressible flow features which change in appearance over time as the vortex moves away from the tube. The open-ended shock tube problem is documented in a number of numerical and experimental publications, for example,~\cite{schmidt1985noise, wang1990numerical, batten1997choice}. The current test case is a modified form of this published work due to the presence of the moving piston, which replaces the driver gas as the source of the shock wave development. This modification to the classic open-ended shock tube problem is also used by Yang et al.~\cite{yang1997cartesian} to validate a cell merging cut-cell method with moving rigid boundaries. The characteristic flow pattern that forms outside of the shock tube following the ejection of the shock wave provides both qualitative and quantitative validation through comparison with experimental shadowgraph visualisation and pressure measurements. Figure~\ref{PistonSketch} shows an overview of the computational set-up for this test case.}

\revision{The current test case is a two-dimensional test case, that models the movement of a disc-shaped rigid body along a cylindrical shock tube by employing the axisymmetric form of the Euler equations. As the resulting flow pattern downstream of the shock tube has rotational symmetry, a reflective boundary can be defined along the centre-line of the shock tube. This forms the lower boundary of the computational domain. Transmissive boundaries are defined along the other three outer boundaries of the computational domain. The shock tube diameter, $D$, is used as the characteristic length. The shock tube defined in this test case is therefore $6D$ in length, starting at the left-hand computational boundary. The computational domain is $10D$ long, in a plane normal to the shock tube exit, and is $4D$ wide. The air ahead of the shock wave has density $\rho =$1.225 kgm$^{-3}$, pressure, $p=$101325Pa, and is initially at rest. The moving rigid disc is initially placed at a distance of $5.5D$ from the shock tube exit, at the start of the computation. This disc is imposed with a constant velocity in the direction of the shock tube exit equal in magnitude to the post-shock state for a Mach 1.76 shock wave, i.e., $u_{piston}=$337.974m/s. The post-shock fluid properties, between the moving disc and the shock wave, are then $\rho=$2.812kgm$^{-3}$, $u =$337.974m/s and $p=$349288Pa. Qualitative and quantitative validation is achieved through comparison with the shadowgraph visualisation and pressure measurements of Schmidt \& Duffy~\cite{schmidt1985noise}, who use a driver gas to initiate the shock wave. To synchronise results from the current test case with the experimental results of Schmidt \& Duffy~\cite{schmidt1985noise}, a common `zero' reference time is defined, $t_{ref}=0$. This zero reference time is defined as the time at which the shock wave is level with the exit plane of the shock tube. In the current simulation, this occurs at $t=1.388$ms after the start of the computation. Numerical results from three distinct mesh resolutions are compared with the pressure measurements of Schmidt \& Duffy~\cite{schmidt1985noise}. \revisionAB{These are a global non-AMR case with cell size $\Delta x=\Delta y=$ 1.6$\times$ 10$^{-3}$m, a non-AMR case with cell size  $\Delta x=\Delta y=$ 0.8$\times$10$^{-3}$m, and a two-level (base mesh and one refined level) AMR case with a cell size $\Delta x=\Delta y=$ 0.4$\times$ 10$^{-3}$m at the most refined level, using density as the refinement criterion, with a threshold value of  $\rho =$1$\times$ 10$^{-3}$kgm$^{-3}$ in order to refine across the shock wave, contact surface and ring vortex. A CFL value 0.5 is specified.}}

\begin{figure}
\begin{center}
\includegraphics[width=10cm]{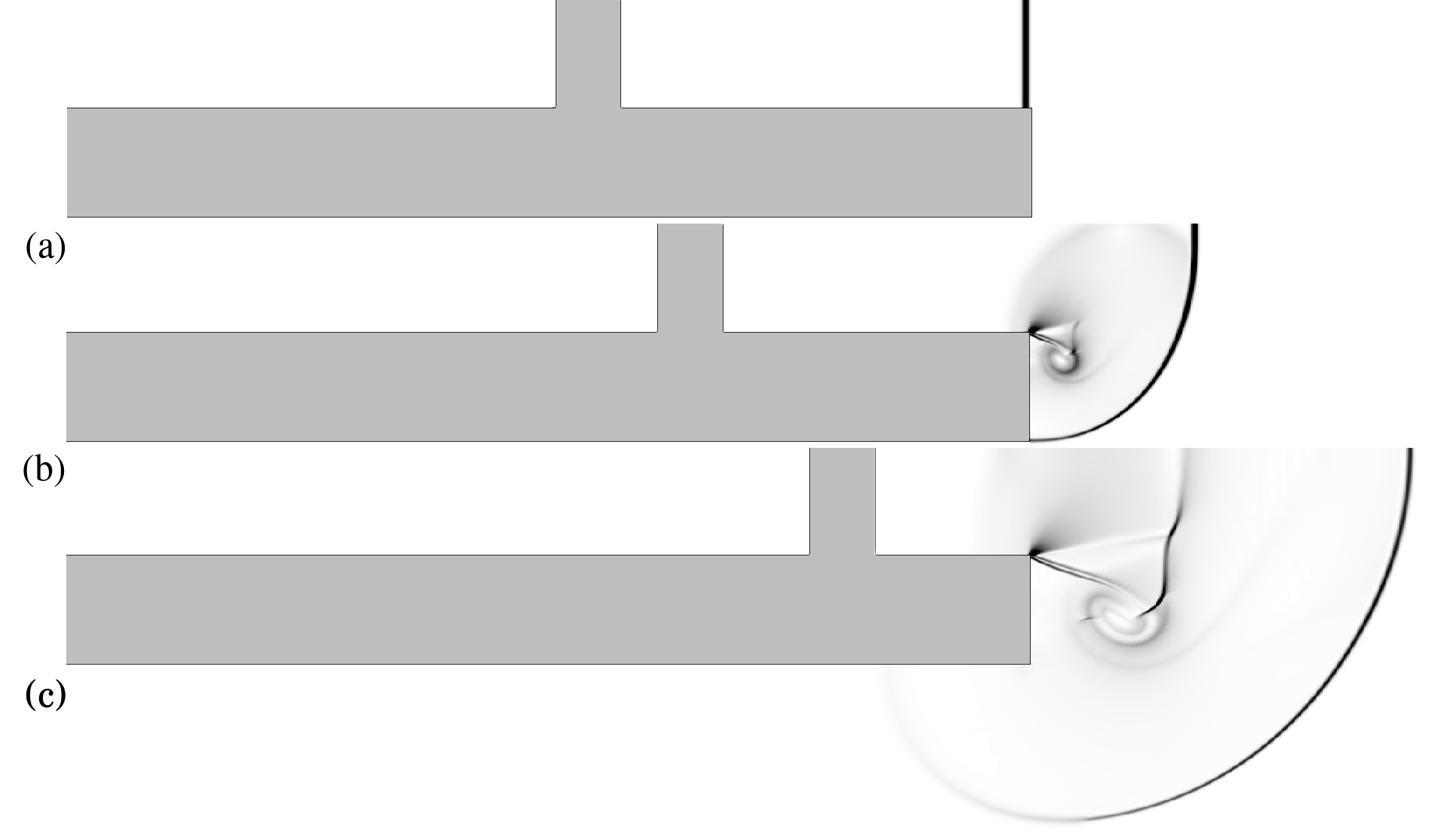} \\
\caption{\revision{Numerical Schlieren pseudo-colour plot showing the development of the flow structure at the exit of the shock tube. (a) Zero reference time ($t_{ref}=0$) when the shock wave has reached the end of the shock tube ($t=$1.388ms). (b) $t_{ref}=$0.2ms. (c) $t_{ref}=$1.466ms.}}
 \label{Piston_SchlierenAll}
\end{center}
\end{figure}

\revision{The development of the shock induced flow, downstream of the shock tube exit, is shown at $t_{ref}=$0, $t_{ref}=$0.2 ms and $ t_{ref}=$0.4933ms in Figure~\ref{Piston_SchlierenAll}.}

\begin{figure}
\begin{center}
\includegraphics[width=5cm]{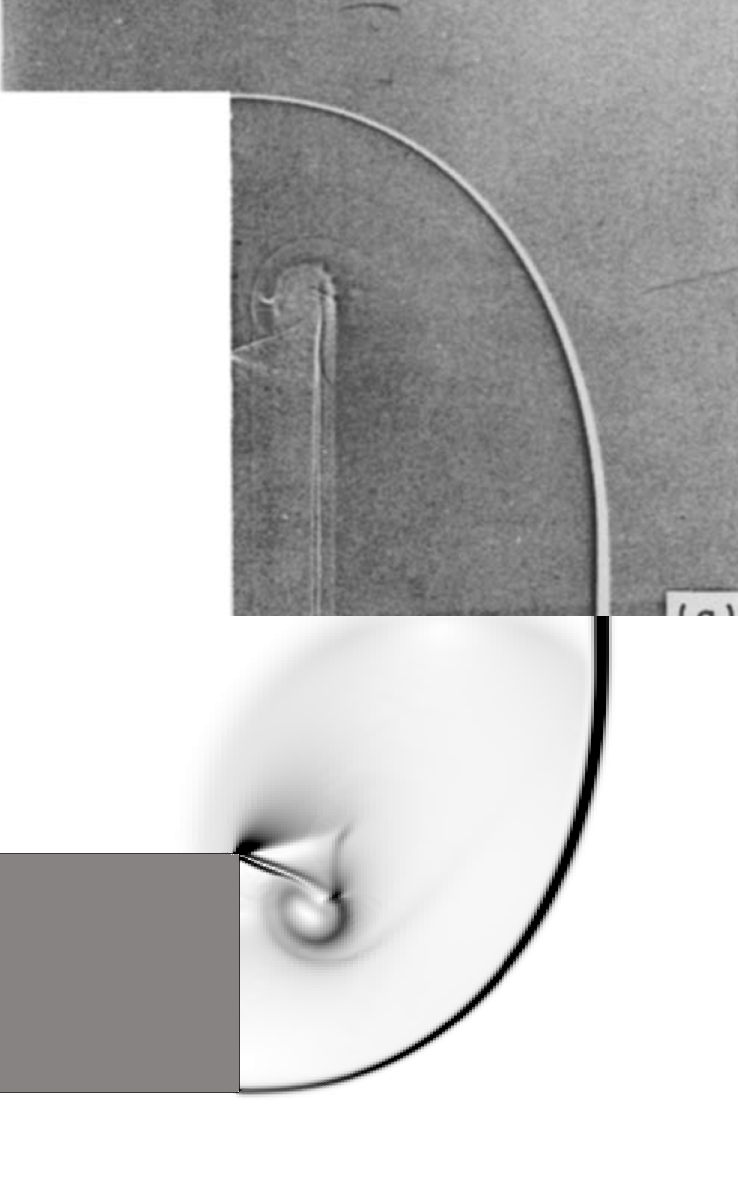} 
\caption{\revision{Comparison of the shadowgraph visualisation of Schmidt \& Duffy~\cite{schmidt1985noise} with the current numerical Schlieren pseudo-colour plot at time $t=0.2$ms. Computational domain view truncation to the shock tube exit region for clarity. The grey block shows the open end of the shock tube.}}
 \label{Piston_SchlierenComparisonTime02ms}
\end{center}
\end{figure}

\begin{figure}
\begin{center}
\includegraphics[width=6.5cm]{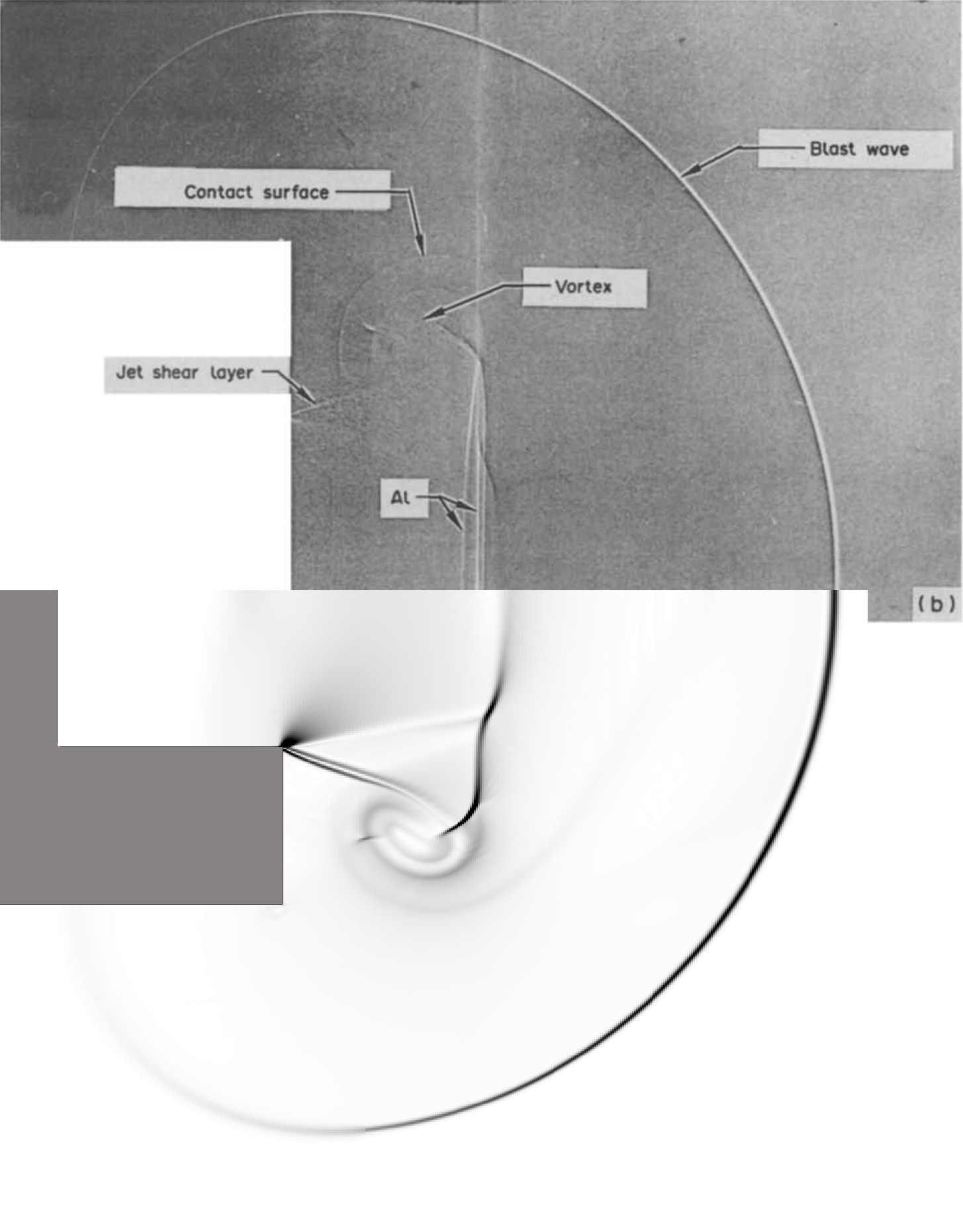}
\caption{\revision{Comparison of the shadowgraph visualisation of Schmidt \& Duffy~\cite{schmidt1985noise} with the current numerical Schlieren pseudo-colour plot at time $t=0.4933$ms. Computational domain view truncation to the shock tube exit region for clarity.  The grey block shows the open end of the shock tube. The end of the piston is also visible at this time. The label `Al' in the experimental results highlight additional spurious flow features only visible in this figure due to the side location of the camera looking through the axisymmetric flow.}}
 \label{Piston_SchlierenComparisonTime0p4ms}
\end{center}
\end{figure}

\revision{The numerical Schlieren plots are taken from the AMR computation with a refined mesh cell size of $\Delta x = \Delta y = $ 0.4$\times$10$^{-3}$m. The disc, which is shown moving along the tube in this sequence, and the stationary shock tube are uniform grey rectangles. In addition to the primary shock wave, which moves radially outwards from the shock tube exit in this sequence, the ring vortex and the embedded compressible flow features are present in the outer regions of the vortex in Figure~\ref{Piston_SchlierenAll}(c). The two latter times in this sequence coincide with the times at which experimental shadowgraph visualisation are available in Schmidt \& Duffy~\cite{schmidt1985noise}. These are directly compared with the current numerical Schlieren plots in Figure~\ref{Piston_SchlierenComparisonTime02ms} and Figure~\ref{Piston_SchlierenComparisonTime0p4ms}.}

\revision{Figure~\ref{Piston_SchlierenComparisonTime02ms} directly compares the numerical Schlieren pseudo-colour plot from the present  computation with experimental shadowgraph visualisation by Schmidt \& Duffy~\cite{schmidt1985noise} at a common reference time of $t_{ref} = 0.2$ms. At this reference time, the primary shock wave has reached the outer edge of the shock tube. The location of the primary shock wave, the vortex and the shear layer compare well, as does the size of diameter of the vortex at this time. 
Further qualitative comparison of the present results with the experiment of Schmidt \& Duffy~\cite{schmidt1985noise} at a later time of $t_{ref}=$ 0.4933ms is given in Figure~\ref{Piston_SchlierenComparisonTime0p4ms}.} 

\revision{At this later time, the vortex has moved downstream compared with Figure~\ref{Piston_SchlierenComparisonTime02ms} and is larger in size. Compressible flow features are evident embedded either side of the vortex in both the numerical and experimental results. The location and shape of these compressible flow features compare well between the numerical and experimental results.}

\begin{figure}
\begin{center}
\includegraphics[width=8cm]{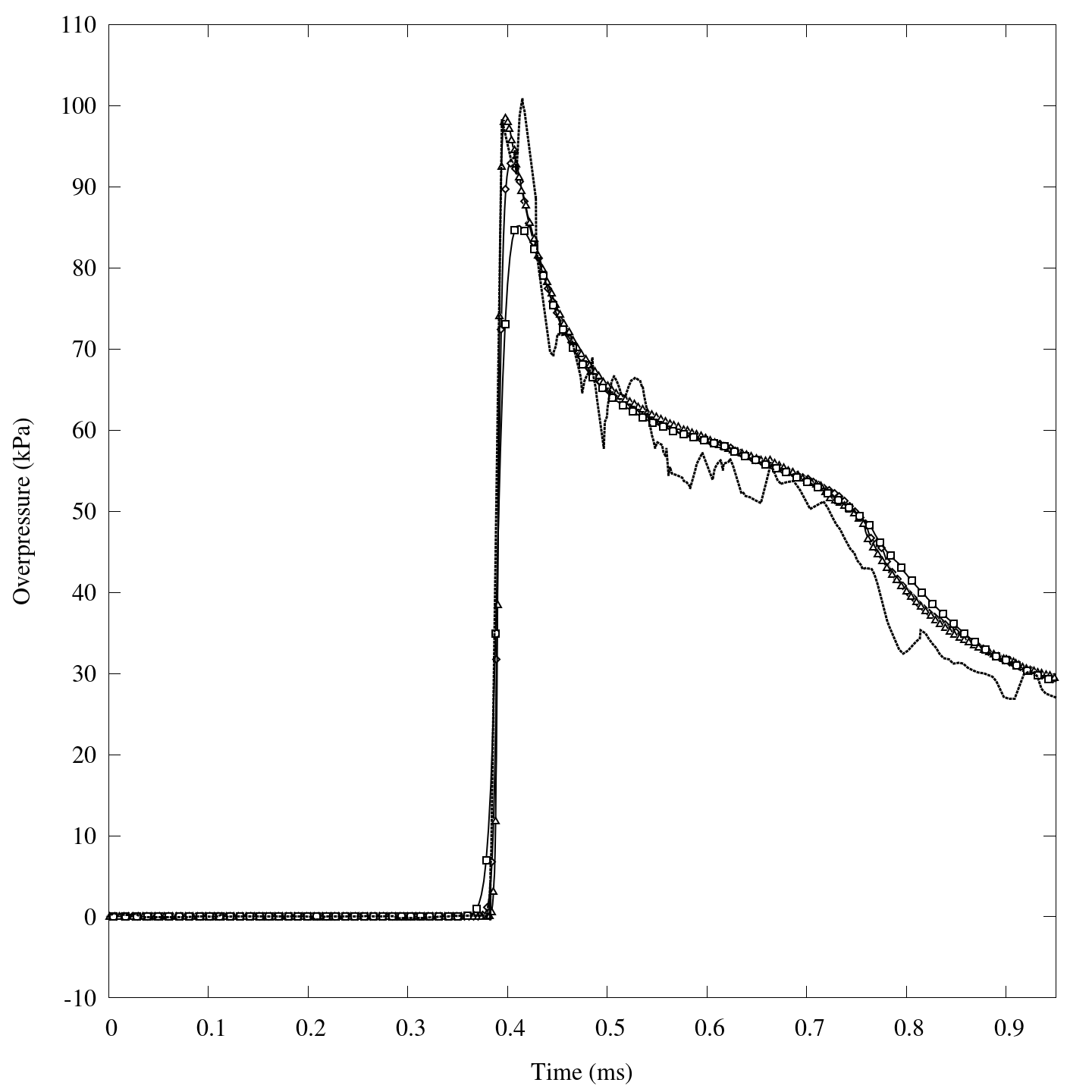}
\caption{\revision{Comparison of overpressure at a distance of 1.5 diameters downstream of the tube exit. Solid line: Measurements of Schmidt \& Duffy~\cite{schmidt1985noise}. Square symbols: Single level (non-AMR) mesh with computational cell dimensions, $\Delta x = \Delta y = $ 1.6$\times$10$^{-3}$m. Diamond symbols: Single level (non-AMR) mesh with computational cell dimensions  $\Delta x = \Delta y = $ 0.8$\times$10$^{-3}$m. Triangle symbols: Two level AMR computation (baseline mesh plus one refined level). Refined level cell dimensions $\Delta x = \Delta y = $ 0.4$\times$10$^{-3}$m.}}
 \label{Piston_PressureTimeHistory}
\end{center}
\end{figure}

\revision{Figure~\ref{Piston_PressureTimeHistory} shows the time history in over-pressure at a fixed location along the shock tube centre-line of 1.5 diameters distance downstream of the tube exit. Three separate mesh resolutions are compared against the measurements of Schmidt \& Duffy~\cite{schmidt1985noise} showing a convergence of the magnitude and location of the peak over-pressure as the mesh resolution across the shock wave increases. Specifically, a global, single level (non-AMR) mesh with computational cell dimensions, $\Delta x = \Delta y = $ 1.6$\times$10$^{-3}$m, provides the coarsest resolution. Increasing the resolution, a global, single level mesh with computational cell dimensions  $\Delta x = \Delta y = $ 0.8$\times$10$^{-3}$m provides the second mesh refinement, identified in Figure~\ref{Piston_PressureTimeHistory} using diamond symbols. Finally, the most refined mesh, identified in Figure~\ref{Piston_PressureTimeHistory} using triangular symbols, adds one additional refinement level to this mesh, resulting in computational cells of dimension $\Delta x = \Delta y = $ 0.4$\times$10$^{-3}$m on the most refined level. The measured over-pressure, shown in Figure~\ref{Piston_PressureTimeHistory} as a dotted line, shows a double peak in over-pressure, with peak over-pressures of $p-p_{\infty}=$98.341kPa at $t_{ref}=$0.394ms and $p-p_{\infty}=$100.9252kPa at $t_{ref}=$0.415ms. The numerical over-pressure peak converges with increasing mesh resolution to the earlier of these measured peaks, reaching a peak over-pressure of $p-p_{\infty} =$98.544kPa at $t_{ref}=$0.398ms. The difference in the first measured over-pressure peak and the most-refined simulation is 0.206\% of the peak measured over-pressure. At later times, the predicted over-pressure follows the overall shape of the measured over-pressure curve with a better fit to the measurement evident in the higher resolution simulations, between $t_{ref}=$0.75 and $t_{ref}=$0.85.}

\subsection{Three-Dimensional Schardin Experiment}

This test case extends the two-dimensional rigid body/shock wave interaction case in Section~\ref{sec:SchardinWedge2D} to a three-dimensional cone moving through a stationary shock wave.
For comparison with the two-dimensional results in Section~\ref{sec:SchardinWedge2D}, the cone used here is an equilateral triangle of base length 20 mm when viewed in cross-section.
We model a stationary shock wave, matching the pressure and density rise across a moving Mach 1.34 shock wave, in order to simulate the practical situation 
of a fixed cone in a fluid that is initially at rest and subject to a passing shock wave. An ideal gas is used in this test case, with a specific heat capacity ratio, $\gamma=1.4$, and specific gas constant, $R=287.058$ J/kgK.

The fluid initially surrounding the cone at time $t=0$ has a static density and static pressure of $\rho=0.595$ kgm$^{-3}$ and $p=0.05$ MPa respectively, with a 
post-shock state of density, $\rho=0.944$ kgm$^{-3}$, and pressure, $p=0.0964$ MPa.

The computational domain size is $0.2 \times 0.2 \times 0.2$ m, with transmissive boundaries defining the outer domain extent. 
As the case is radially symmetric, we reduce the computational effort here by terminating the domain along the cone centre-line.

\revisionAB{Adaptive Mesh Refinement (AMR) is used to refine across the shock wave using density as the refinement criterion with a threshold value of $\rho = 1 \times 10^{-2}$, defining two levels of mesh refinement above the baseline mesh of $100 \times 100 \times 50$ cells.} 
A refinement factor of two is used to half the cell length between AMR levels, leading to the equivalent mesh resolution of a $400 \times 400 \times 200$ cell mesh on the most refined level. 
The cell dimensions of the most refined AMR level are $\Delta x = \Delta y = \Delta z = 5 \times 10^{-4}$ m. A CFL number of 0.5 is used.

\begin{figure}
\begin{center}
\includegraphics[width=12cm]{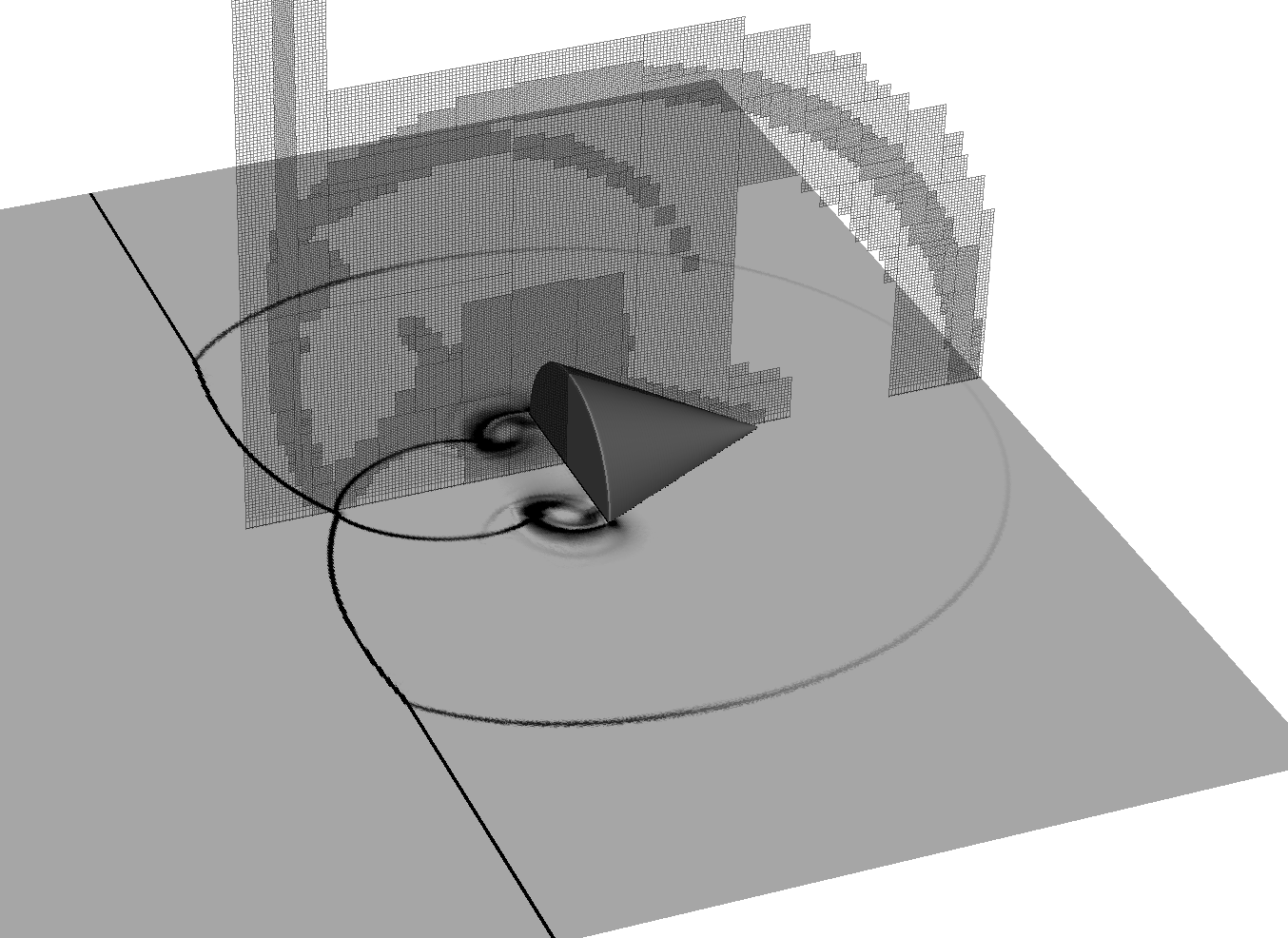}\\
\caption{Numerical Schlieren plot at a time of 102 $\mu$s after the initial contact of the cone with the shock wave. The most refined two mesh levels of are shown.}
 \label{schardin_3D_snapsnots}
\end{center}
\end{figure}

Figure~\ref{schardin_3D_snapsnots} shows the development of the cone at a time of $t=102~\mu$s after the initial contact of the cone with the stationary shock wave.
A two-dimensional slice across the centre of the cone is plotted, with a numerical Schlieren pseudo-colour plot showing compressible and vortical flow features. 
The two finest mesh levels are plotted in this figure along a plane normal to the pseudo-colour plot.

As expected, the two-dimensional slice is qualitatively similar to the triangular wedge results in Figure~\ref{schardin_snapsnots}~(d). 
The passage of the three-dimensional cone, of course, creates a shock and vortex pattern with rotational symmetry.
The numerical Schlieren shows that the main flow features are captured by the current method, in particular, the reflected shock, 
Mach stem, downstream vortex ring (shown as two distinct vortices in the two-dimensional slice) and the shock triple points.
These flow features are shown as corresponding regions of mesh refinement.


\section{Conclusions}
\label{sec:Conclusions}

A new moving boundary cut-cell method is presented for flows involving rigid bodies moving in a surrounding single-phase inviscid fluid.
This cut-cell method uses an explicit directional operator splitting approach to solve the time-dependent hyperbolic conservation laws, extending the static boundary cut-cell method of Klein et al.~\cite{Klein09}. A flux stabilization approach is used to permit cut-cells with small cell volumes to be advanced in a stable manner using a global time step based on the regular Cartesian cell dimensions.
The movement of the rigid body, leading to a transient interface between the cut-cell boundary and the stationary Cartesian mesh, introduces a number of unique challenges in formulating a moving boundary flux stabilization method. 
Specifically, the continuous decrease in some cut-cell fluid volume fractions and the disappearance of fluid cut-cells during the rigid body movement have implications for the conservation and stability properties of the scheme. 
Conversely, solid cells are permitted to change to cut-cells with no initial fluid state as the moving rigid body uncovers computational cells. 
These uncovered cells may subsequently increase in volume to become regular fluid cells over time. 
An additional consistency condition requires that an exact solution is produced for a solid body moving at the same velocity as the surrounding fluid.
The cut-cell method outlined in this paper replaces the regular cell interface flux in cut-cells with a stable flux computed using a linear weighted combination of the regular flux and the cut-cell boundary 
flux. The movement of non-porous rigid cut-cell boundaries is accounted for through boundary pressure terms in the momentum and energy equations. 
A consideration of the time averaged cut-cell boundary movement during each time step is used to help maintain conservation in the directional operator splitting scheme.
A volume weighted averaging procedure is introduced to populate newly uncovered cells with conservative state variables. 
We attempt to maintain global conservation following cell transformation to solid cells through a conservative redistribution to neighbouring cells. 

The moving boundary cut-cell method is applied to six compressible flow test cases in this paper to assess the solution accuracy and stability when rigid boundaries move through smooth and discontinuous flows. 
The numerical results compare favourably with published references. This is shown qualitatively through comparison of experimental Schlieren visualisation, and quantitatively by reference to pressure and force measurements or exact analytical solution.
A further development to the moving boundary cut-cell method described in this paper is currently under way to extend the method to two-phase flows, consisting of both rigid body cut-cell boundaries in two-phase flows and 
cut-cell boundaries defining fluid-fluid interfaces.


\section*{ACKNOWLEDGEMENTS}

Rupert Klein acknowledges support by Deutsche Forschungsgemeinschaft through Grants SFB 1029 ``TurbIn'', Project C01 and SFB 1114 ``Scaling
Cascades in Complex Systems'', Project C01. Dr W.P. Bennett was kindly supported by AWE during the development of this work. 

\bibliographystyle{unsrt}
\bibliography{MovingBoundaryPaper_WPBennett_RevisedVersion_20180424}

\begin{thebibliography}{10}

\bibitem{mittal2005immersed}
R.~Mittal and G.~Iaccarino.
\newblock Immersed boundary methods.
\newblock {\em Annu. Rev. Fluid Mech.}, 37:239--261, 2005.

\bibitem{Sotiropoulos20141}
Fotis Sotiropoulos and Xiaolei Yang.
\newblock {Immersed boundary methods for simulating fluid--structure
  interaction}.
\newblock {\em Progress in Aerospace Sciences}, 65:1--21, 2014.

\bibitem{maxey2017simulation}
Martin Maxey.
\newblock Simulation methods for particulate flows and concentrated
  suspensions.
\newblock {\em Annual Review of Fluid Mechanics}, 49:171--193, 2017.

\bibitem{peskin1972flow}
C.~S. Peskin.
\newblock Flow patterns around heart valves: a numerical method.
\newblock {\em Journal of Computational Physics}, 10(2):252--271, 1972.

\bibitem{peskin2002immersed}
C.~S. Peskin.
\newblock The immersed boundary method.
\newblock {\em Acta numerica}, 11:479--517, 2002.

\bibitem{goldstein1993modeling}
D.~Goldstein{,}~R. Handler and L.~Sirovich.
\newblock Modeling a no-slip flow boundary with an external force field.
\newblock {\em Journal of Computational Physics}, 105(2):354--366, 1993.

\bibitem{tyagi2005large}
M.~Tyagi and S.~Acharya.
\newblock Large eddy simulation of turbulent flows in complex and moving rigid
  geometries using the immersed boundary method.
\newblock {\em International Journal for Numerical Methods in Fluids},
  48(7):691--722, 2005.

\bibitem{Wang2008283}
Zeli Wang, Jianren Fan, and Kun Luo.
\newblock {Combined multi-direct forcing and immersed boundary method for
  simulating flows with moving particles}.
\newblock {\em International Journal of Multiphase Flow}, 34(3):283--302, 2008.

\bibitem{luo2017direct}
Kun Luo, Chenshu Hu, Fan Wu, and Jianren Fan.
\newblock Direct numerical simulation of turbulent boundary layer with fully
  resolved particles at low volume fraction.
\newblock {\em Physics of Fluids}, 29(5):053301, 2017.

\bibitem{Ren2013694}
Weiwei Ren, Chang Shu, and Wenming Yang.
\newblock {An efficient immersed boundary method for thermal flow problems with
  heat flux boundary conditions}.
\newblock {\em International Journal of Heat and Mass Transfer}, 64:694--705,
  2013.

\bibitem{Xia2014302}
Junjie Xia, Kun Luo, and Jianren Fan.
\newblock {A ghost-cell based high-order immersed boundary method for
  inter-phase heat transfer simulation}.
\newblock {\em International Journal of Heat and Mass Transfer}, 75:302--312,
  2014.

\bibitem{Glowinski1994283}
Roland Glowinski, Tsorng-Whay Pan, and Jacques Periaux.
\newblock {A fictitious domain method for Dirichlet problem and applications}.
\newblock {\em Computer Methods in Applied Mechanics and Engineering},
  111(3):283--303, 1994.

\bibitem{Angot1999}
Philippe Angot, Charles-Henri Bruneau, and Pierre Fabrie.
\newblock {A penalization method to take into account obstacles in
  incompressible viscous flows}.
\newblock {\em Numerische Mathematik}, 81(4):497--520, Feb 1999.

\bibitem{khadra2000fictitious}
Khodor Khadra, Philippe Angot, Sacha Parneix, and Jean-Paul Caltagirone.
\newblock Fictitious domain approach for numerical modelling of navier--stokes
  equations.
\newblock {\em International journal for numerical methods in fluids},
  34(8):651--684, 2000.

\bibitem{randrianarivelo2005numerical}
TN~Randrianarivelo, G~Pianet, S~Vincent, and JP~Caltagirone.
\newblock Numerical modelling of solid particle motion using a new penalty
  method.
\newblock {\em International Journal for Numerical Methods in Fluids},
  47(10-11):1245--1251, 2005.

\bibitem{angot2010fictitious}
Philippe Angot.
\newblock A fictitious domain model for the stokes/brinkman problem with jump
  embedded boundary conditions.
\newblock {\em Comptes Rendus Mathematique}, 348(11-12):697--702, 2010.

\bibitem{angot2012fast}
Philippe Angot, Jean-Paul Caltagirone, and Pierre Fabrie.
\newblock A fast vector penalty-projection method for incompressible
  non-homogeneous or multiphase navier--stokes problems.
\newblock {\em Applied Mathematics Letters}, 25(11):1681--1688, 2012.

\bibitem{ducassou2017fictitious}
B~Ducassou, J~Nu{\~n}ez, M~Cruchaga, and S~Abadie.
\newblock A fictitious domain approach based on a viscosity penalty method to
  simulate wave/structure interaction.
\newblock {\em Journal of Hydraulic Research}, pages 1--16, 2017.

\bibitem{mohd1997simulations}
J~Mohd-Yusof.
\newblock For simulations of flow in complex geometries.
\newblock {\em Annual Research Briefs}, 317, 1997.

\bibitem{fadlun2000combined}
E.~A. Fadlun{,} R. Verzicco{,}~P. Orlandi and J.~Mohd-Yusof.
\newblock Combined immersed-boundary finite-difference methods for
  three-dimensional complex flow simulations.
\newblock {\em Journal of Computational Physics}, 161(1):35--60, 2000.

\bibitem{uhlmann2005immersed}
Markus Uhlmann.
\newblock An immersed boundary method with direct forcing for the simulation of
  particulate flows.
\newblock {\em Journal of Computational Physics}, 209(2):448--476, 2005.

\bibitem{breugem2012second}
Wim-Paul Breugem.
\newblock A second-order accurate immersed boundary method for fully resolved
  simulations of particle-laden flows.
\newblock {\em Journal of Computational Physics}, 231(13):4469--4498, 2012.

\bibitem{qiu2016boundary}
YL~Qiu, C~Shu, J~Wu, Y~Sun, LM~Yang, and TQ~Guo.
\newblock A boundary condition-enforced immersed boundary method for
  compressible viscous flows.
\newblock {\em Computers \& Fluids}, 136:104--113, 2016.

\bibitem{saurel2018diffuse}
Richard Saurel and Carlos Pantano.
\newblock Diffuse-interface capturing methods for compressible two-phase flows.
\newblock {\em Annual Review of Fluid Mechanics}, 50(1), 2018.

\bibitem{fedkiw1999non}
R.~P. Fedkiw{,} T. Aslam{,}~B. Merriman and S.~Osher.
\newblock A non-oscillatory {E}ulerian approach to interfaces in multimaterial
  flows (the ghost fluid method).
\newblock {\em Journal of Computational Physics}, 152(2):457--492, 1999.

\bibitem{fedkiw2002coupling}
R.~P. Fedkiw.
\newblock Coupling an {E}ulerian fluid calculation to a {L}agrangian solid
  calculation with the ghost fluid method.
\newblock {\em Journal of Computational Physics}, 175(1):200--224, 2002.

\bibitem{terashima2009front}
Hiroshi Terashima and Gr{\'e}tar Tryggvason.
\newblock A front-tracking/ghost-fluid method for fluid interfaces in
  compressible flows.
\newblock {\em Journal of Computational Physics}, 228(11):4012--4037, 2009.

\bibitem{liu2003ghost}
TG~Liu, BC~Khoo, and KS~Yeo.
\newblock Ghost fluid method for strong shock impacting on material interface.
\newblock {\em Journal of Computational Physics}, 190(2):651--681, 2003.

\bibitem{kaboudian2015ghost}
Abouzar Kaboudian, Peyman Tavallali, and BC~Khoo.
\newblock The ghost solid methods for the elastic--plastic solid--solid
  interface and the $\vartheta$-criterion.
\newblock {\em Journal of Computational Physics}, 302:618--652, 2015.

\bibitem{kaboudian2014ghost}
Abouzar Kaboudian and BC~Khoo.
\newblock The ghost solid method for the elastic solid--solid interface.
\newblock {\em Journal of Computational Physics}, 257:102--125, 2014.

\bibitem{feng2017simulation}
ZW~Feng, A~Kaboudian, JL~Rong, and BC~Khoo.
\newblock The simulation of compressible multi-fluid multi-solid interactions
  using the modified ghost method.
\newblock {\em Computers \& Fluids}, 154:12--26, 2017.

\bibitem{Pember95}
R.~B. Pember{,} J.~B. Bell{, } P. Colella{,} W.~Y. Crutchfield and M.~L.
  Welcome.
\newblock An adaptive {C}artesian grid method for unsteady compressible flow in
  irregular regions.
\newblock {\em Journal of Computational Physics}, 120(2):278--304, 1995.

\bibitem{Colella06}
P.~Colella{,} D.~T. Graves{,} Benjamin~J. Keen and D.~Modiano.
\newblock A {C}artesian grid embedded boundary method for hyperbolic
  conservation laws.
\newblock {\em Journal of Computational Physics}, 211(1):347--366, 2006.

\bibitem{HuEtAl2006}
X.~Y. Hu, B.~C. Khoo, N.~A. Adams, and F.~L. Huang.
\newblock A conservative interface method for compressible flows.
\newblock {\em Journal of Computational Physics}, 219:553--578, 2006.

\bibitem{schneiders2013accurate}
L.~Schneiders{,} D. Hartmann{,}~M. Meinke and W.~Schr{\"o}der.
\newblock An accurate moving boundary formulation in cut-cell methods.
\newblock {\em Journal of Computational Physics}, 235:786--809, 2013.

\bibitem{Colella1990}
P.~Colella.
\newblock Multidimensional upwind methods for hyperbolic conservation laws.
\newblock {\em Journal of Computational Physics}, 87:171, 1990.

\bibitem{Klein09}
R.~Klein{,} K.~R. Bates and N.~Nikiforakis.
\newblock Well-balanced compressible cut-cell simulation of atmospheric flow.
\newblock {\em Philosophical Transactions of the Royal Society A: Mathematical,
  Physical and Engineering Sciences}, 367(1907):4559--4575, 2009.

\bibitem{Clarke85}
D.~K. Clarke{,} H.~A. Hassan and M.~D. Salas.
\newblock Euler calculations for multielement airfoils using {C}artesian grids.
\newblock In {\em American Institute of Aeronautics and Astronautics, Aerospace
  Sciences Meeting, 23 rd, Reno, NV}, 1985.

\bibitem{Quirk94}
J.~J. Quirk.
\newblock An alternative to unstructured grids for computing gas dynamic flows
  around arbitrarily complex two-dimensional bodies.
\newblock {\em Computers \& fluids}, 23(1):125--142, 1994.

\bibitem{Berger03}
M.~J. Berger{,}~C. Helzel and R.~J. LeVeque.
\newblock H-box methods for the approximation of hyperbolic conservation laws
  on irregular grids.
\newblock {\em SIAM Journal on Numerical Analysis}, 41(3):893--918, 2003.

\bibitem{Ingram03}
D.~M. Ingram{,} D.~M. Causon and C.~G. Mingham.
\newblock Developments in {C}artesian cut cell methods.
\newblock {\em Math. Comput. Simul.}, 61(3-6):561--572, 2003.

\bibitem{Xu97}
S.~Xu{,}~T. Aslam and D.~S. Stewart.
\newblock {High resolution numerical simulation of ideal and non-ideal
  compressible reacting flows with embedded internal boundaries }.
\newblock {\em Combustion Theory Modelling}, 1:113--142, March 1997.

\bibitem{yang1997cartesian}
G.~Yang{,} D.~M. Causon{,} D.~M. Ingram{,}~R. Saunders and P.~Batten.
\newblock A {C}artesian cut cell method for compressible flows part {A}:
  {S}tatic body problems.
\newblock {\em Aeronautical Journal}, 101(1002):47--56, 1997.

\bibitem{barton2011conservative}
P.~T. Barton{,}~B. Obadia and D.~Drikakis.
\newblock A conservative level-set based method for compressible solid/fluid
  problems on fixed grids.
\newblock {\em Journal of Computational Physics}, 230(21):7867--7890, 2011.

\bibitem{falcovitz1997two}
J.~Falcovitz{,}~G. Alfandary and G.~Hanoch.
\newblock A two-dimensional conservation laws scheme for compressible flows
  with moving boundaries.
\newblock {\em Journal of Computational Physics}, 138(1):83--102, 1997.

\bibitem{HartmannEtAl2011}
D.~Hartmann, M.~Meinke, and W.~Schr{\"o}der.
\newblock A strictly conservative {C}artesian cut-cell method for compressible
  viscous flows on adaptive grids.
\newblock {\em Methods Appl. Mech. Eng.}, 200:1038--1052, 2011.

\bibitem{Helzel05}
C.~Helzel{,} M.~J. Berger and R.~J. Leveque.
\newblock A high-resolution rotated grid method for conservation laws with
  embedded geometries.
\newblock {\em SIAM Journal on Scientific Computing}, 26(3):785--809, 2005.

\bibitem{LeVequeBook2002}
R.~J. LeVeque.
\newblock {\em Finite Volume Methods for Hyperbolic Problems}.
\newblock Cambridge University Press, Basel, 2002.

\bibitem{Roe1981}
P.~L. Roe.
\newblock Approximate {R}iemann solvers, parameter vectors, and difference
  schemes.
\newblock {\em Journal of Computational Physics}, 43:357--372, 1981.

\bibitem{BergerHelzel2012}
M.~J. Berger and C.~Helzel.
\newblock A simplified h-box method for embedded boundary grids.
\newblock {\em SIAM Journal on Scientific Computing}, 34(2):A861--A888, 2012.

\bibitem{muralidharan2016high}
Balaji Muralidharan and Suresh Menon.
\newblock A high-order adaptive cartesian cut-cell method for simulation of
  compressible viscous flow over immersed bodies.
\newblock {\em Journal of Computational Physics}, 321:342--368, 2016.

\bibitem{muralidharan2018simulation}
Balaji Muralidharan and Suresh Menon.
\newblock Simulation of moving boundaries interacting with compressible
  reacting flows using a second-order adaptive cartesian cut-cell method.
\newblock {\em Journal of Computational Physics}, 357:230--262, 2018.

\bibitem{krause2017incompressible}
Dennis Krause and Florian Kummer.
\newblock An incompressible immersed boundary solver for moving body flows
  using a cut cell discontinuous galerkin method.
\newblock {\em Computers \& Fluids}, 153:118--129, 2017.

\bibitem{kummer2017extended}
Florian Kummer.
\newblock Extended discontinuous galerkin methods for two-phase flows: the
  spatial discretization.
\newblock {\em International Journal for Numerical Methods in Engineering},
  109(2):259--289, 2017.

\bibitem{aftosmis2000parallel}
M.~J. Aftosmis{,} M.~J. Berger and G.~Adomavicius.
\newblock A parallel multilevel method for adaptively refined {C}artesian grids
  with embedded boundaries.
\newblock {\em AIAA paper}, 808:2000, 2000.

\bibitem{murman2003implicit}
S.~M. Murman{,} M.~J. Aftosmis and M.~J. Berger.
\newblock Implicit approaches for moving boundaries in a 3-{D} {C}artesian
  method.
\newblock {\em AIAA paper}, 1119:2003, 2003.

\bibitem{MeinkeEtAl2013}
M.~Meinke{,} L. Schneiders{,}~C. G{\"u}nther and W.~Schr{\"o}der.
\newblock A cut-cell method for sharp moving boundaries in {C}artesian grids.
\newblock {\em Comput. Fluids}, 85:135--142, 2013.

\bibitem{lin2017simulation}
Jian-Yu Lin, Yi~Shen, Hang Ding, Nan-Sheng Liu, and Xi-Yun Lu.
\newblock Simulation of compressible two-phase flows with topology change of
  fluid--fluid interface by a robust cut-cell method.
\newblock {\em Journal of Computational Physics}, 328:140--159, 2017.

\bibitem{deng2018simulating}
Xiao-Long Deng and Maojun Li.
\newblock Simulating compressible two-medium flows with sharp-interface
  adaptive runge--kutta discontinuous galerkin methods.
\newblock {\em Journal of Scientific Computing}, 74(3):1347--1368, 2018.

\bibitem{patel2018dual}
Tejas Patel and Absar Lakdawala.
\newblock A dual grid, dual level set based cut cell immersed boundary approach
  for simulation of multi-phase flow.
\newblock {\em Chemical Engineering Science}, 177:180--194, 2018.

\bibitem{strang1968construction}
G.~Strang.
\newblock On the construction and comparison of difference schemes.
\newblock {\em SIAM Journal on Numerical Analysis}, 5(3):506--517, 1968.

\bibitem{schneiders2016efficient}
L.~Schneiders{,}~C. G{\"u}nther, M.~Meinke, and W.~Schr{\"o}der.
\newblock An efficient conservative cut-cell method for rigid bodies
  interacting with viscous compressible flows.
\newblock {\em Journal of Computational Physics}, 311:62--86, 2016.

\bibitem{Toro}
E.~F. Toro.
\newblock {\em Riemann Solvers and Numerical Methods for Fluid Dynamics}.
\newblock Springer-Verlag, 1999.

\bibitem{ben2003generalized}
M.~Ben-Artzi and J.~Falcovitz.
\newblock {\em Generalized Riemann Problems in Computational Fluid Dynamics}.
\newblock Cambridge University Press, 2003.

\bibitem{Gokhale14}
N.~Gokhale.
\newblock Dynamic mesh generation for fluid structure interaction.
\newblock {\em MPhil Thesis}, 2014.

\bibitem{zdravkovich1997flow}
M.~M. Zdravkovich.
\newblock Flow around circular cylinders.
\newblock {\em Fundamentals}, 1:566--571, 1997.

\bibitem{bryson1961diffraction}
A.~E. Bryson and R.~W.~F. Gross.
\newblock Diffraction of strong shocks by cones, cylinders, and spheres.
\newblock {\em Journal of Fluid Mechanics}, 10(01):1--16, 1961.

\bibitem{Schardin57}
H.~Schardin.
\newblock High frequency cinematography in the shock tube.
\newblock {\em Journal of Photographic Science}, 5:19, 1957.

\bibitem{chang2000shock}
S-M. Chang and K-S. Chang.
\newblock On the shock--vortex interaction in {S}chardin's problem.
\newblock {\em Shock Waves}, 10(5):333--343, 2000.

\bibitem{venkatakrishnan1995implicit}
V.~Venkatakrishnan and D.~J. Mavriplis.
\newblock {\em Implicit method for the computation of unsteady flows on
  unstructured grids}.
\newblock Institute for Computer Applications in Science and Engineering, NASA
  Langley Research Center, 1995.

\bibitem{kirshman2006flutter}
D.~J Kirshman and F.~Liu.
\newblock Flutter prediction by an {E}uler method on non-moving {C}artesian
  grids with gridless boundary conditions.
\newblock {\em Computers \& Fluids}, 35(6):571--586, 2006.

\bibitem{landon1982compendium}
R.~H. Landon.
\newblock Compendium of unsteady aerodynamic measurements.
\newblock {\em AGARD Report}, 702, 1982.

\bibitem{schmidt1985noise}
EM~Schmidt and SJ~Duffy.
\newblock Noise from shock tube facilities.
\newblock In {\em 23rd AIAA Aerospace Sciences Meeting}, 1985.

\bibitem{wang1990numerical}
JCT Wang and GF~Widhopf.
\newblock Numerical simulation of blast flowfields using a high resolution tvd
  finite volume scheme.
\newblock {\em Computers \& Fluids}, 18(1):103--137, 1990.

\bibitem{batten1997choice}
Paul Batten, Nicholas Clarke, Claire Lambert, and Derek~M Causon.
\newblock On the choice of wavespeeds for the hllc riemann solver.
\newblock {\em SIAM Journal on Scientific Computing}, 18(6):1553--1570, 1997.

\end{thebibliography}

\end{document}